\documentclass{article}

\usepackage{arxiv}

\usepackage[utf8]{inputenc} 
\usepackage[T1]{fontenc}    
\usepackage{hyperref}       
\usepackage{url}            
\usepackage{amsmath}   
\usepackage{booktabs}       
\usepackage{amsfonts}       
\usepackage{nicefrac}       
\usepackage{microtype}      
\usepackage{cleveref}       
\usepackage{lipsum}         
\usepackage{graphicx}
\usepackage{natbib}
\usepackage{doi}
\usepackage{arydshln}
\usepackage{caption}
\usepackage{float}

\title{A Novel Approach to Identifying Open Star Cluster Members in {\it Gaia} DR3: Integrating MST and GMM Techniques}

\newif\ifuniqueAffiliation
\uniqueAffiliationtrue

\ifuniqueAffiliation 
\author{Rafe Sharif \\
	Amirkabir University of Technology\\
	Tehran \\
	\texttt{rafe.sharif@aut.ac.ir} \\
	\And
	\href{https://orcid.org/0000-0003-0206-4668}{\includegraphics[scale=0.06]{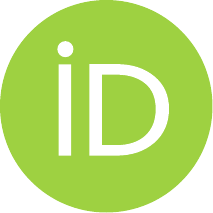}\hspace{1mm}M. Khakian Ghomi} \\
	Department of Physcis and Energy Engineering\\
	Amirkabir University of Technology\\
	Tehran \\
	\texttt{khakian@aut.ac.ir} \\
	 \AND
	 M. Taefi \\
	 Amirkabir University of Technology \\
	 Tehran \\
	 \texttt{m.a.taefi.g@aut.ac.ir} \\
}
\else
\usepackage{authblk}

\newbox{\orcid}\sbox{\orcid}{\includegraphics[scale=0.06]{orcid.pdf}} 
\affil[1]{Department of Physcis and Energy Engineering, Amirkabir University of Technology, Tehran}
\fi

\renewcommand{\shorttitle}{\textit{arXiv} Template}

\hypersetup{
pdftitle={MST-GMM techniques on Gaia DR3},
pdfauthor={Rafe Sharif, M. Khakian Ghomi},
pdfkeywords={Open Clusters, Gaia DR3, MST, GMM},
}

\begin{document}
\maketitle

\begin{abstract}
	We present a novel approach for identifying members of open star clusters using Gaia DR3 data by combining Minimum Spanning Tree (MST) and Gaussian Mixture Model (GMM) techniques. Our method employs a three-step process: initial filtering based on astrometric parameters, MST analysis for spatial distribution filtering, and GMM for final membership probability determination. We tested this methodology on 12+1 open clusters of varying ages, distances, and richness. The method demonstrates superior performance in distinguishing cluster members from field stars, particularly in regions with overlapping populations, as evidenced by its application to clusters like NGC 7790. By effectively reducing the number of probable field stars through MST analysis before applying GMM, our approach enhances both computational efficiency and membership determination accuracy. The results show strong agreement with previous studies while offering improved precision in member identification. This method provides a robust framework for analyzing the extensive datasets provided by Gaia DR3, addressing the challenges of processing large-scale astronomical data while maintaining high accuracy in cluster membership determination.
\end{abstract}


\section{Introduction}
	\label{sec1}
	Open star clusters (OCs) are fundamental astrophysical systems that serve as the birthplaces of many stars within our Galaxy\citep{miller_birthplaces_1978}. These gravitationally bound systems typically comprise anywhere from a few dozen to several thousand stars, making them ideal laboratories for studying stellar evolution and dynamics \citep{portegies_zwart_star_2001}. Identifying the members of OCs is crucial for determining their fundamental properties, such as distance, metallicity, age, and mass distributions \citep{kalirai_cfht_2003}. To achieve accurate membership determinations, we require precise astrometric and photometric data for each star within the cluster's vicinity. The {\it Gaia} Data Release 3 (DR3) provides an unprecedented wealth of reliable astrometric and photometric measurements for approximately 1.8 billion celestial objects \citep{gaia_collaboration_gaia_2023, babusiaux_gaia_2023}. This extensive dataset not only enables more precise identification of OC members than ever before but also facilitates the detection of subgroups within these clusters, enhancing our understanding of their formation and evolutionary processes.
	
	The vast volume of data provided by {\it Gaia} DR3 renders traditional methods for analyzing astronomical data practically inefficient \citep{eyer_gaia_2023}. This is where Machine Learning (ML) and Data Mining techniques become essential \citep{bissekenov_cluster_2024}. ML offers effective approaches for determining the membership of star clusters. For instance, the Density-Based Spatial Clustering of Applications with Noise (DBSCAN) algorithm has been applied by \citet{raja_membership_2024} and \citet{gao_membership_2014} to identify cluster members. Another unsupervised ML method, Gaussian Mixture Modeling (GMM), was explored by \citet{mahmudunnobe_using_2024}. Furthermore, \citep{hunt_improving_2023} utilized HDBSCAN to identify open clusters and their members within the {\it Gaia} DR3 data set. 
	
	In contrast to unsupervised methods, some researchers have employed supervised techniques for determining membership. \citet{castro-ginard_hunting_2020} trained an Artificial Neural Network (ANN) to analyze 1,867 clusters in {\it Gaia} DR2, while \citet{gao_machine-learning-based_2018} implemented Random Forest algorithms on the same dataset to investigate M67. Combining multiple ML algorithms in sequence is also common practice. For example in \citet{noormohammadi_membership_2023}, we used both DBSCAN and GMM to investigate 12 open clusters in {\it Gaia} EDR3. Similarly, \citep{castro-ginard_hunting_2020} first applied DBSCAN to identify dense regions of the Galaxy before employing a deep neural network to determine cluster members based on isochrone patterns in color-magnitude diagrams. \citet{gao_membership_2019} used GMM and Random Forest to investigate the membership of Praesepe. In \citet{noormohammadi_membership_2024}, we utilized DBSCAN and GMM algorithms to determine the membership of open star clusters and developed a dataset to train a Random Forest model for identifying cluster members beyond the tidal radius.
	
	In this study, we introduce a novel approach for determining the membership of open star clusters using a sequential application of Minimum Spanning Tree (MST) and GMM, applied to the selected clusters from the latest {\it Gaia} data release \citep{gaia_collaboration_gaia_2023}. This method offers several advantages over existing techniques. Firstly, the MST algorithm effectively reduces the number of probable field stars by analyzing their spatial distribution, which enhances the initial filtering process. This step is crucial for minimizing noise and improving the accuracy of subsequent analyses.
	
	The GMM then leverages multiple parameters to estimate membership probabilities, providing a robust probabilistic framework that distinguishes cluster members from field stars with high confidence. This dual-step approach not only improves the precision of membership determination but also enhances computational efficiency by systematically narrowing down the dataset before applying complex modeling techniques.
	
	Our method demonstrates superior performance in accurately identifying cluster members, even in regions with overlapping populations, as evidenced by its application to clusters like NGC 7790 (for a detailed discussion, see Subsection \ref{subsec:ngc7790}). This efficiency and accuracy make our approach a valuable tool for analyzing the extensive datasets provided by {\it Gaia} DR3, offering a significant improvement over traditional and other machine learning methods.
	
	A comprehensive description of the data utilized in this study is provided in Section \ref{sec:data}. The methodologies employed, including our novel approach, are detailed in Section \ref{sec:method}. In Section \ref{sec:results}, we present and compare the results of our method with those obtained by \citet{hunt_improving_2024}, incorporating analyses such as the King profile and color-magnitude diagrams (CMDs). Finally, Section \ref{sec:conc} offers a discussion of the results, along with concluding remarks and potential implications for future research.The analysis was conducted using custom scripts available on GitHub\footnote{The code used for this research is available at \url{https://github.com/Astrolab-AUT/MST-GMM-membersip}.}.

\section{Data}
	\label{sec:data}
	
	To test the method described in Section \ref{sec:method}, we selected 12 + 1 open clusters (OCs) spanning a wide range of ages, distances from Earth, and membership richness. The properties of these clusters are summarized in Table \ref{table:1}, with parameters estimated by \citet{hunt_improving_2024}. Using the right ascension ($\alpha$) and declination ($\delta$) of each cluster, we queried the {\it Gaia} DR3 database to extract proper motions ($\mu_{\alpha}$ and $\mu_{\delta}$) and parallaxes ($\varpi$), refining our initial dataset as described in Section \ref{sec:method}. The query excluded stars fainter than magnitude 20 and those with negative parallaxes. Additionally, it selected objects with complete astrometric data ($\alpha$, $\delta$, $\mu_{\alpha}$, $\mu_{\delta}$, and $\varpi$) and photometric data (G-band magnitude and BP-RP color index) to ensure high accuracy. The selected clusters cover ages ranging from $log(t)=6.61$ to $9.43$, distances from $173$ to $3928$ parsecs, and membership counts from $67$ to $3543$ stars.
	
	We present visualizations and detailed results for 6 + 1 OCs in the main text, highlighting the robustness and effectiveness of our methodology. The results for the remaining 6 OCs are provided in the appendix for completeness and to support further exploratory analyses.

	\begin{table*}[h!]
		\centering
		\caption{Properties of Selected Open Clusters reported by \citet{hunt_improving_2024}}
		\label{table:1}
		\begin{tabular}{lccccccc} 
			\hline
			Name & N & $r_{\text{tot}}$ &$\mu_{\alpha}$ & $\mu_{\delta}$ & $\varpi$ & Distance & logAge \\ 
			&     (Prob>0.5)       &            (deg)              & (mas/yr)      & (mas/yr)      & (mas)        & (pc)            &                 \\ \hline
			NGC6231      & 646  & 0.56  & -0.58  & -2.18 & 0.61  & 1555.74 & 6.73  \\ 
			NGC6561      & 89   & 0.25  & -0.00  & -0.79 & 0.67  & 1415.54 & 7.35  \\ 
			NGC7788      & 98   & 0.16  & -3.10  & -1.81 & 0.34  & 2721.01 & 7.47  \\ 
			Alessi34     & 490  & 6.45  & -5.01  & 5.74  & 2.05  & 485.76  & 7.51  \\ 
			NGC2422      & 510 & 3.17  & -7.07  & 1.04  & 2.10  & 468.52  & 8.29  \\ 
			NGC2682      & 807 & 1.68  & -10.97 & -2.91 & 1.15  & 837.90  & 9.42  \\
			NGC2243      & 585 & 0.52  & -1.27  & 5.49  & 0.22  & 3951.94 & 9.62  \\ 
			IC4756       & 397  & 3.25  & 1.28   & -4.97 & 2.11  & 467.01  & 9.13  \\ 
			Melotte20    & 616  & 11.42 & 22.92  & -25.46 & 5.74 & 173.62  & 7.97  \\ 
			NGC2477      & 2081 & 1.21  & -2.43  & 0.91  & 0.69  & 1384.68 & 9.02  \\
			NGC7429      & 67   & 1.50  & 4.77   & -2.24 & 2.37  & 417.84  & 8.14  \\  
			NGC2287      & 598  & 0.99  & -4.37  & -1.36 & 1.37  & 624.17  & 8.14  \\ \hdashline \addlinespace[0.08cm]
			NGC 7790      & 143  & 0.23  & -3.24  & -1.73  & 0.29  & 3127.01 & 7.83  \\
			\hline
		\end{tabular}
	\end{table*}

\section{Method}
	\label{sec:method}

	Our analysis comprises three primary steps: (1) an initial filtering of field stars based on mean values of key astronomical parameters for open clusters, (2) construction and analysis of a MST to further filter field stars, and (3) membership probability estimation using a GMM. The GMM leverages multiple parameters to identify stars with a high likelihood of being cluster members. Each step is described in detail below.
	
	\subsection{Initial Filtering}
	\label{subsec:initial}
	The initial filtering step aims to remove a significant fraction of field stars while retaining all open cluster (OC) stars. This is achieved by applying constraints based on predetermined mean values of three astronomical parameters: $\mu_{\alpha}$, $\mu_{\delta}$, and $\varpi$. These parameter ranges are derived from the work of \citet{hunt_improving_2024}.
	
	For each parameter, \(x\), a filtering criterion is defined as:
	\begin{equation}
		\lvert x_i - \mu_x \rvert \leq \lvert \delta_x \cdot \mu_x \rvert
	\end{equation}
	where \(\mu_x\) is the derived mean value of \(x\), and \(\delta_x\) represents the chosen interval. The interval is calibrated to ensure that all potential cluster stars remain within the sample while excluding the majority of field stars. Typical values for \(\delta_x\) are set to 0.2 for \(\mu_x > 1\) and 0.5 for \(\mu_x < 1\). This step effectively reduces the dataset size and enhances computational efficiency for subsequent analysis.
	
	\subsection[MST]{Minimum Spanning Tree}
	\label{subsec:mst}
	The MST is a fundamental concept in graph theory that provides an efficient representation of connectivity in a dataset. It is defined as a subset of edges in a connected, undirected, weighted graph that connects all nodes without forming cycles and minimizes the total edge weight.
	
	In our approach, stars are treated as nodes in a graph \(\mathbf{G} = (\mathbf{V}, \mathbf{E})\), where \(\mathbf{V}\) represents the set of stars and \(\mathbf{E}\) denotes the set of edges. The edge weights are computed as pairwise distances between stars in a normalized parameter space comprising right ascension, declination, and parallax.
	
	To normalize the data, we employ the \texttt{StandardScaler} module from \texttt{ Scikit-learn}, transforming each parameter to have zero mean and unit variance. The distance between two stars, \(x_i, x_j \in \mathbb{R}^3\), is calculated using the Minkowski distance:
	\begin{equation}
		d(x_i, x_j) = \left( \sum_{k=1}^{3} \lvert x_{i,k} - x_{j,k} \rvert^p \right)^{1/p},
	\end{equation}
	where \(p\) is the Minkowski parameter, typically set to \(p = 2\) for the Euclidean distance. The resulting graph is constructed so that each node is connected to its nearest neighbors and the edge weights represent the distances between them.
	
	From the constructed graph, we compute the MST using Kruskal's algorithm \citep{kruksal_zhang_2023} to minimize the total edge weight:
	\begin{equation}
		\text{Weight}(T) = \sum_{(i,j) \in \mathbf{E}_T} w_{i,j}, \quad \text{where } \mathbf{E}_T \subseteq \mathbf{E}.
	\end{equation}

	The MST is analyzed to identify anomalously long edges, indicative of weak connections or outliers in the data. A threshold for significant edge weights, \(\tau\), is defined as:
	
	\begin{equation}
		\tau = \mu_w + \lambda \sigma_w ,
	\end{equation}
	where \(\mu_w\) and \(\sigma_w\) are the mean and standard deviation of the MST edge weights, and \(\lambda\) is a scaling factor, typically set to 3. Edges with weights exceeding \(\tau\) are removed, and the associated nodes are flagged as field stars. The pruned MST thus delineates the connectivity structure of the OC stars, which are retained for subsequent analysis.
	
	\subsection{Gaussian Mixture Model}
	\label{subsec:gmm}
	The GMM is a probabilistic machine learning model that assumes the data is generated from a mixture of Gaussian distributions, each characterized by its mean and covariance matrix. This method is particularly effective for clustering datasets where the underlying distributions are well-represented by Gaussian components. In our analysis, the GMM is applied to the subset of stars retained after the MST-based filtering step to differentiate cluster members from field stars.
	
	We model the data with two Gaussian components, corresponding to the expected cluster members and field stars, respectively. Previous studies, such as \citet{gao_machine-learning-based_2018} and \cite{agarwal_ml-moc_2021}, have demonstrated the effectiveness of GMM in handling datasets with relatively small sample sizes. Despite the randomness in the distribution of field stars, the inherent normal distribution exhibited by cluster members within the parameter space ensures that GMM can confidently separate the two populations. This observation is supported by \cite{Cabrera-Cano1990} and \cite{de_lichtbuer_astrometric_1971}, who highlight the importance of a high ratio of cluster members to field stars in ensuring robust clustering.
	
	The parameters used for the GMM include $\alpha$, $\delta$, $\mu_{\alpha}$, $\mu_{\delta}$, and $\varpi$. The data is first normalized using a standard scaler, ensuring uniform contribution from each parameter. The GMM is then fitted to the normalized data, and the resulting probabilities are used to assign each star a membership label and a confidence score.

\section{Results}
	\label{sec:results}
\begin{figure*}[htbp] 
		\centering
		\begin{minipage}{0.3\textwidth}
			\includegraphics[width=\linewidth]{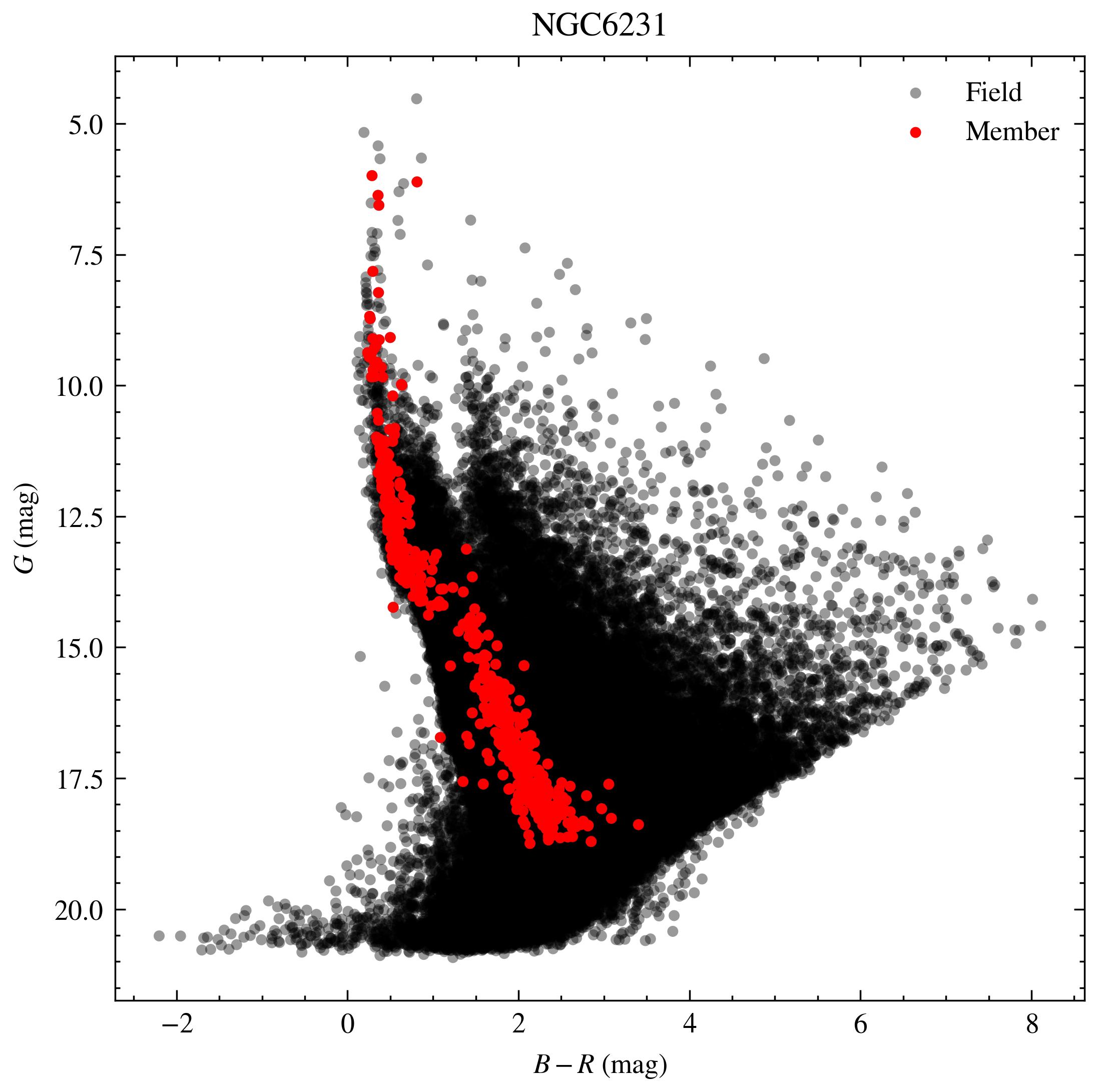} 
			
		\end{minipage}\hfill
		\begin{minipage}{0.3\textwidth}
			\includegraphics[width=\linewidth]{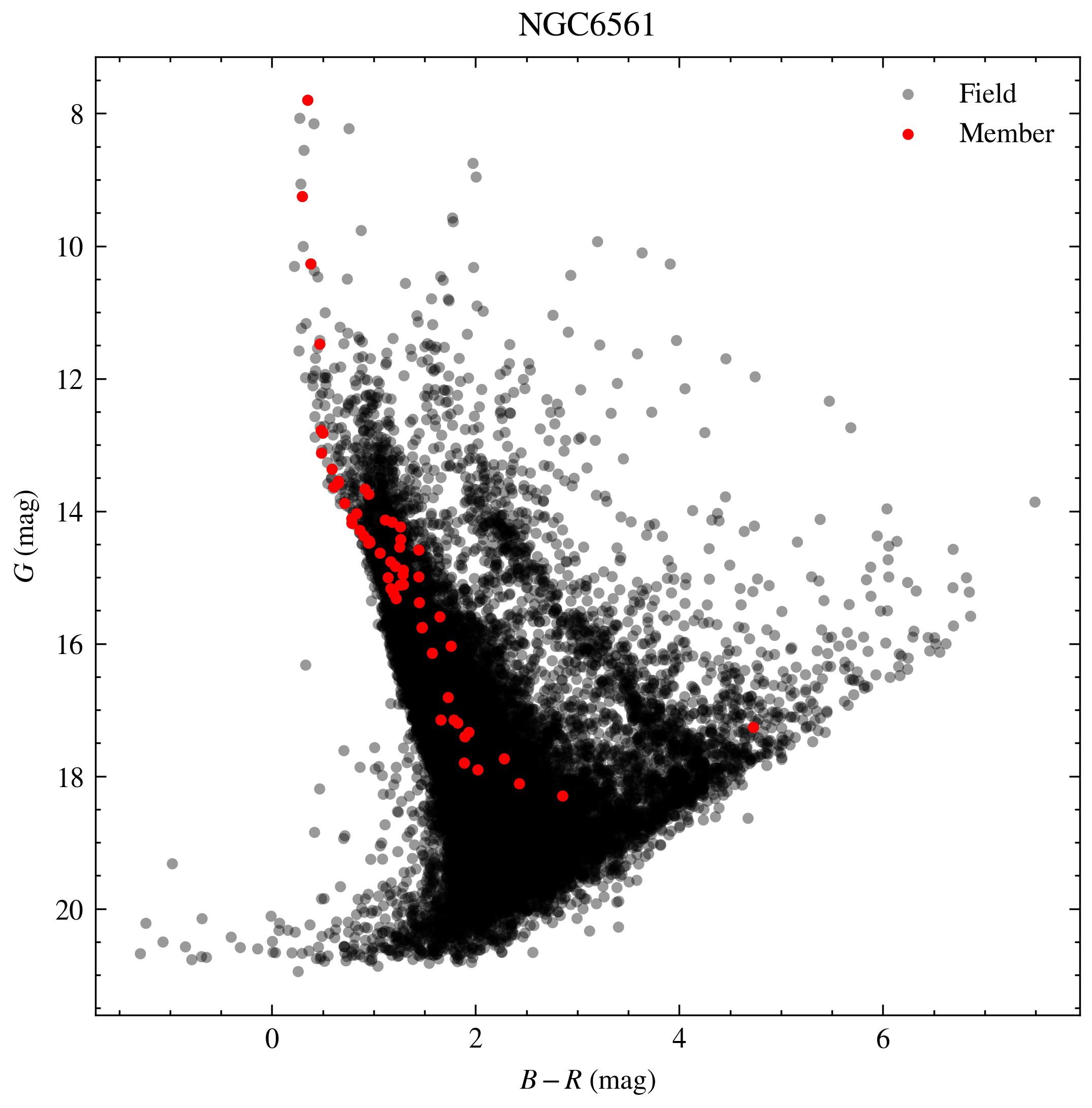} 
			
		\end{minipage}\hfill
		\begin{minipage}{0.3\textwidth}
			\includegraphics[width=\linewidth]{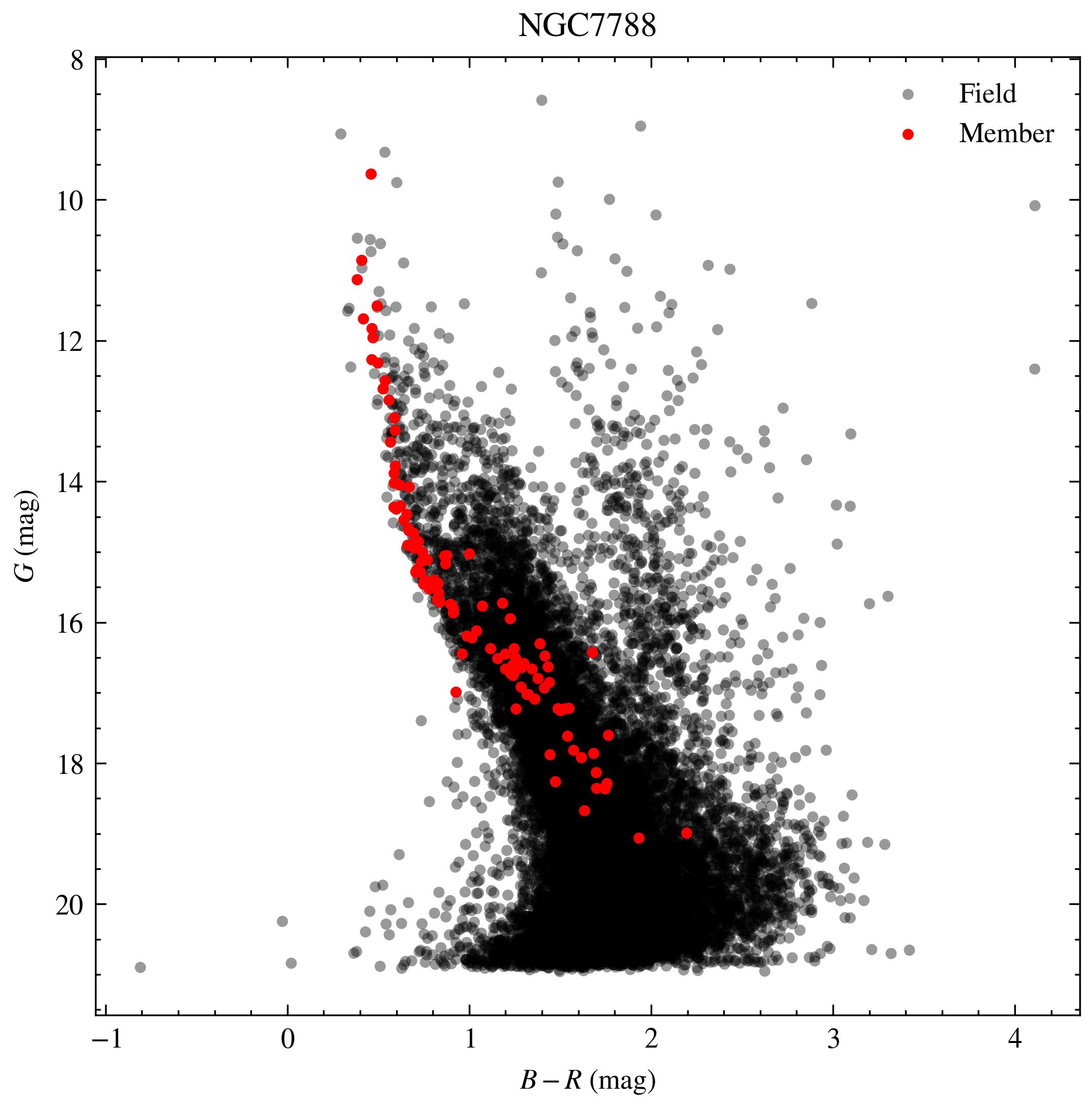} 
			
		\end{minipage}
		
		\vspace{0.5cm} 
		\begin{minipage}{0.3\textwidth}
			\includegraphics[width=\linewidth]{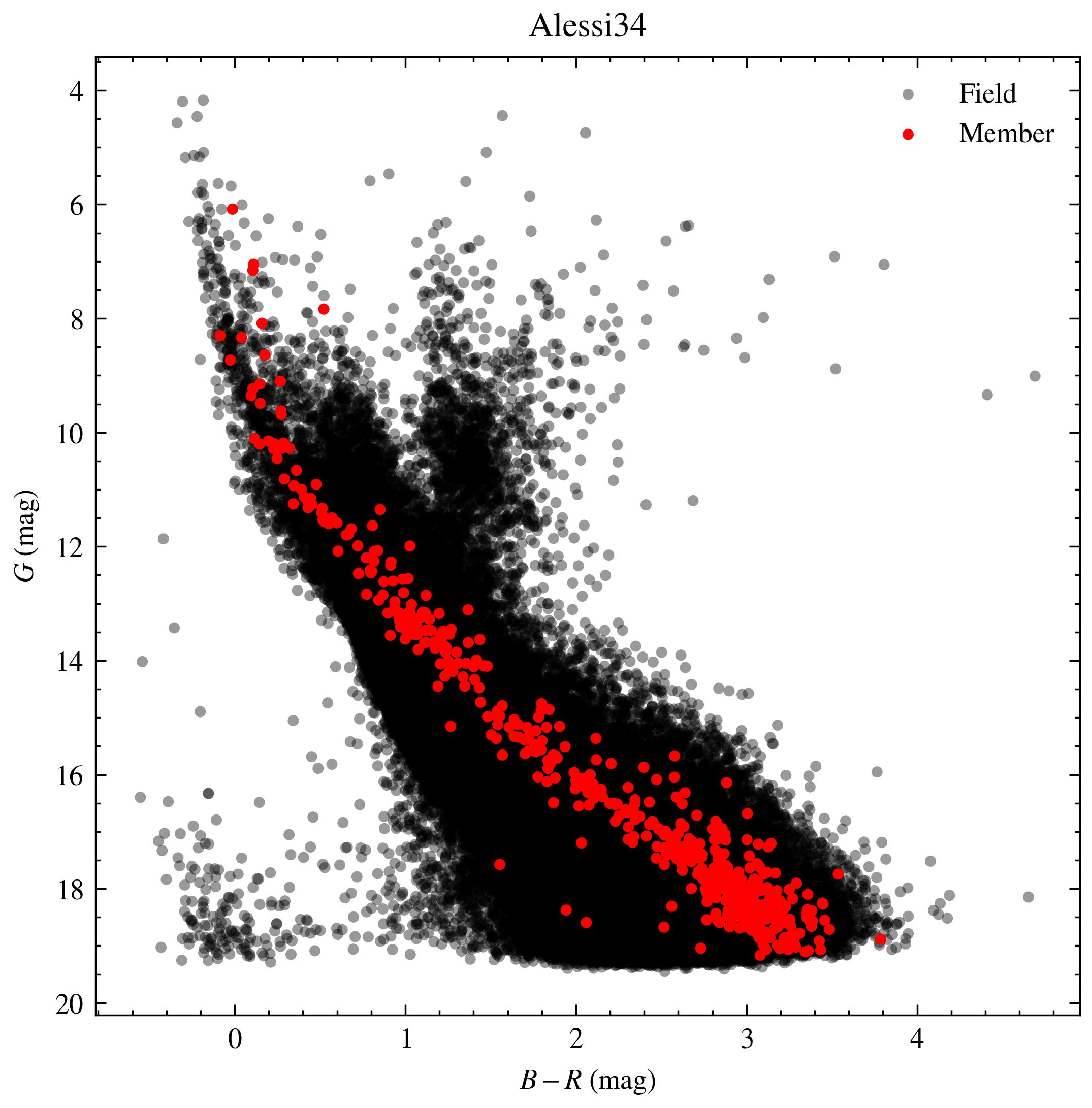} 
			
		\end{minipage}\hfill
		\begin{minipage}{0.3\textwidth}
			\includegraphics[width=\linewidth]{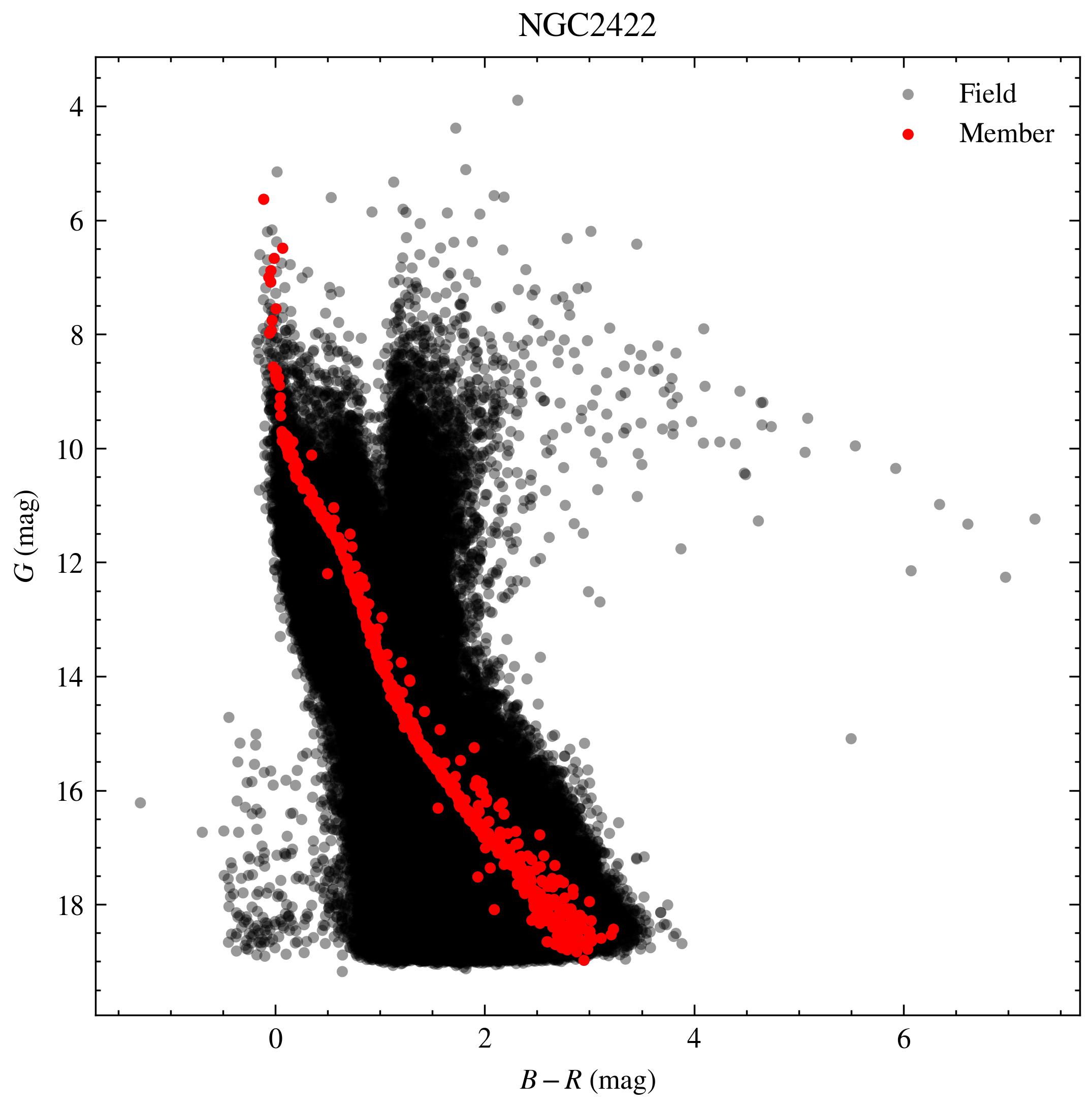} 
			
		\end{minipage}\hfill
		\begin{minipage}{0.3\textwidth}
			\includegraphics[width=\linewidth]{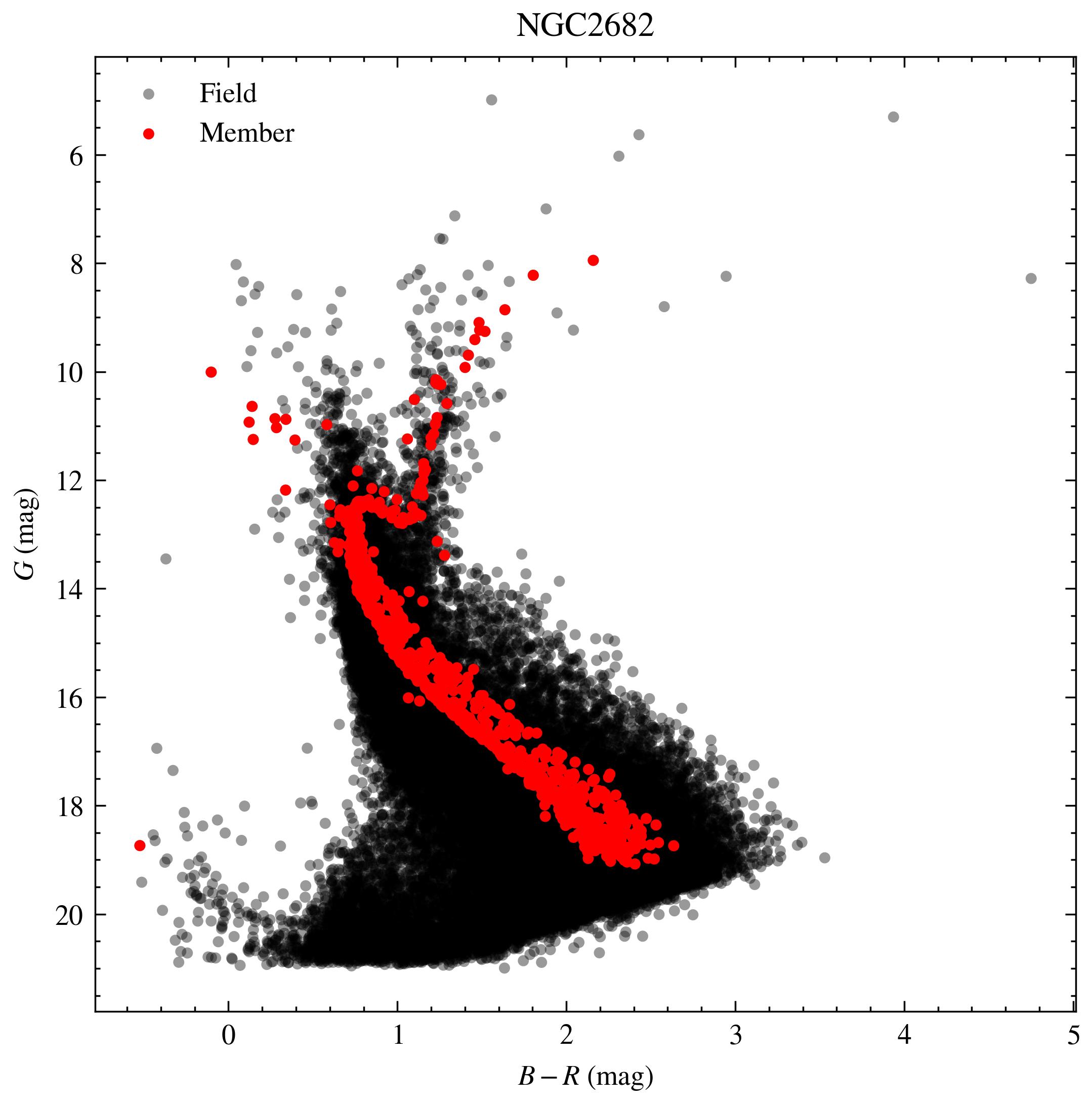} 

		\end{minipage}
		
		\caption{Color-magnitude diagrams illustrating the distribution of cluster members (red dots) and field stars (black dots) within the field of view of six open clusters: NGC 6231, NGC 6561, NGC 7788, Alessi 34, NGC 2422, and NGC 2682.}
		\label{fig:cmdinfield}
		
	\end{figure*}

\begin{figure*}[htbp] 
		\centering
		\begin{minipage}{0.3\textwidth}
			\includegraphics[width=\linewidth]{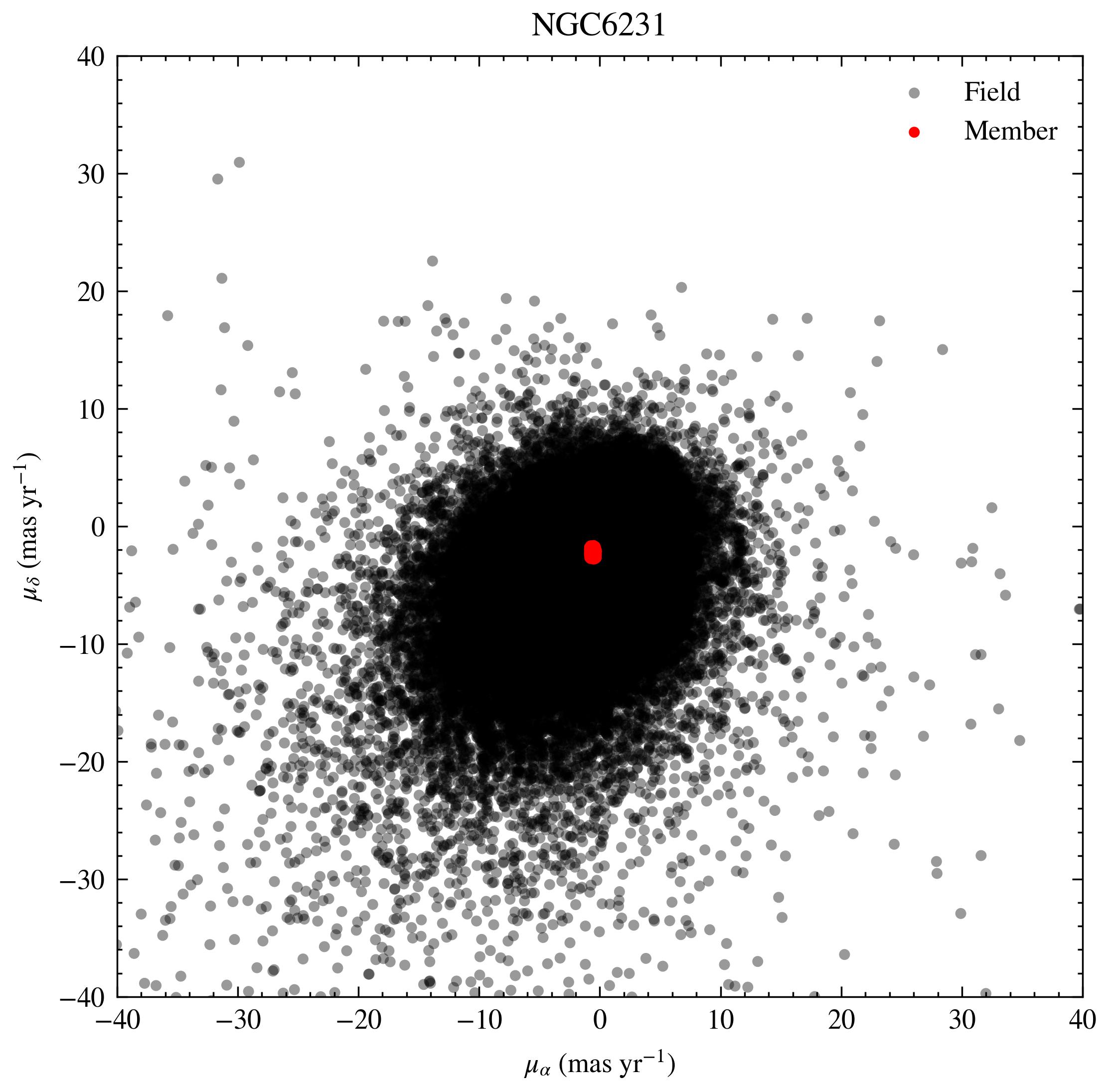} 
			
		\end{minipage}\hfill
		\begin{minipage}{0.3\textwidth}
			\includegraphics[width=\linewidth]{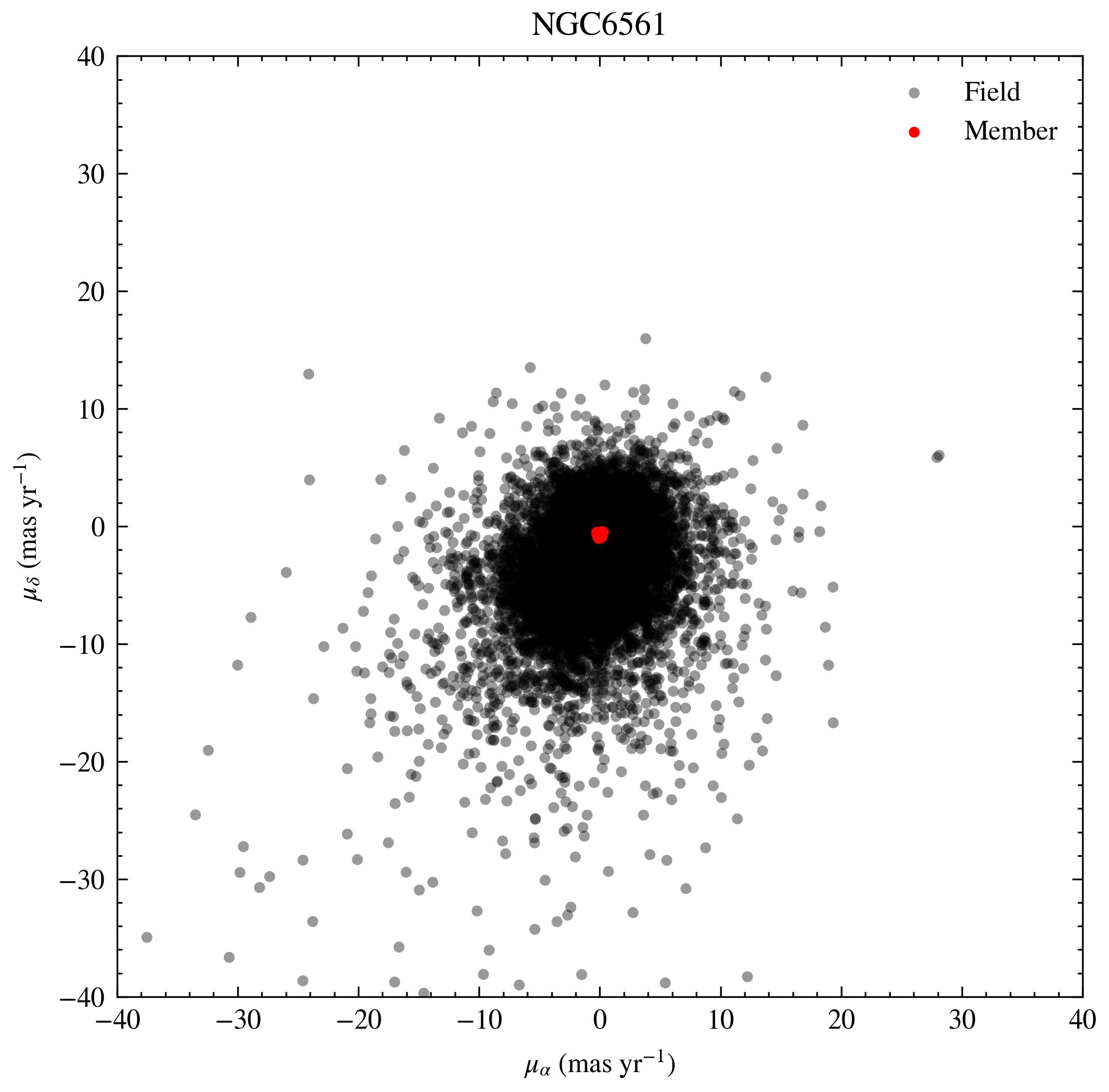} 
			
		\end{minipage}\hfill
		\begin{minipage}{0.3\textwidth}
			\includegraphics[width=\linewidth]{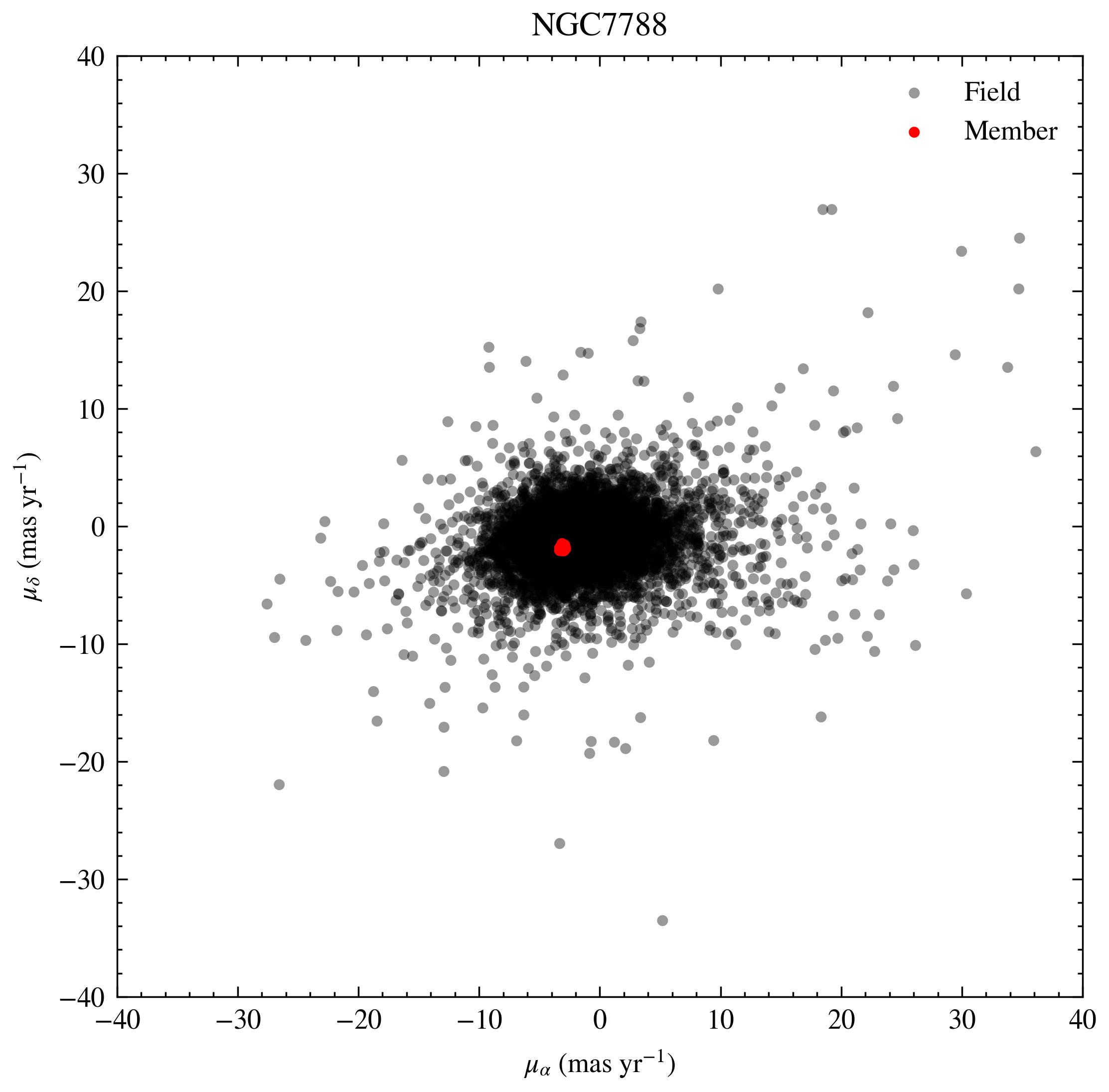} 
			
		\end{minipage}
		
		\vspace{0.5cm} 
		\begin{minipage}{0.3\textwidth}
			\includegraphics[width=\linewidth]{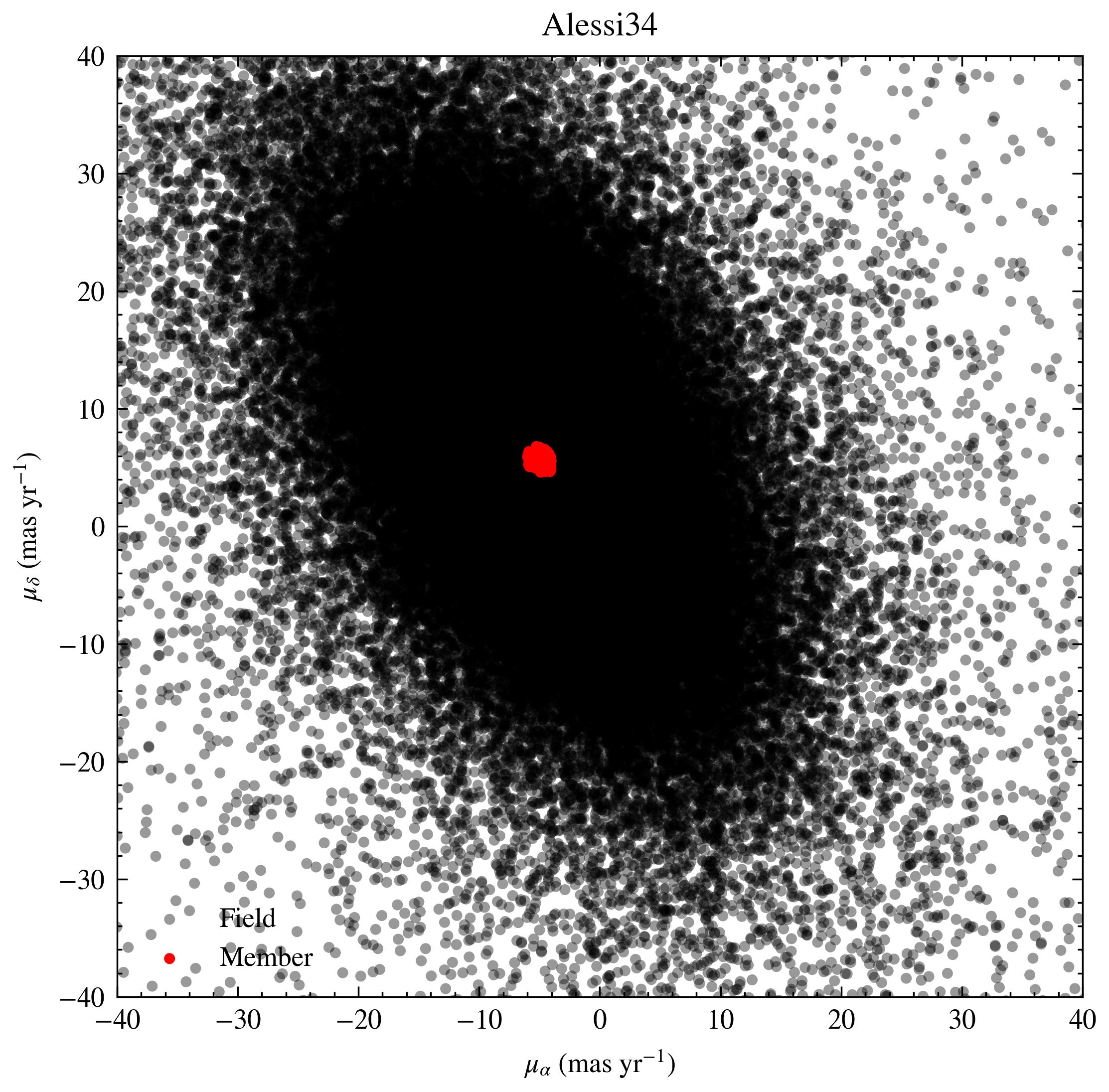} 
			
		\end{minipage}\hfill
		\begin{minipage}{0.3\textwidth}
			\includegraphics[width=\linewidth]{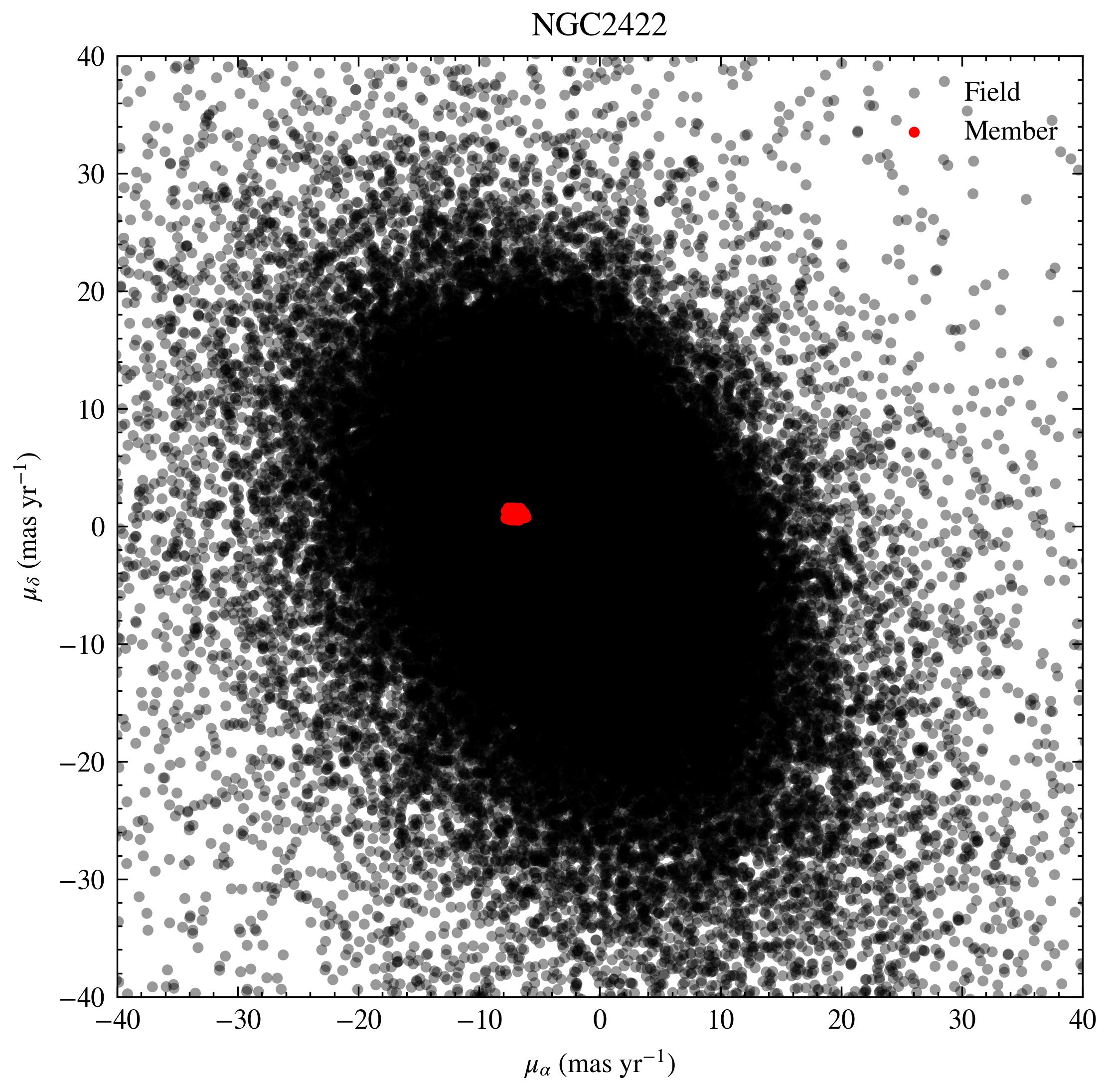} 
			
		\end{minipage}\hfill
		\begin{minipage}{0.3\textwidth}
			\includegraphics[width=\linewidth]{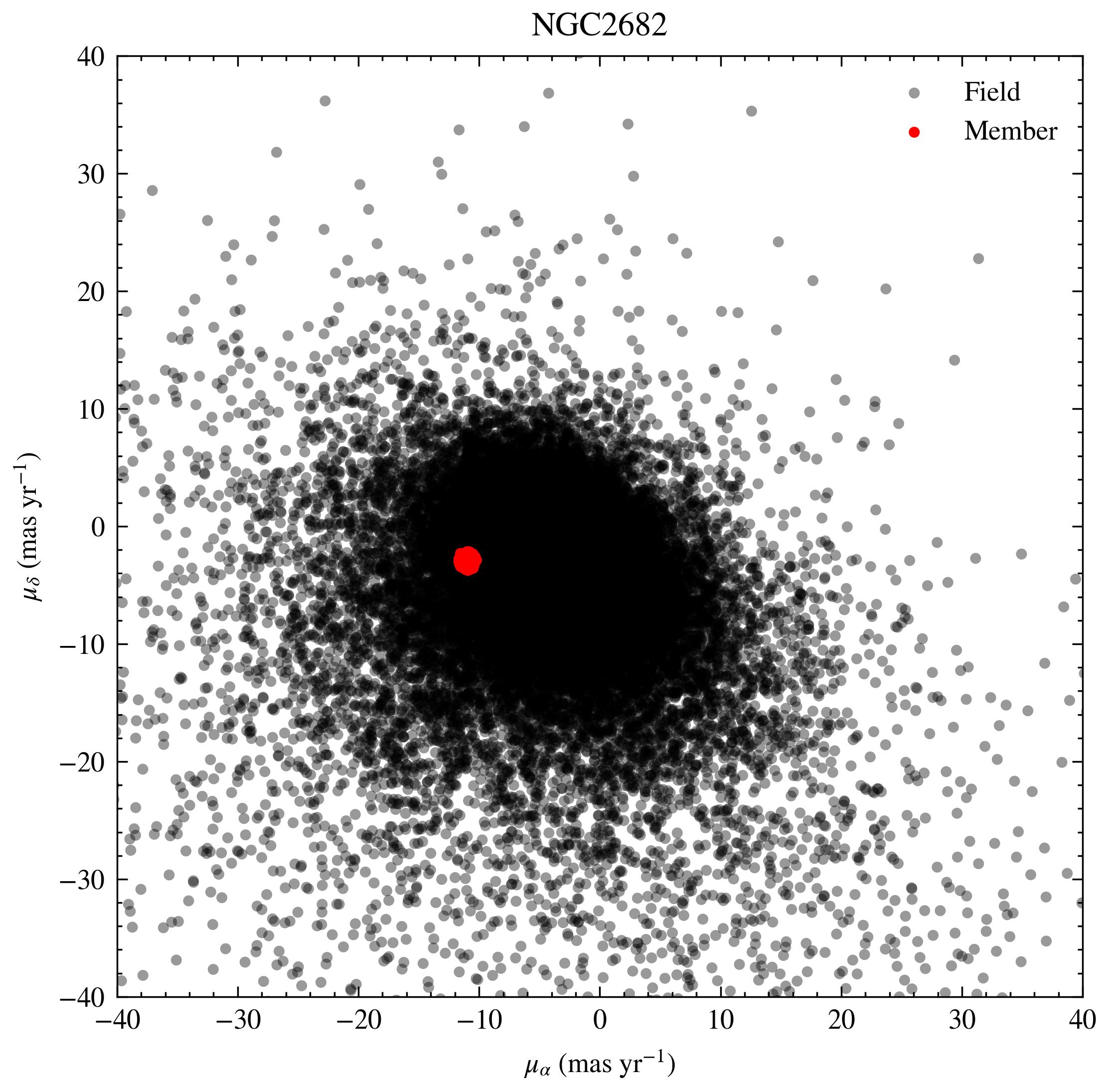} 

		\end{minipage}
		
		\caption{Proper motion diagrams illustrating the spatial distribution of cluster members (red dots) and field stars (black dots) within the field of view of six open clusters: NGC 6231, NGC 6561, NGC 7788, Alessi 34, NGC 2422, and NGC 2682.}
		\label{fig:motioninfield}
		
	\end{figure*}

	As the first step, we applied an initial filter to the proper motion and parallax data to exclude stars unlikely to be cluster members. Using the MST algorithm on the remaining stars, we further reduced the number of probable field stars by analyzing their right ascension, declination, and parallax values. Building on this refined dataset, the GMM was applied to classify stars into cluster members and non-members utilizing five astrometric parameters ($\alpha$, $\delta$, $\mu_{\alpha}$, $\mu_{\delta}$, $\varpi$). Table \ref{table:2} represents the number of stars at each step with the GMM results. Table \ref{table:3} also summarizes the physical parameters estimated for each cluster when GMM applied. Figs. \ref{fig:cmdinfield} and \ref{fig:motioninfield} illustrate color-magnitude diagrams and proper motions of stars for six clusters, respectively. For other six clusters see \ref{sec:app1}.
	
\begin{figure*}[htbp] 
		\centering
		\begin{minipage}{0.3\textwidth}
			\includegraphics[width=\linewidth]{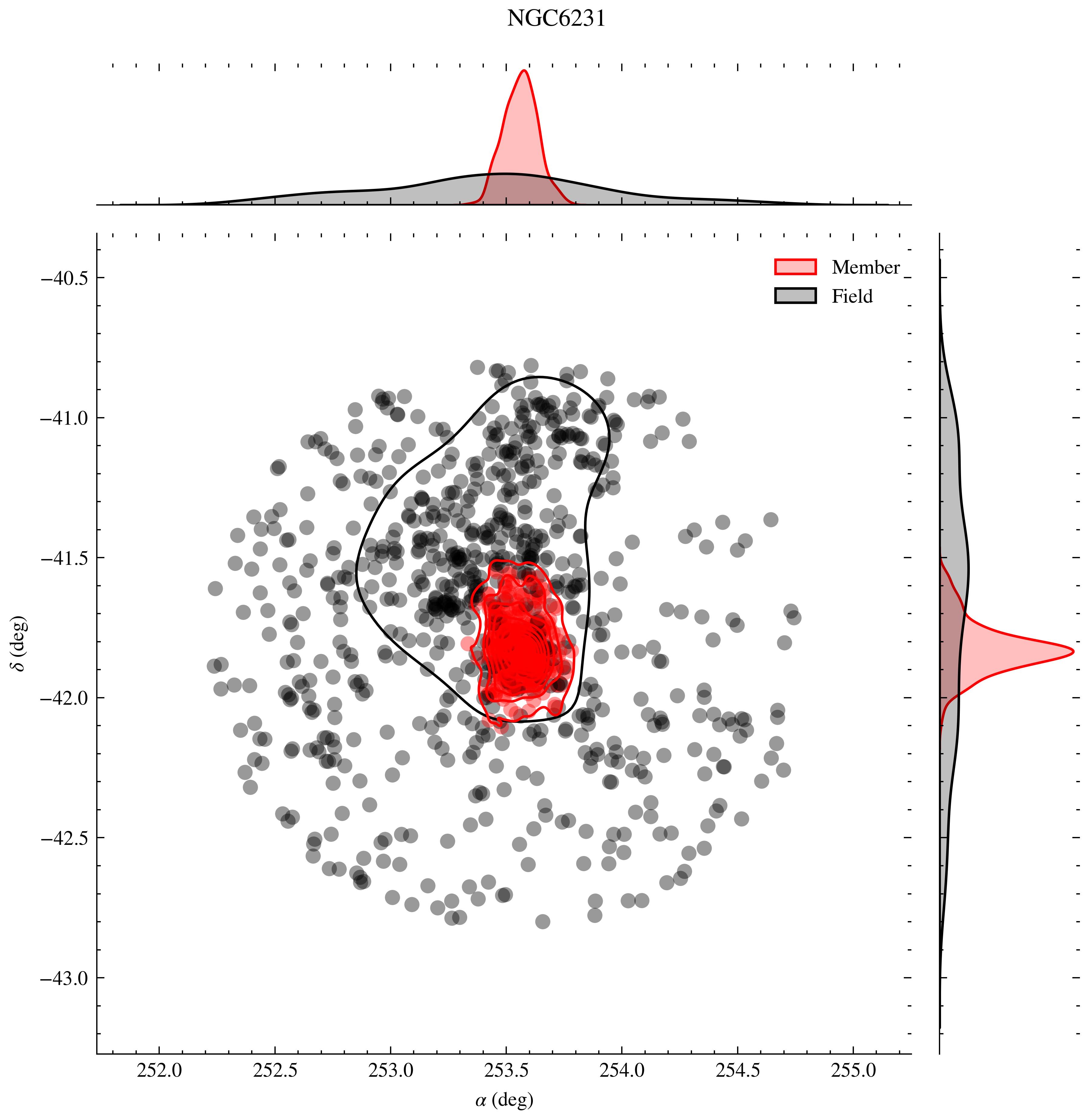} 
			
		\end{minipage}\hfill
		\begin{minipage}{0.3\textwidth}
			\includegraphics[width=\linewidth]{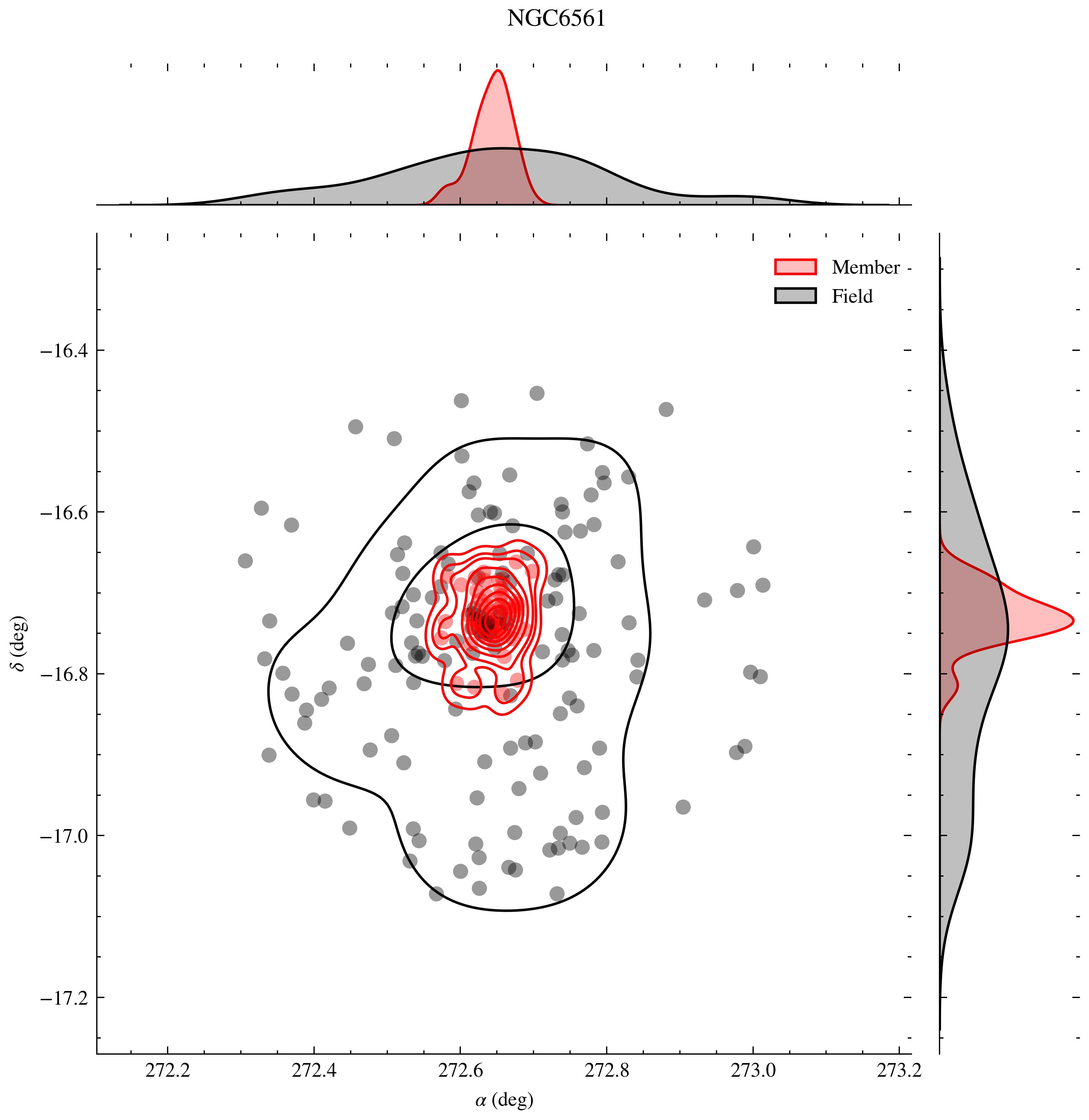} 
			
		\end{minipage}\hfill
		\begin{minipage}{0.3\textwidth}
			\includegraphics[width=\linewidth]{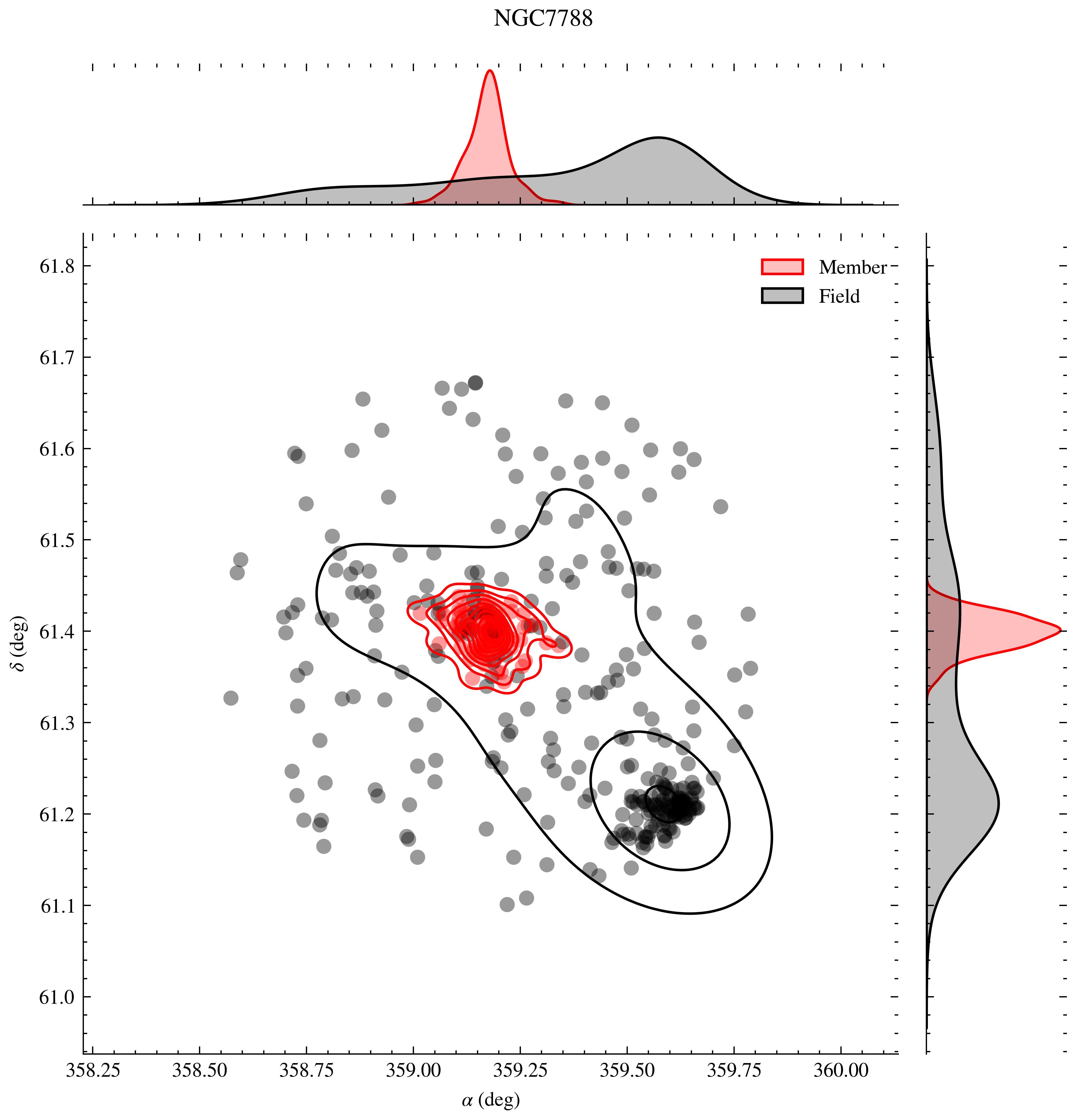} 
			
		\end{minipage}
		
		\vspace{0.5cm} 
		\begin{minipage}{0.3\textwidth}
			\includegraphics[width=\linewidth]{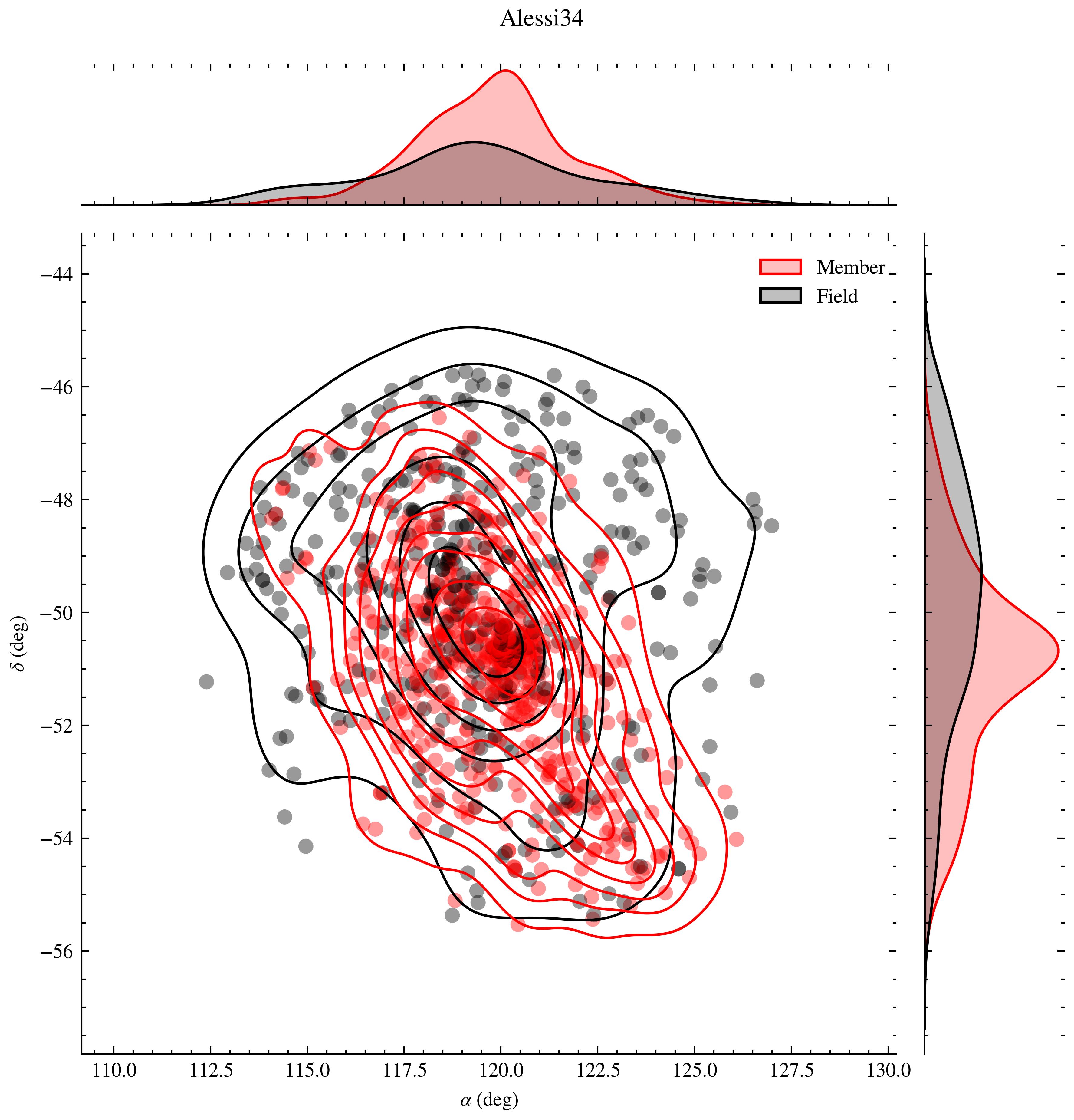} 
			
		\end{minipage}\hfill
		\begin{minipage}{0.3\textwidth}
			\includegraphics[width=\linewidth]{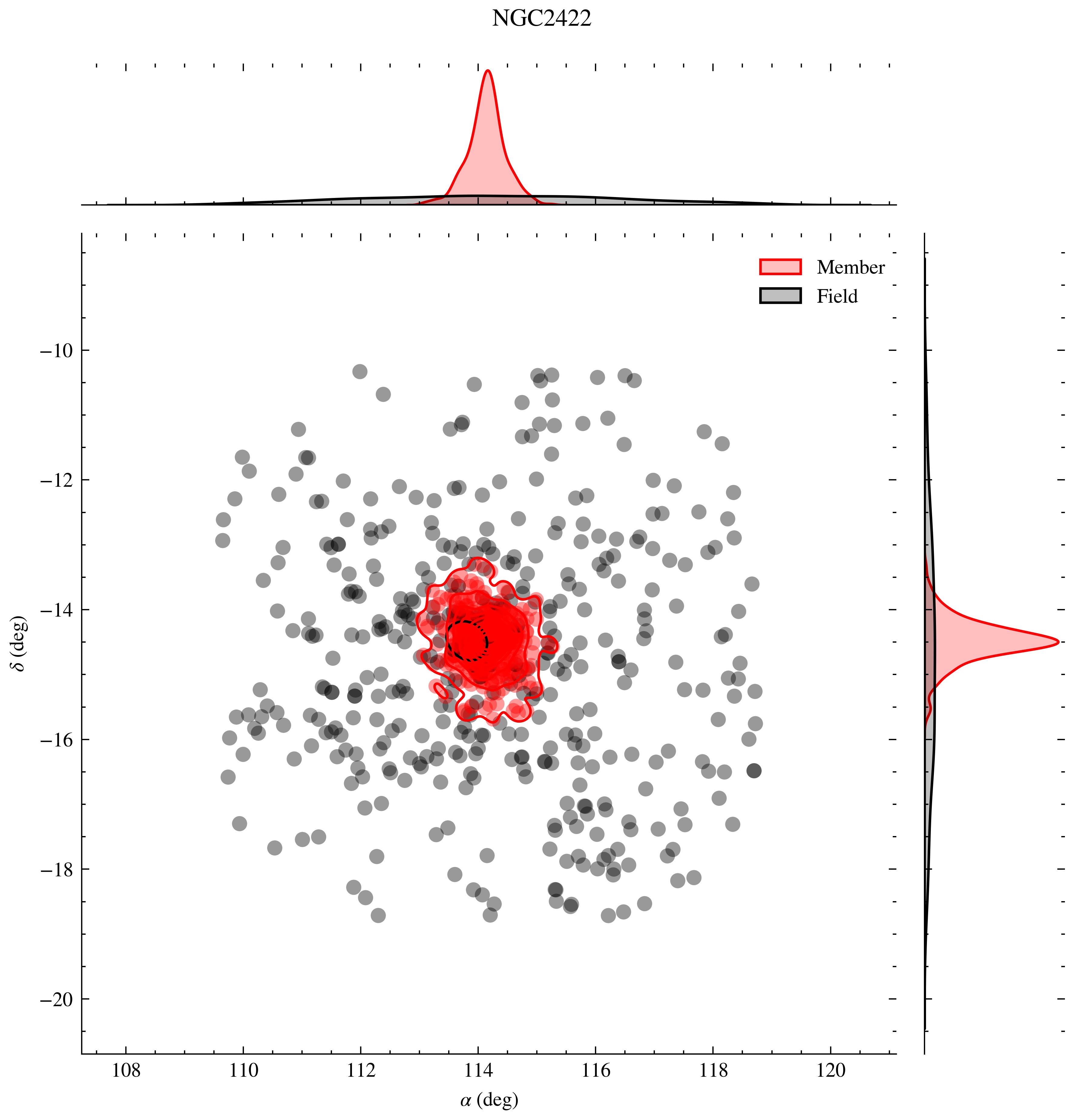} 
			
		\end{minipage}\hfill
		\begin{minipage}{0.3\textwidth}
			\includegraphics[width=\linewidth]{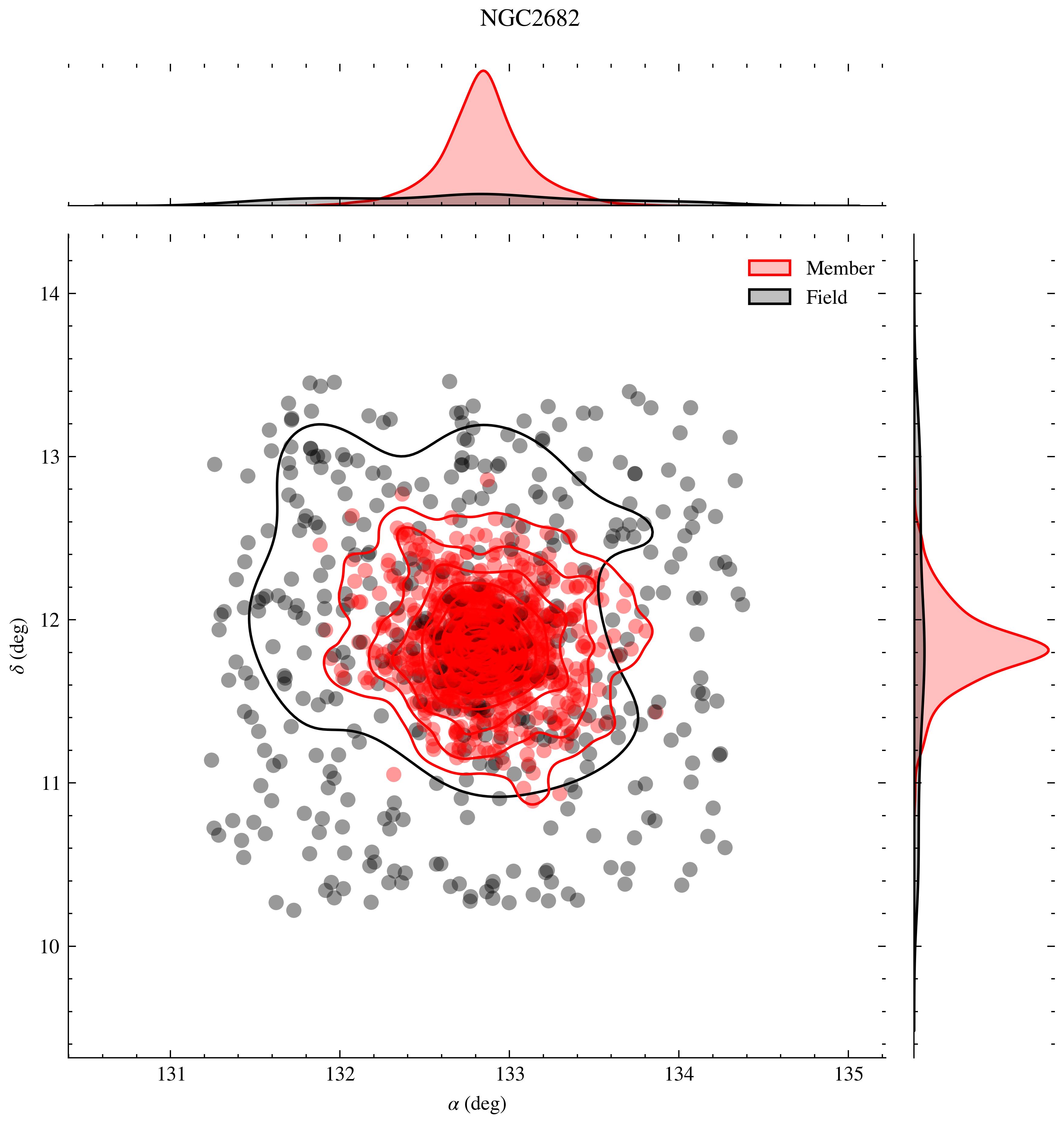} 

		\end{minipage}
		
		\caption{Kernel Density Estimator (KDE) plots illustrating the distribution of cluster members within the field of view of six open clusters: NGC 6231, NGC 6561, NGC 7788, Alessi 34, NGC 2422, and NGC 2682.}
		\label{fig:kde}
		
	\end{figure*}

	After classifying stars into cluster members and non-members using GMM, based on their kinematic properties, Kernel Density Estimation (KDE) plots (shown in Figure \ref{fig:kde} and \ref{fig:kde2}) were employed to visualize the distributions of these two groups within the parameter space. KDE plots reveal distinct peaks for cluster members, indicating a concentrated distribution that aligns with the expected characteristics of the clusters. In contrast, non-members exhibit a more dispersed distribution, reflecting their varied origins and properties. This differentiation underscores the effectiveness of GMM in accurately classifying stars within the cluster environment. Additionally, KDE plots provide visual confirmation of the GMM results while offering deeper insights into the cluster's structure and the spatial distribution of its members. These findings contribute to a refined characterization of the cluster, facilitating further astrophysical analyses and interpretations.
	
	\begin{table*}[!ht]
		\begin{minipage}{1\linewidth}
			\centering
			\caption{Initial Data and Analysis Results for Various Star Clusters: Metrics Including MST, GMM, and Probability Thresholds}
			\label{table:2}
			\begin{tabular}{llllll}
				\toprule
				Name & Init Data & Init Filter & MST & GMM & Prob$>$0.8 \\ 
				\midrule  
				NGC6231 & 225870 & 5318 & 1246 & 499 & 444 \\ 
				NGC6561 & 22417 & 366 & 199 & 58 & 51 \\ 
				NGC7788 & 18835 & 886 & 411 & 111 & 99 \\ 
				Alessi34 & 148224 & 1211 & 924 & 538 & 451 \\ 
				NGC2422 & 521586 & 1336 & 1087 & 618 & 591 \\ 
				NGC2682 & 31288 & 2279 & 1826 & 1284 & 1246 \\ 
				NGC2243 & 287394 & 1535 & 1234 & 776 & 762 \\ 
				IC4756 & 1991520 & 962 & 795 & 469 & 435 \\ 
				Melotte20 & 103171 & 1777 & 1393 & 763 & 733 \\ 
				NGC2477 & 1579626 & 5438 & 4497 & 3142 & 3009 \\ 
				NGC7429 & 204697 & 122 & 108 & 73 & 70 \\ 
				NGC2287 & 418836 & 995 & 917 & 697 & 662 \\ \hdashline \addlinespace[0.08cm] 
				NGC7790 & 8932 & 594 & 392 & 146 & 122 \\
				\bottomrule
			\end{tabular}
		\end{minipage}
		
	\end{table*}
	
	\begin{table*}[h!]
		\begin{minipage}{1\linewidth}
			\centering
			\caption{Estimated astrometric parameters for selected clusters.}
			\label{table:3}

			\begin{tabular}{lcccccc}
				\toprule
				Name & $\alpha$ (deg) & $\delta$ (deg) & $\varpi$ (mas) & $\mu_{\alpha}$ (mas/yr) & $\mu_{\delta}$ (mas/yr) \\ \midrule
				NGC6231 & $253.56 \pm 0.07$ & $-41.82 \pm 0.05$ & $0.61 \pm 0.08$ & $-0.58 \pm 0.10$ & $-2.19 \pm 0.08$ \\
				NGC6561 & $272.64 \pm 0.07$ & $-16.73 \pm 0.06$ & $0.67 \pm 0.08$ & $0.01 \pm 0.08$ & $-0.78 \pm 0.07$ \\
				NGC7788 & $359.17 \pm 0.05$ & $61.40 \pm 0.05$ & $0.34 \pm 0.06$ & $-3.09 \pm 0.06$ & $-1.81 \pm 0.06$ \\
				Alessi34 & $119.78 \pm 0.07$ & $-51.04 \pm 0.07$ & $2.03 \pm 0.07$ & $-4.93 \pm 0.09$ & $5.63 \pm 0.09$ \\
				NGC2422 & $114.14 \pm 0.05$ & $-14.49 \pm 0.05$ & $2.11 \pm 0.07$ & $-7.06 \pm 0.06$ & $1.03 \pm 0.06$ \\
				NGC2682 & $132.85 \pm 0.06$ & $11.84 \pm 0.03$ & $1.15 \pm 0.07$ & $-10.96 \pm 0.07$ & $-2.91 \pm 0.06$ \\
				NGC2243 & $97.39 \pm 0.06$ & $-31.28 \pm 0.06$ & $0.22 \pm 0.07$ & $-1.26 \pm 0.08$ & $5.50 \pm 0.08$ \\
				IC4756 & $279.66 \pm 0.05$ & $5.44 \pm 0.05$ & $2.12 \pm 0.06$ & $1.28 \pm 0.06$ & $-4.97 \pm 0.06$ \\
				Melotte20 & $51.82 \pm 0.05$ & $48.89 \pm 0.05$ & $5.74 \pm 0.07$ & $22.81 \pm 0.07$ & $-25.45 \pm 0.07$ \\
				NGC2477 & $118.05 \pm 0.05$ & $-38.54 \pm 0.05$ & $0.70 \pm 0.06$ & $-2.43 \pm 0.06$ & $0.90 \pm 0.07$ \\
				NGC7429 & $343.99 \pm 0.05$ & $59.84 \pm 0.05$ & $2.38 \pm 0.06$ & $4.74 \pm 0.07$ & $-2.23 \pm 0.06$ \\
				NGC2287 & $101.52 \pm 0.03$ & $-20.71 \pm 0.04$ & $1.37 \pm 0.06$ & $-4.37 \pm 0.04$ & $-1.36 \pm 0.05$ \\  \hdashline \addlinespace[0.08cm] 
				NGC7790 & $359.61 \pm 0.05$ & $61.20 \pm 0.05$ & $0.29 \pm 0.06$ & $-3.24 \pm 0.06$ & $-1.73 \pm 0.06$ \\   \bottomrule
			\end{tabular}
		\end{minipage}
		
	\end{table*}
	\clearpage
	
	To further evaluate the membership of the clusters under study, we computed the King surface density profile as formulated by \citet{King1962}. This profile is instrumental in characterizing the spatial distribution of stars within a cluster, offering critical insights into the cluster's structure and dynamics. The King profile is expressed by the equation:
	\begin{equation}
		f(r) = f_b + \frac{f_0}{1 + \left(\frac{r}{R_c}\right)^2},
	\end{equation}
	where $f(r)$ represents the surface density at a distance $r$ from the cluster center, $f_b$ is the background density, $f_0$ is the central surface density, and $R_c$ is the core radius of the cluster. The core radius, $R_c$, provides a measure of the cluster's concentration, with smaller values of $R_c$ indicating a more tightly bound cluster. The background density, $f_b$, accounts for the contribution of field stars, ensuring that the derived membership is robust against contamination.
	
\begin{figure}[H] 
		\centering
		\begin{minipage}{0.3\textwidth}
			\includegraphics[width=\linewidth]{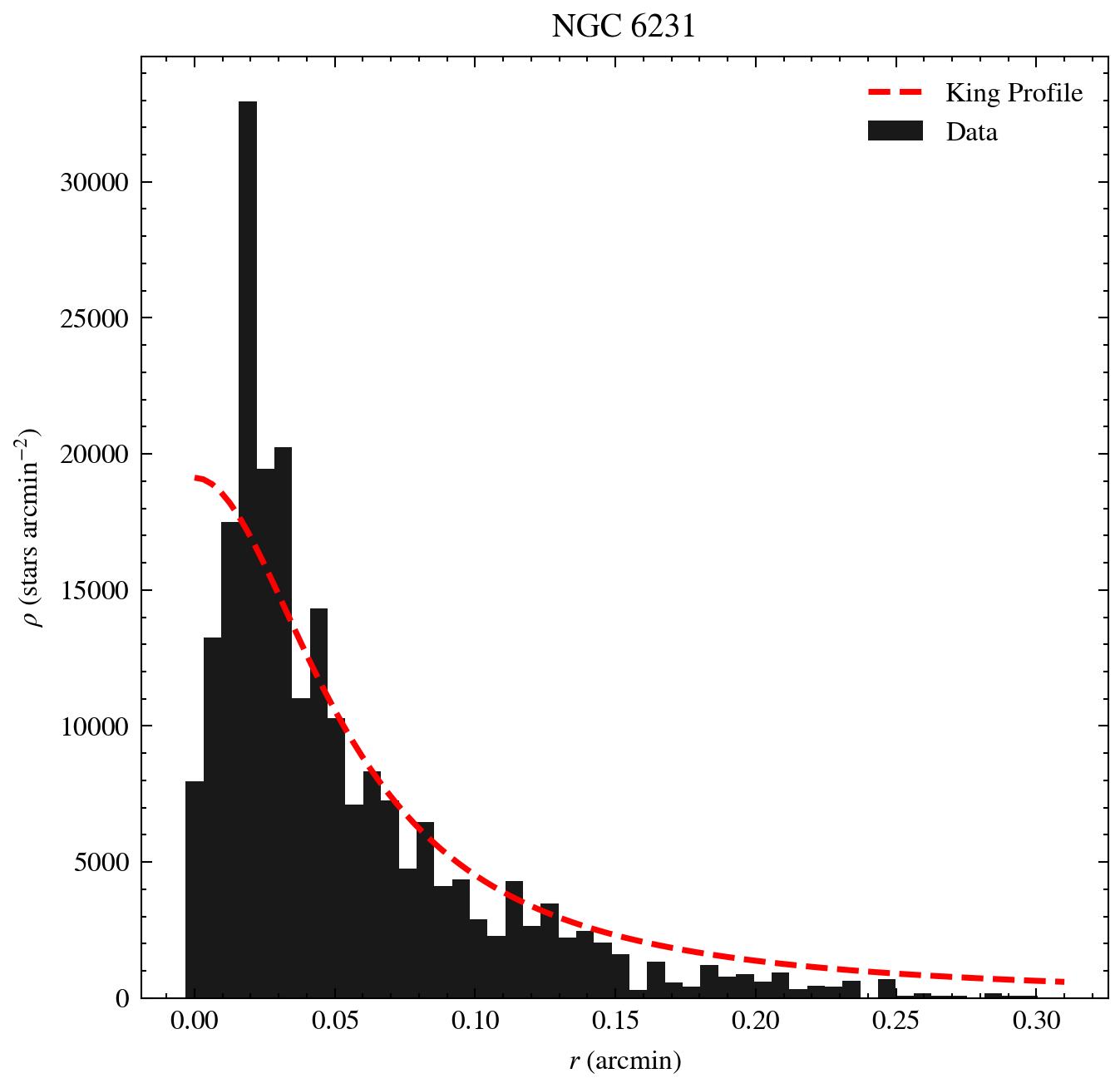} 
			
		\end{minipage}\hfill
		\begin{minipage}{0.3\textwidth}
			\includegraphics[width=\linewidth]{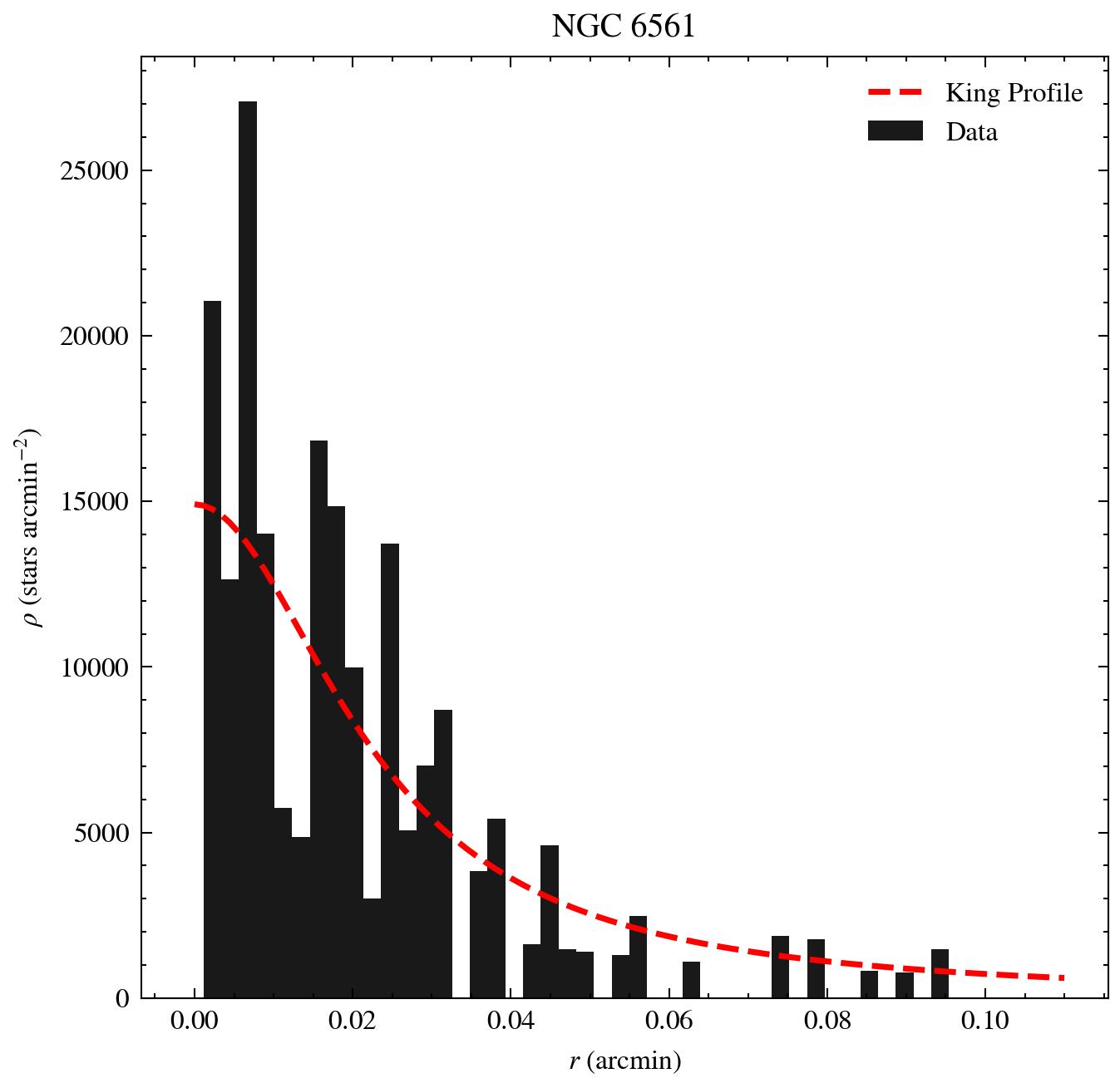} 
			
		\end{minipage}\hfill
		\begin{minipage}{0.3\textwidth}
			\includegraphics[width=\linewidth]{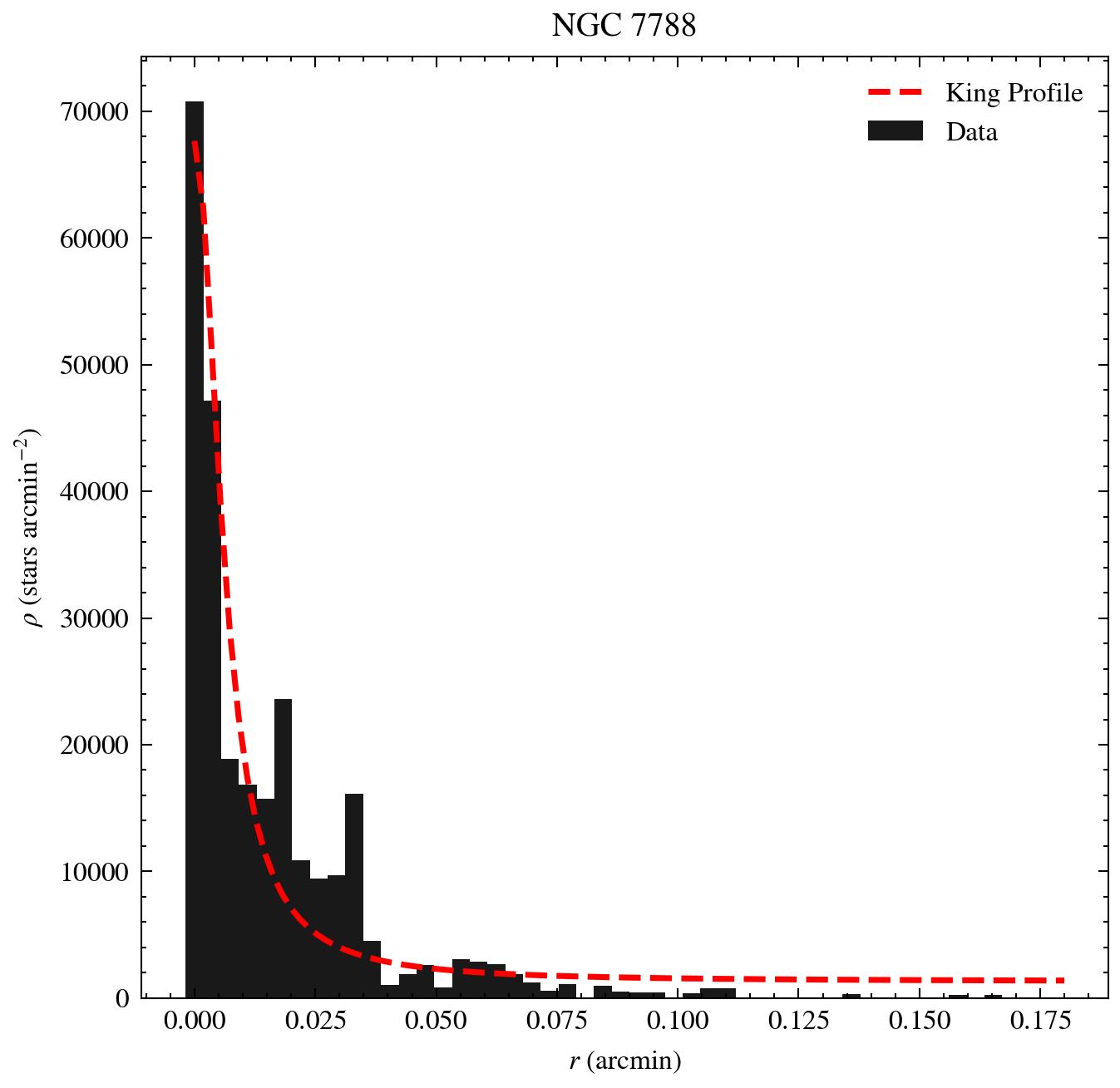} 
			
		\end{minipage}
		
		\vspace{0.5cm} 
		\begin{minipage}{0.3\textwidth}
			\includegraphics[width=\linewidth]{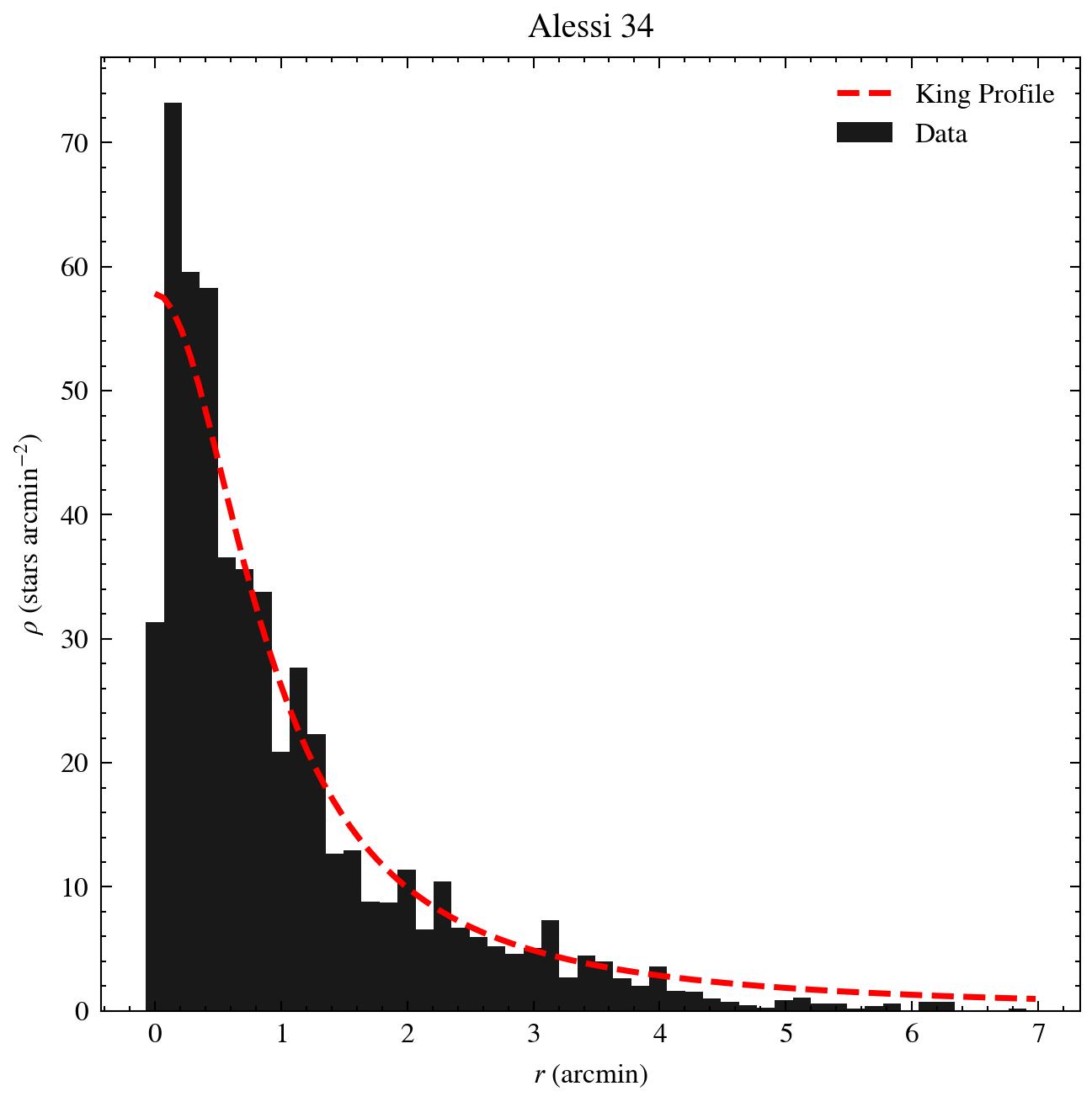} 
			
		\end{minipage}\hfill
		\begin{minipage}{0.3\textwidth}
			\includegraphics[width=\linewidth]{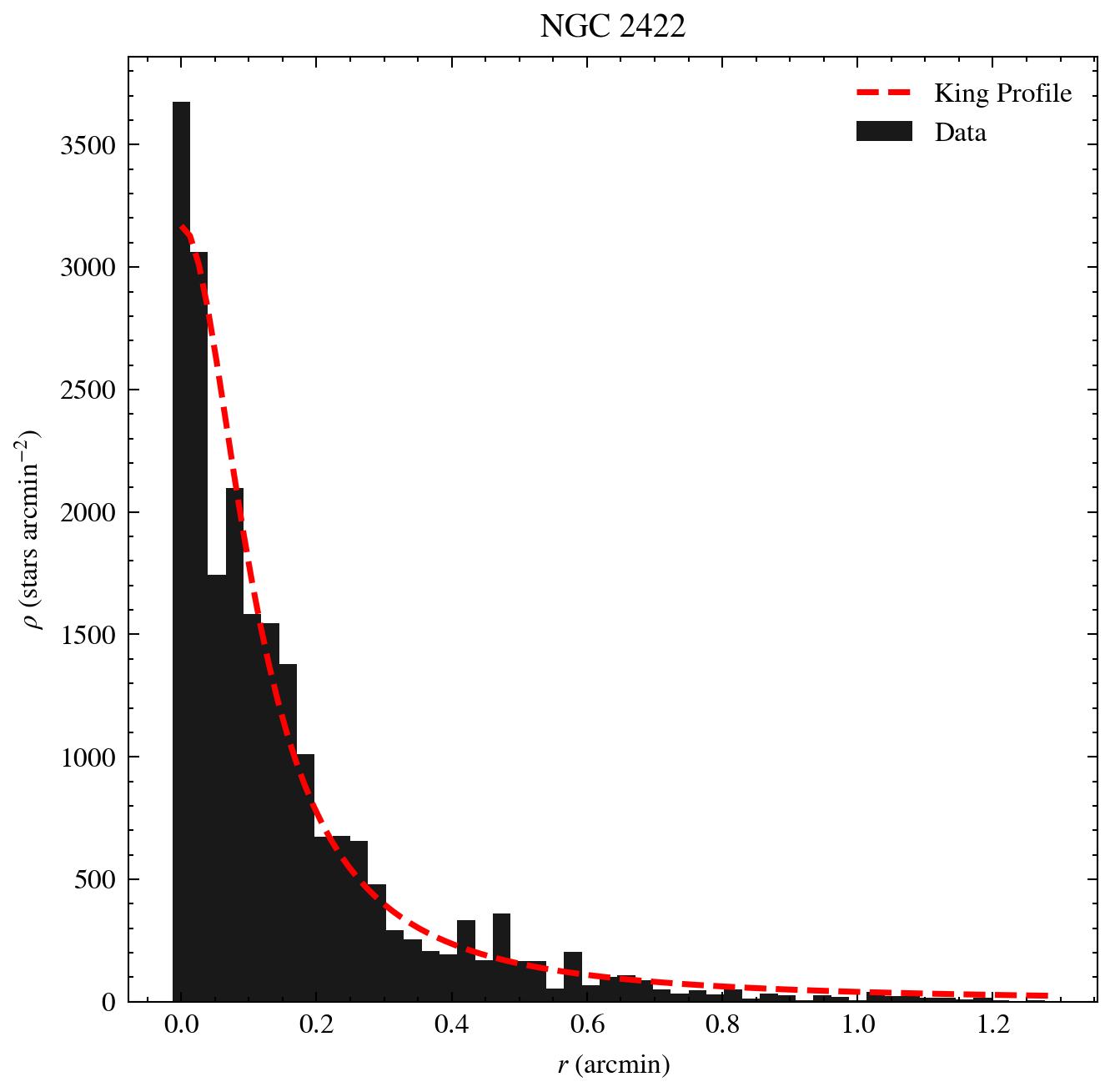} 
			
		\end{minipage}\hfill
		\begin{minipage}{0.3\textwidth}
			
			\includegraphics[width=\linewidth]{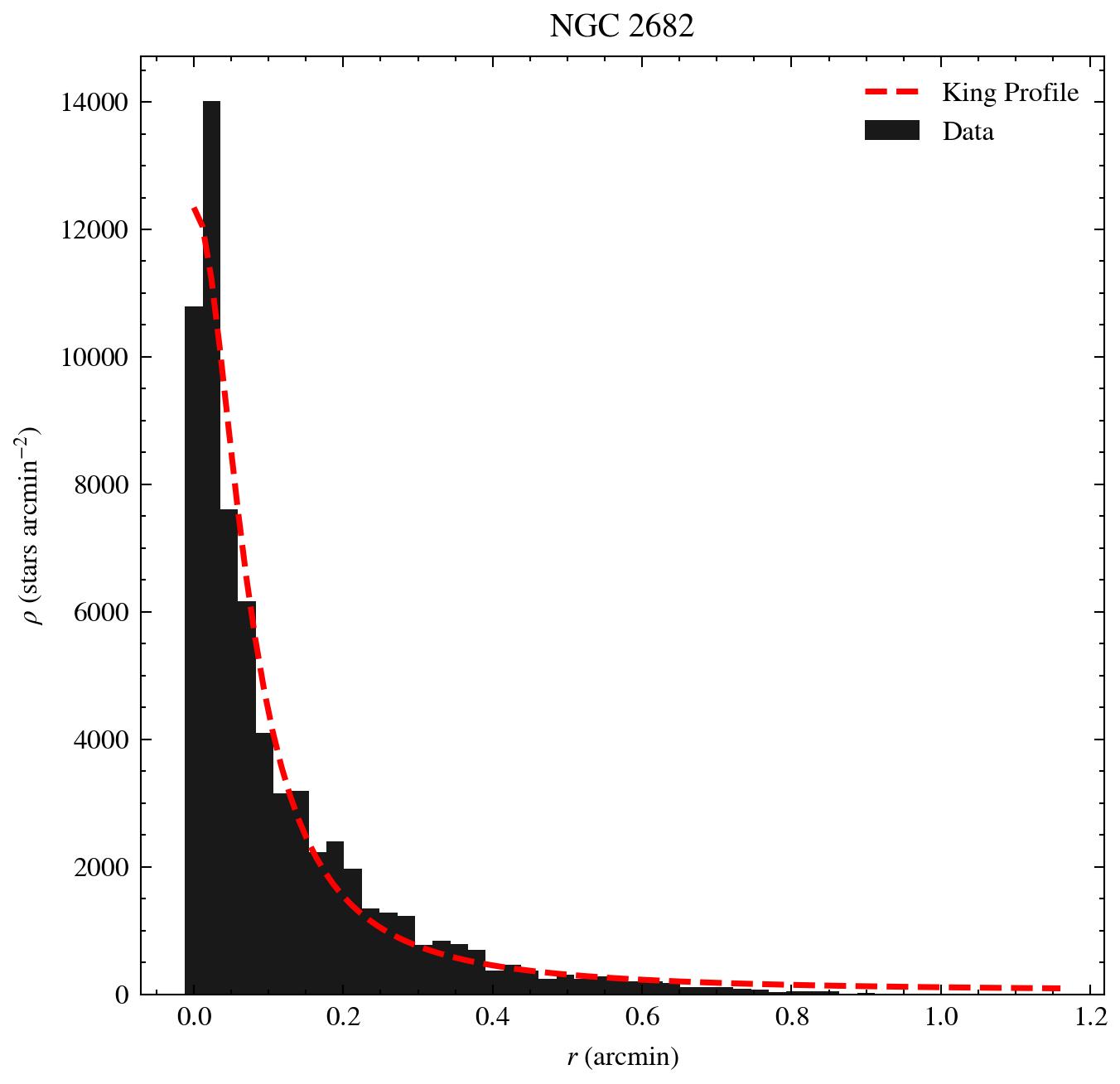} 

		\end{minipage}
		
		\caption{King profiles illustrating the radial density distribution of cluster members for six open clusters: NGC 6231, NGC 6561, NGC 7788, Alessi 34, NGC 2422, and NGC 2682.}
		\label{fig:kingprofile}
		
	\end{figure}

	As shown in Figs. \ref{fig:kingprofile} and \ref{fig:kingprofile2}, the King profile was fitted to the observed radial density distribution for each cluster, highlighting the alignment of cluster members with the expected density distribution.
	
	\subsection{Comparison with Previous Work}
	\label{subsec:compare}
	In this study, we compared the detected members and derived parameters of open clusters with the results reported by \citet{hunt_improving_2024}, who applied the HDBSCAN algorithm to GDR3 for OC identification. The primary objective of this comparison is to evaluate the consistency and reliability of our methodology against an established clustering technique.
	
	Comparison with Hunt and Reffert (Figs. \ref{fig:comparison_part1} and \ref{fig:comparison_part2}) reveals discrepancies in their spatial distribution and individual member assignments, although our filtering and modeling techniques effectively isolate cluster members (Figs. \ref{fig:cmdinfield} and \ref{fig:motioninfield}).
	
	Table \ref{table:3} presents our derived physical parameters for each cluster, including central coordinates, proper motion centroids, and parallaxes. These values are compared with the corresponding parameters reported by Hunt (Table \ref{table:1}). Our results demonstrate close alignment with Hunt's, with differences within acceptable margins of error.
	
	\begin{figure*}[htbp]
		\centering
		\begin{minipage}{0.9\textwidth}
			\includegraphics[width=\linewidth]{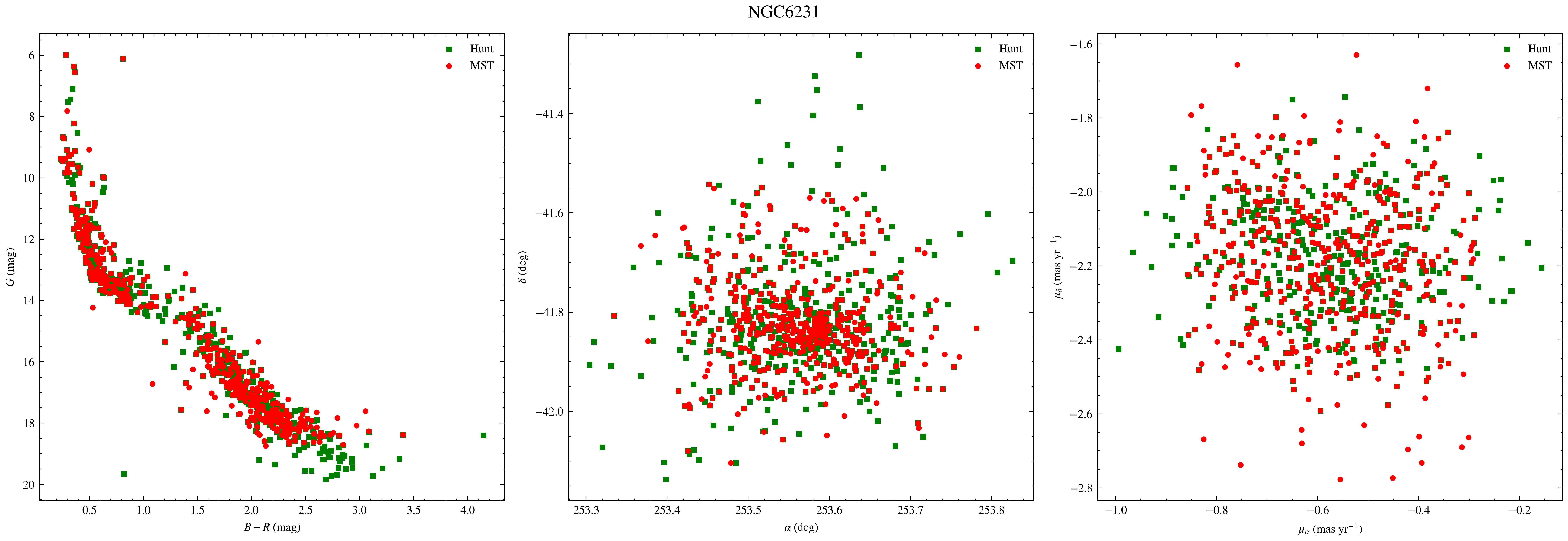} 
		\end{minipage}
		
		\vspace{0.5cm} 
		
		\begin{minipage}{0.9\textwidth}
			\includegraphics[width=\linewidth]{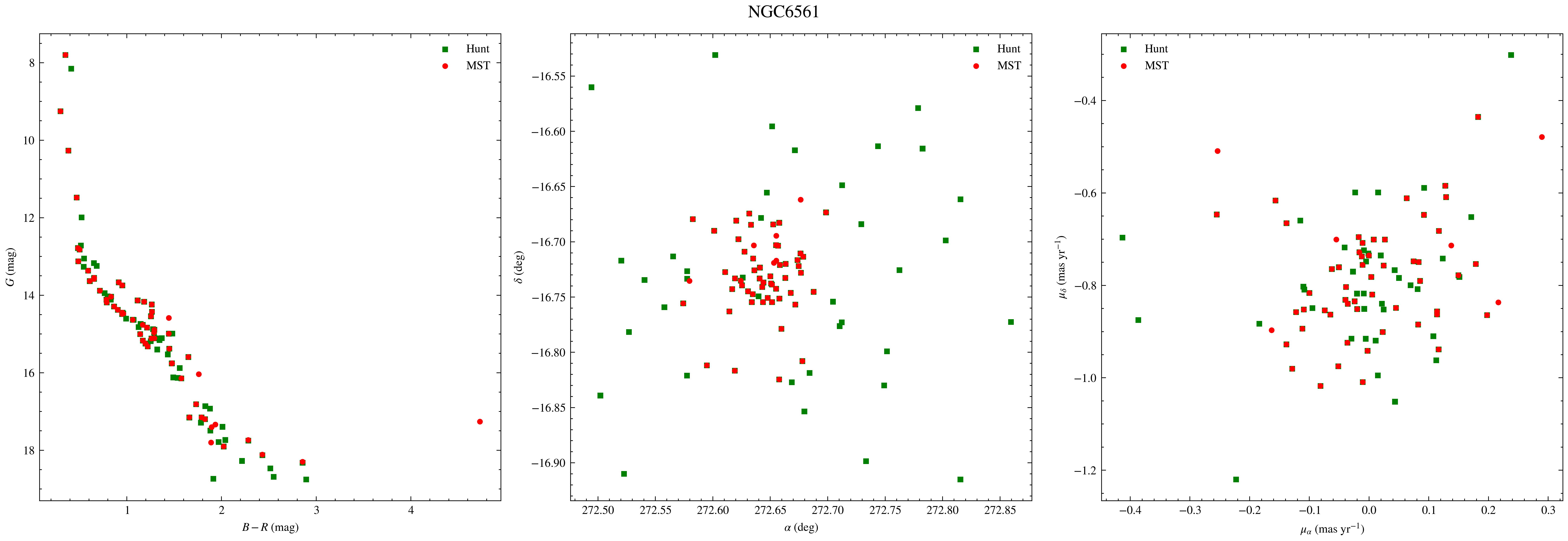} 
		\end{minipage}
		
		\vspace{0.5cm} 
		
		\begin{minipage}{0.9\textwidth}
			\includegraphics[width=\linewidth]{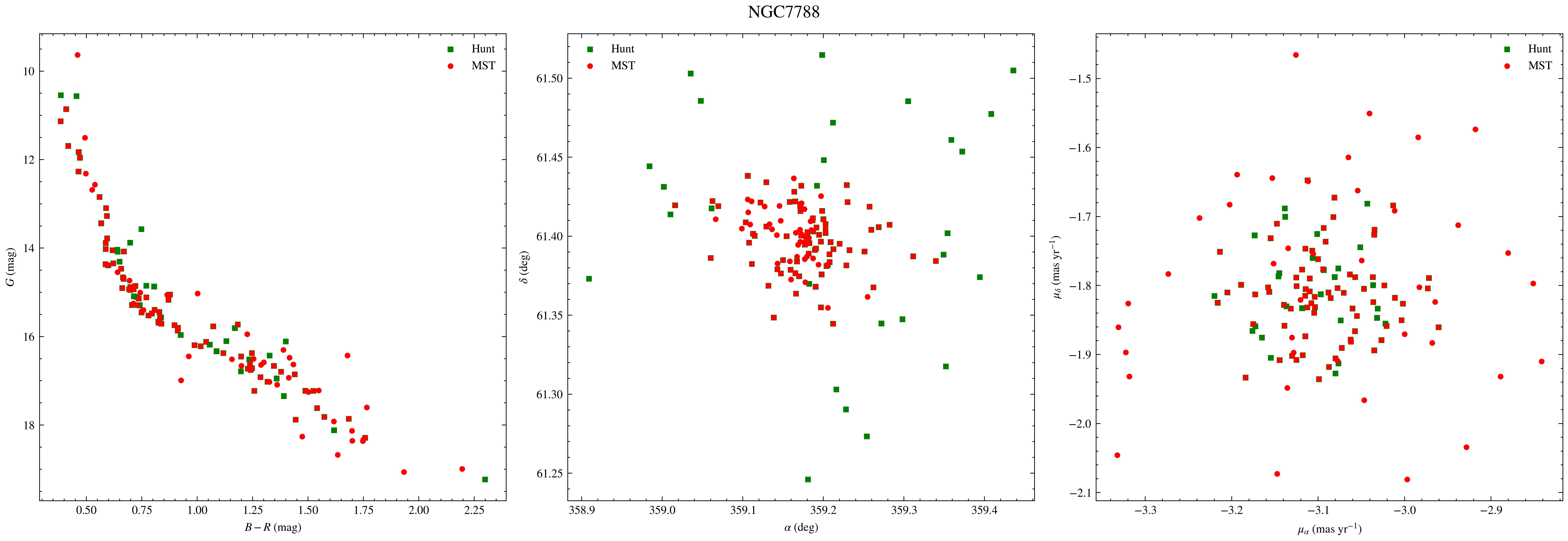} 
		\end{minipage}
		
		\caption{Comparison of cluster membership identification using the proposed Minimum Spanning Tree-Gaussian Mixture Model (MST-GMM) algorithm and the method of Hunt et al. (2024) for three open clusters: NGC 6231, NGC 6561, NGC 7788, highlighting the differences in membership assignment and cluster structure. (Part 1)}
		\label{fig:comparison_part1}
	\end{figure*}
	
	\newpage
	
	\begin{figure}[H]
		\centering
		
		\begin{minipage}{0.9\textwidth}
			\includegraphics[width=\linewidth]{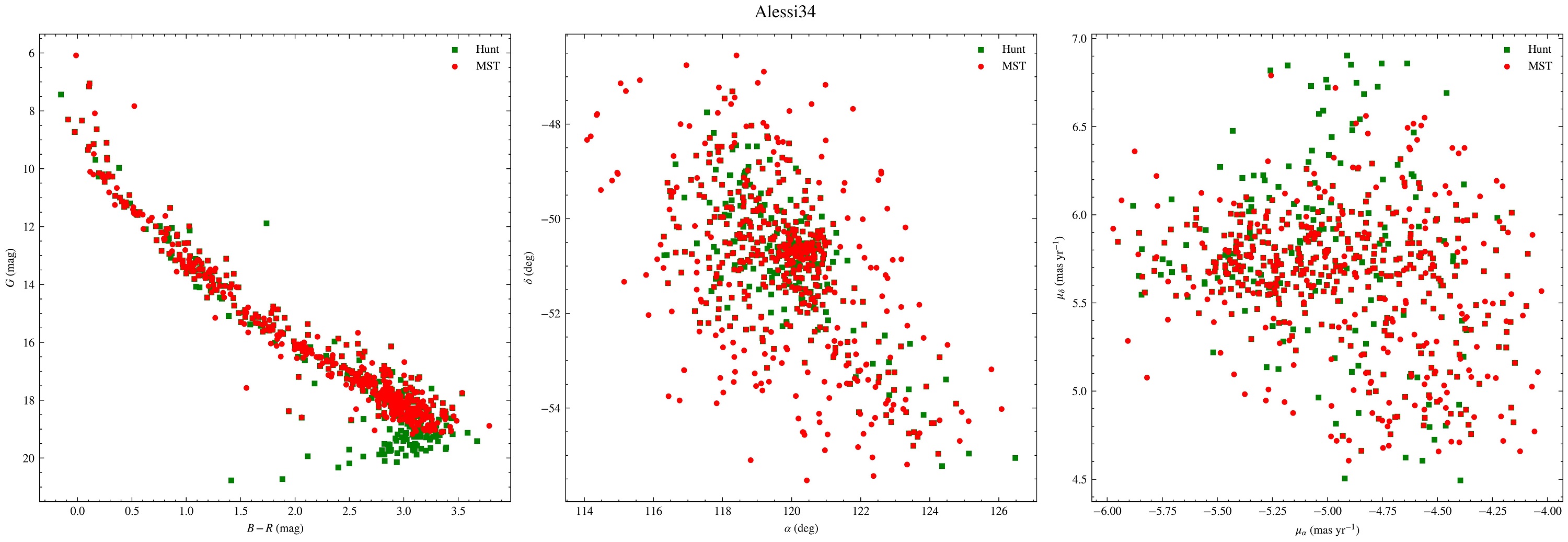} 
		\end{minipage}
		
		\vspace{0.5cm} 
		
		\begin{minipage}{0.9\textwidth}
			\includegraphics[width=\linewidth]{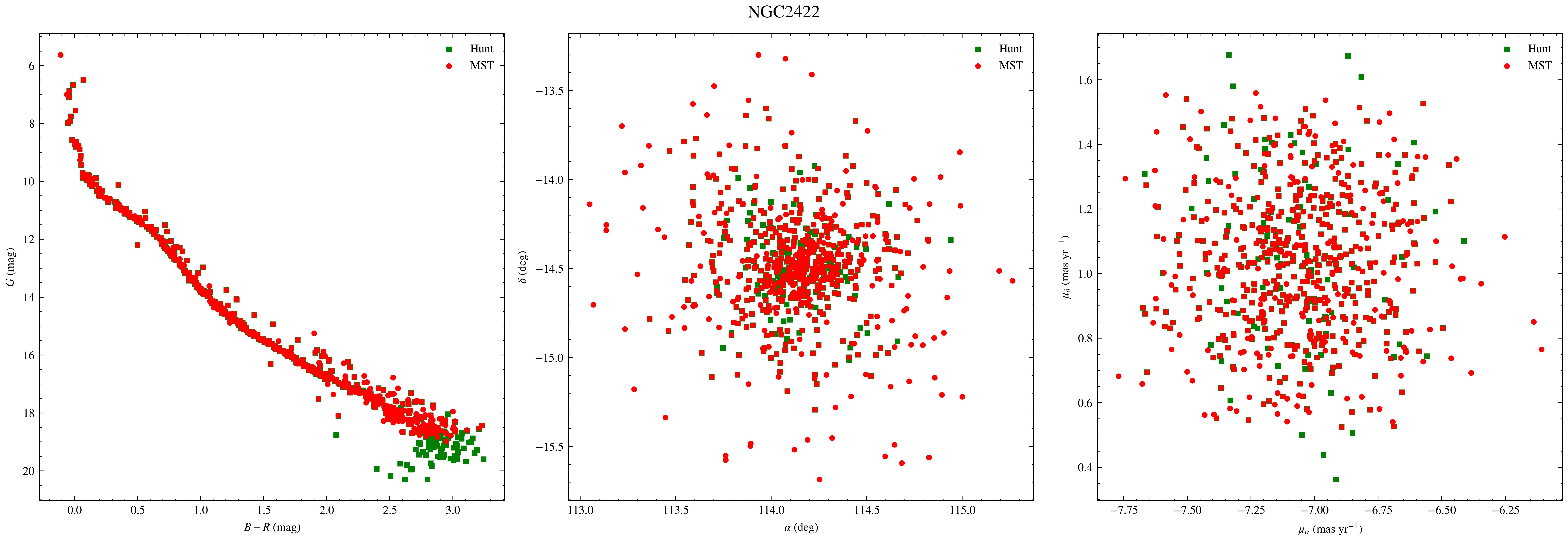} 
		\end{minipage}
		
		\vspace{0.5cm} 
		
		\begin{minipage}{0.9\textwidth}
			\includegraphics[width=\linewidth]{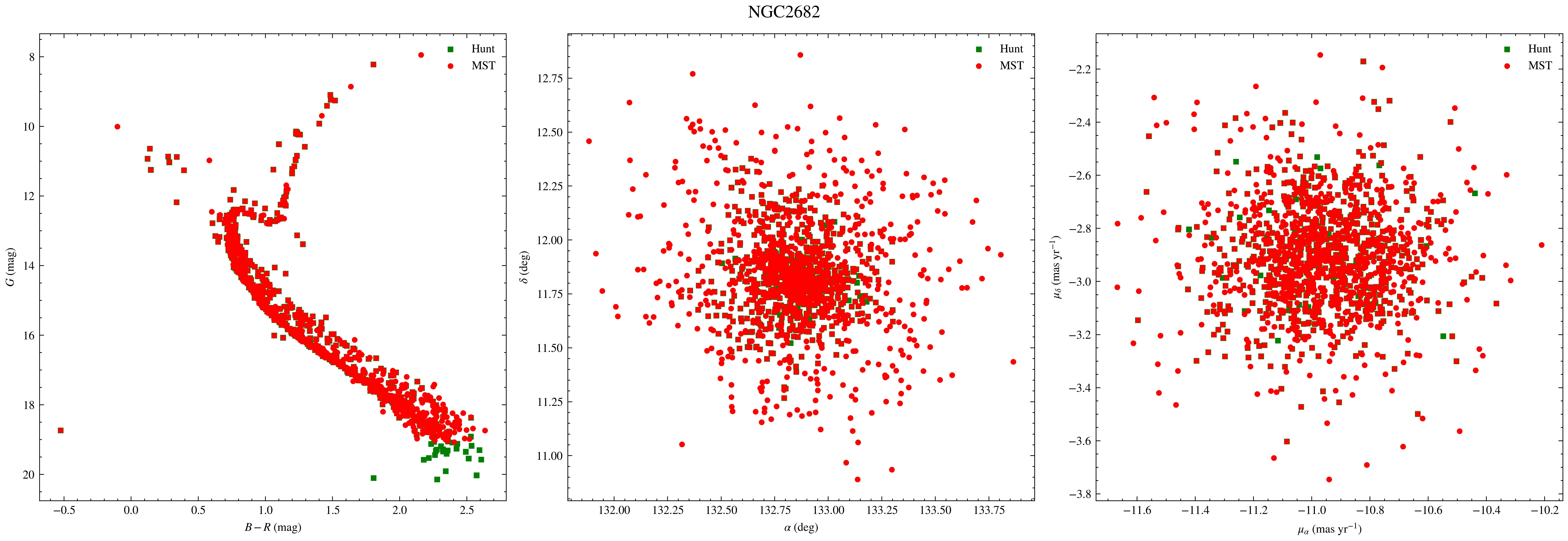} 
		\end{minipage}
		
		\caption{Comparison of cluster membership identification using the proposed Minimum Spanning Tree-Gaussian Mixture Model (MST-GMM) algorithm and the method of Hunt et al. (2024) for three open clusters: Alessi 34, NGC 2422, and NGC 2682, highlighting the differences in membership assignment and cluster structure. (Part 2)}
		\ContinuedFloat
		\caption*{(Continued from Figure \ref{fig:comparison_part2})}
		
		\label{fig:comparison_part2}
	\end{figure}

	\subsection{NGC 7788 and NGC 7790}
	\label{subsec:ngc7790}
\begin{figure}[htbp] 
		\centering
		\includegraphics[width=1\linewidth]{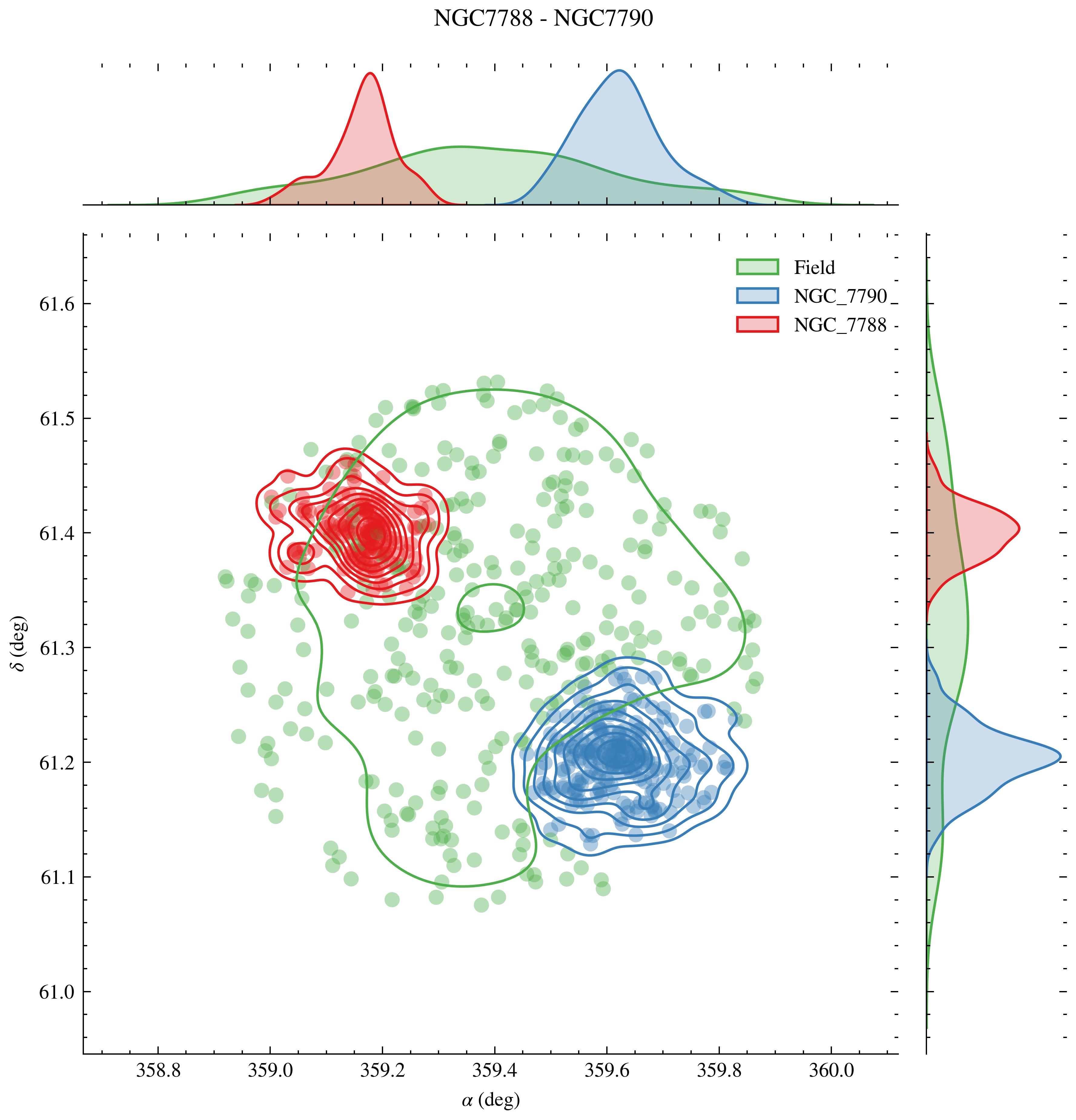}
		
		\caption{Density Estimator (KDE) plot illustrating the density distribution of cluster members (red dots: NGC 7788, blue dots: NGC 7790) and field stars (green dots) within the field of view. }
		\label{fig:kde_NGC7788_7790}
	\end{figure}

	In our analysis of NGC 7788, we identified two distinct populations within the field of view, as illustrated in Fig. \ref{fig:kde}. To further investigate these populations, a three-component GMM was employed (Fig. \ref{fig:kde_NGC7788_7790}). KDE plot in this figure demonstrates that the three-component GMM effectively distinguishes the members of each cluster from field stars, confirming the presence of multiple populations.
	
	Due to the similarity of the proper motion and parallax values between these clusters, leading to overlap in these parameters, the initial filtering step did not exclude any members of the adjacent cluster. However, upon applying MST and GMM to the filtered data, the presence of another cluster within the field, NGC 7790, became evident.
	
\begin{figure}[htbp] 
		\centering
		\includegraphics[width=\linewidth]{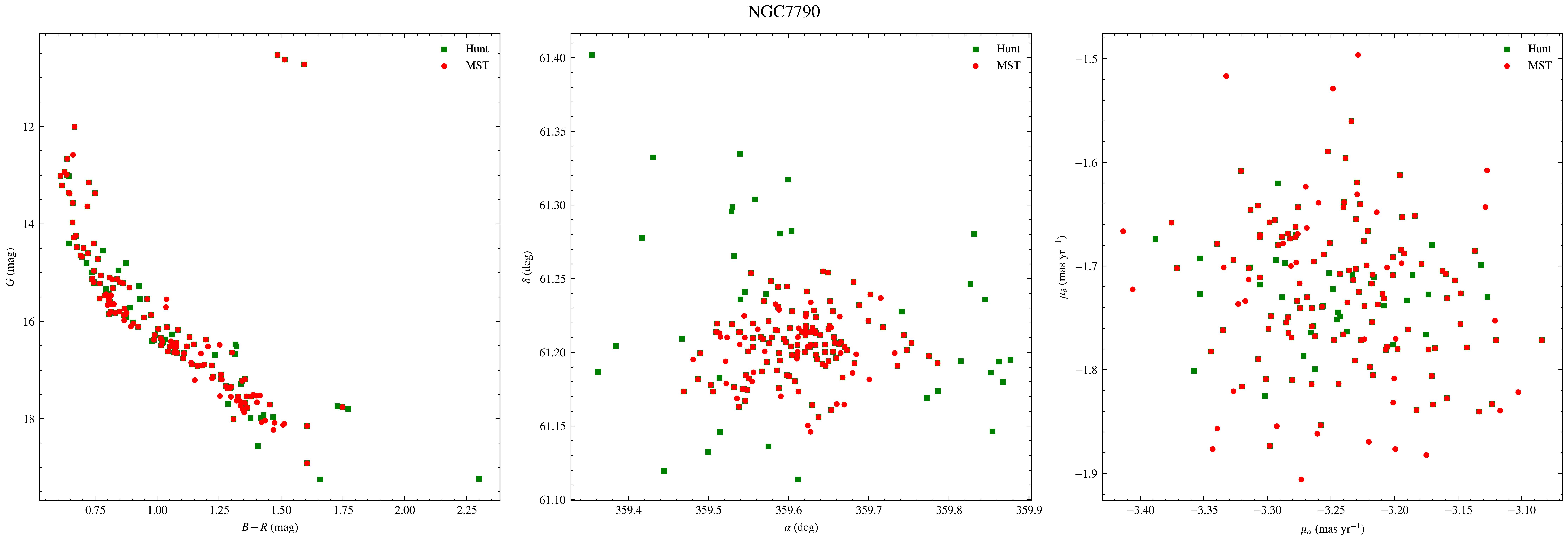}
		
		\caption{Comparison of cluster membership identification using the proposed Minimum Spanning Tree-Gaussian Mixture Model (MST-GMM) algorithm and the method of Hunt et al. (2024) for NGC 7790, highlighting the differences in membership assignment and cluster structure.}
		\label{fig:hunt_mst_ngc7790}
	\end{figure}

	To further validate our findings, we applied the proposed method to the adjacent cluster, NGC 7790. Fig. \ref{fig:hunt_mst_ngc7790} and Table \ref{table:3} illustrate a strong agreement in the membership determination of NGC 7790 compared to the results reported by \citet{hunt_improving_2024}. This agreement highlights the efficacy of our methodology in precisely identifying cluster members, even in complex regions where populations exhibit overlap across multiple dimensions.

\section{Conclusion}
	\label{sec:conc}
	Our study presents a comprehensive analysis of open star clusters using a novel combination of MST and GMM. This approach has demonstrated significant improvements in the accuracy and reliability of cluster member identification compared to methods like the HDBSCAN algorithm used by \citet{hunt_improving_2024}. By utilizing initial filtering and MST, we effectively reduce the influence of field stars and enhance the precision of our results, leading to a clear classification by GMM. The visual evidence from KDE plots, King surface density profiles, and comparison plots further supports the robustness of our methodology, providing clear insights into the spatial distribution and structure of clusters.
	
	In comparing our results with those of Hunt and Reffert, we observe that our model consistently aligns with expected cluster characteristics, with differences falling within acceptable margins of error. Building on these comparative results, our model's computational efficiency and robustness make it highly scalable to larger datasets, paving the way for applications in other astrophysical systems or domains involving large-scale clustering problems. Future datasets such as Gaia DR4, with its richer astrometric information, represent an ideal testing ground for this methodology.
	
	In conclusion, the novel integration of MST and GMM in our study provides a powerful tool for the analysis of open star clusters, offering significant advancements over existing methodologies. Our approach not only matches but in some cases exceeds the accuracy of other models, demonstrating its effectiveness in both typical and complex clustering scenarios. The results of this study offer a robust framework for future investigations into stellar populations, star cluster evolution, and dynamic processes such as stellar migration and cluster dissolution, shaping the broader understanding of our galaxy.
	
\section{Acknowledgment}
	This work has made use of data from the European Space Agency (ESA) mission
	{\it Gaia} (\url{https://www.cosmos.esa.int/gaia}), processed by the {\it Gaia} Data Processing and Analysis Consortium (DPAC, \url{https://www.cosmos.esa.int/web/gaia/dpac/consortium}). Funding for the DPAC has been provided by national institutions, in particular the institutions participating in the {\it Gaia} Multilateral Agreement.
	
	In this study, we utilized the VizieR catalog service for data retrieval and analysis \citep{vizier2000}.
	
\appendix
\section{Figures of More Clusters}
\label{sec:app1}
\begin{figure}[H] 
	\centering
	\begin{minipage}{0.3\textwidth}
		\includegraphics[width=\linewidth]{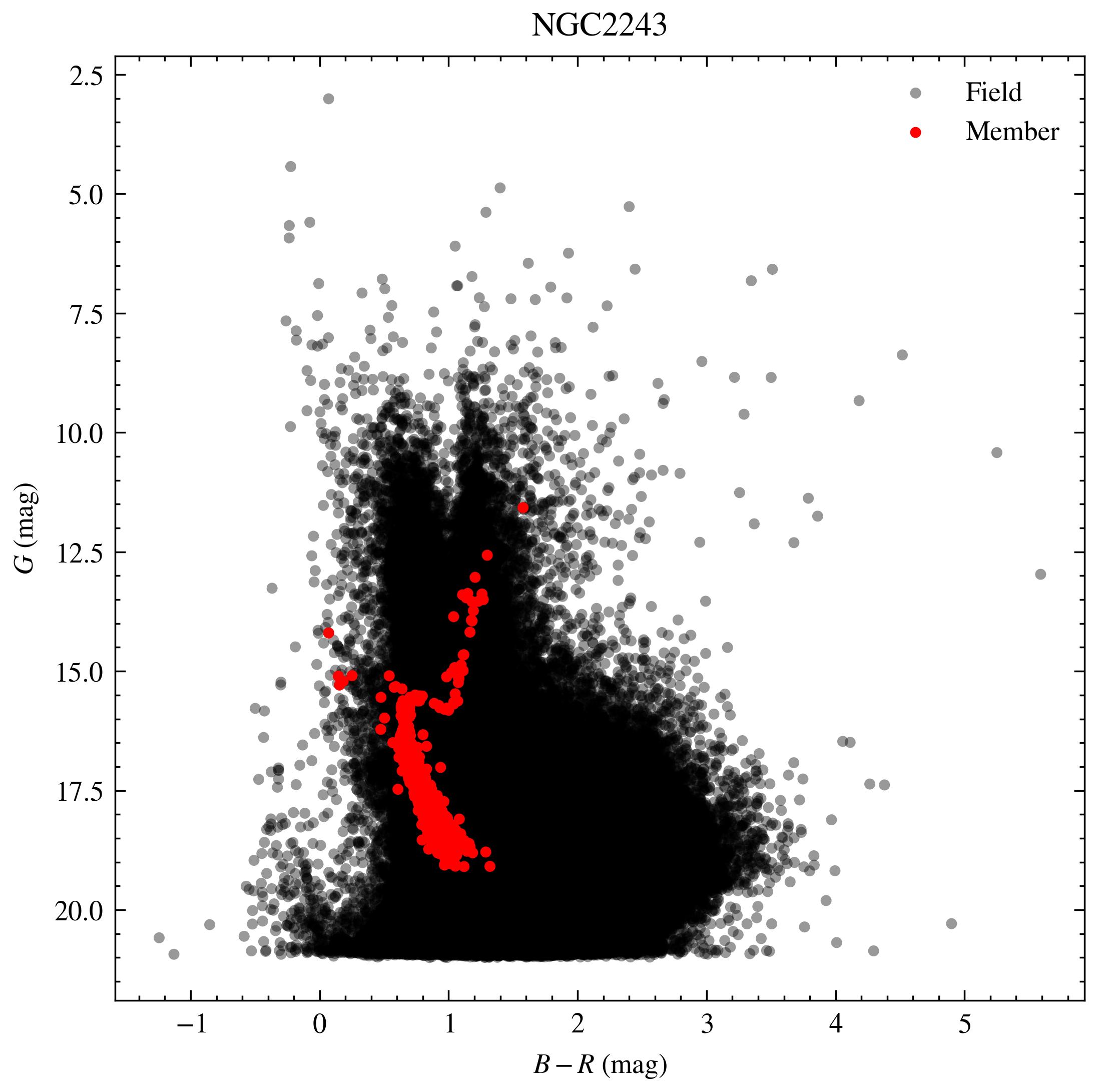} 
		
	\end{minipage}\hfill
	\begin{minipage}{0.3\textwidth}
		\includegraphics[width=\linewidth]{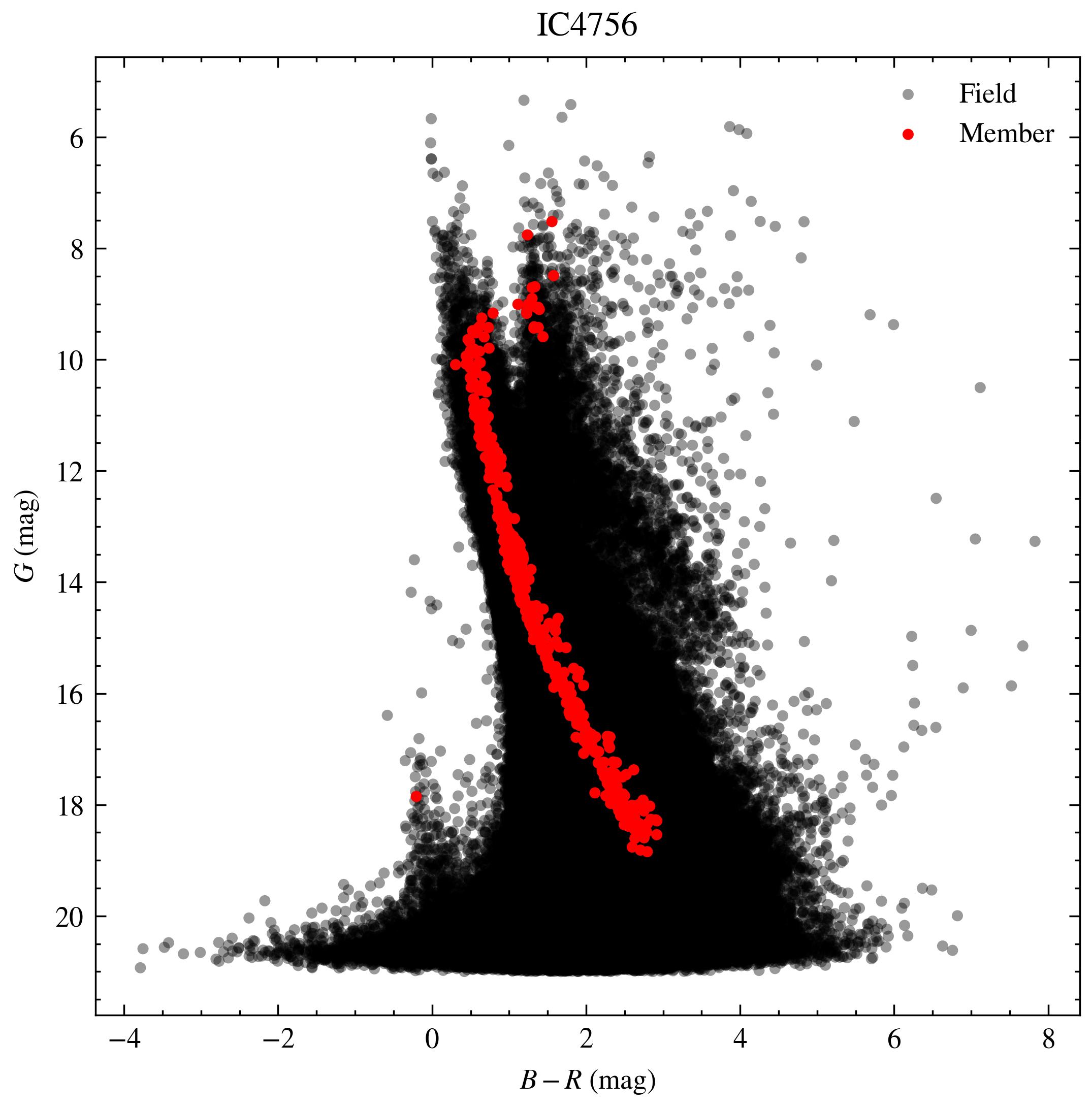} 
		
	\end{minipage}\hfill
	\begin{minipage}{0.3\textwidth}
		\includegraphics[width=\linewidth]{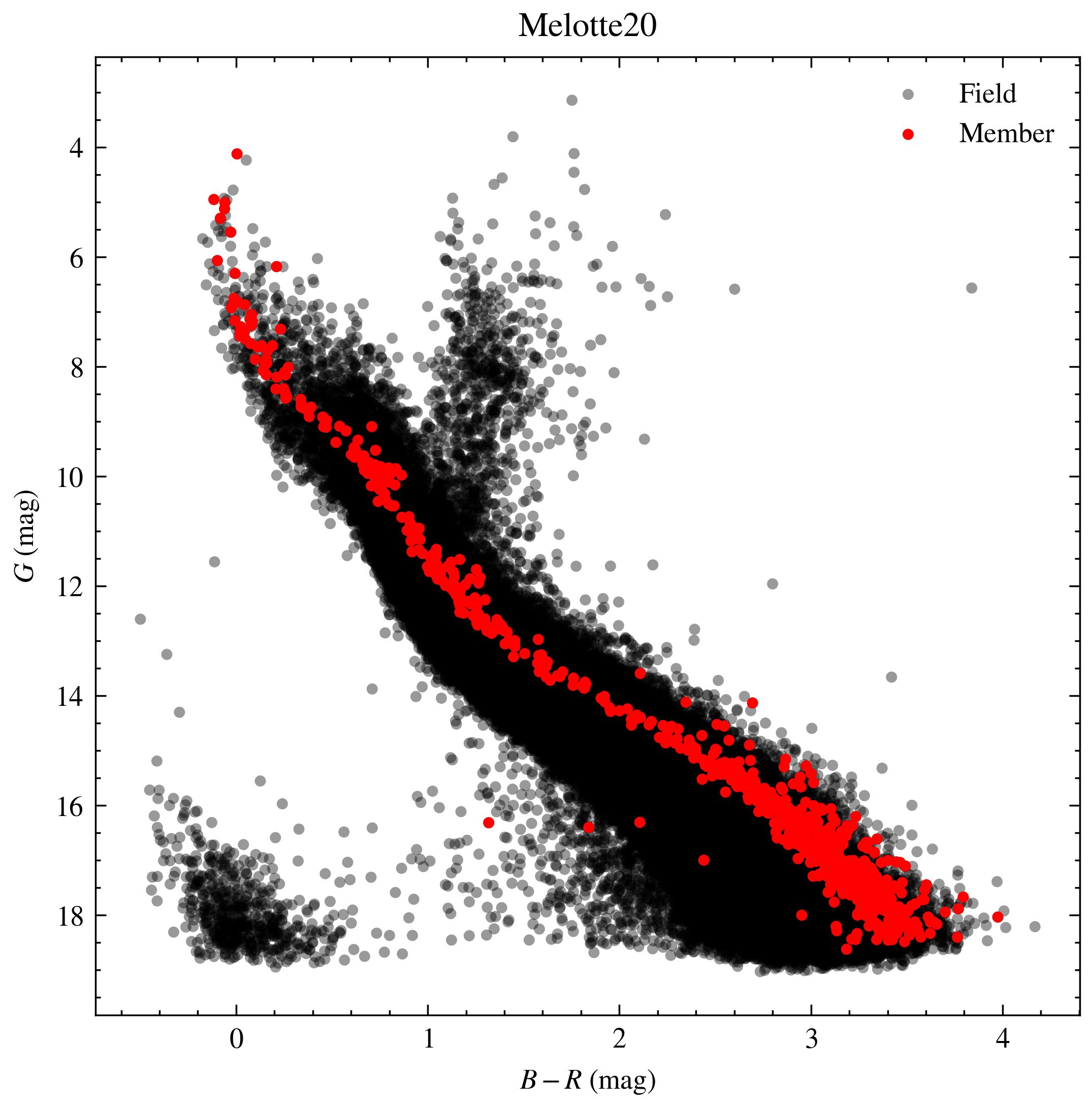} 
		
	\end{minipage}
	
	\vspace{0.5cm} 
	\begin{minipage}{0.3\textwidth}
		\includegraphics[width=\linewidth]{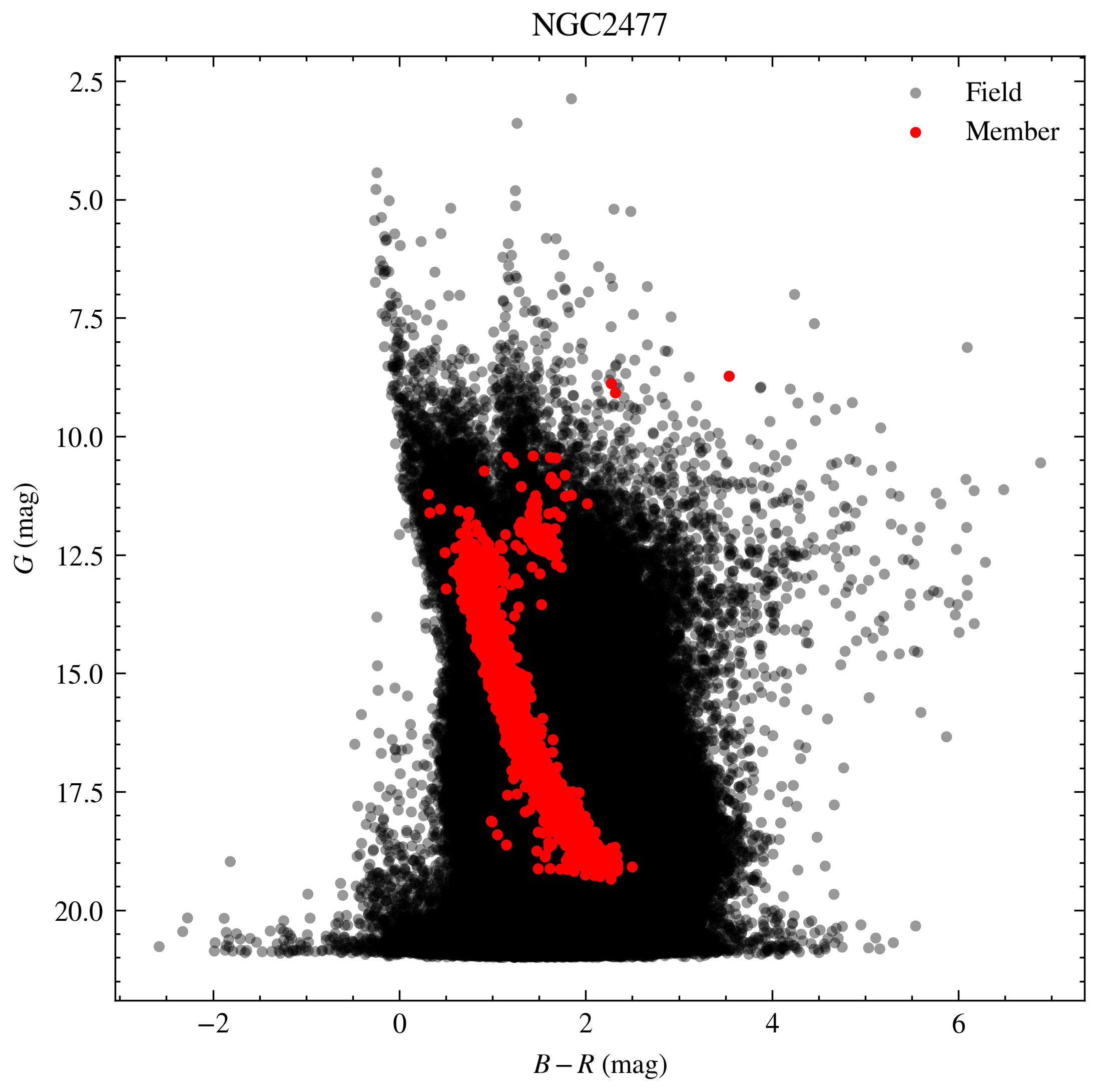} 
		
	\end{minipage}\hfill
	\begin{minipage}{0.3\textwidth}
		\includegraphics[width=\linewidth]{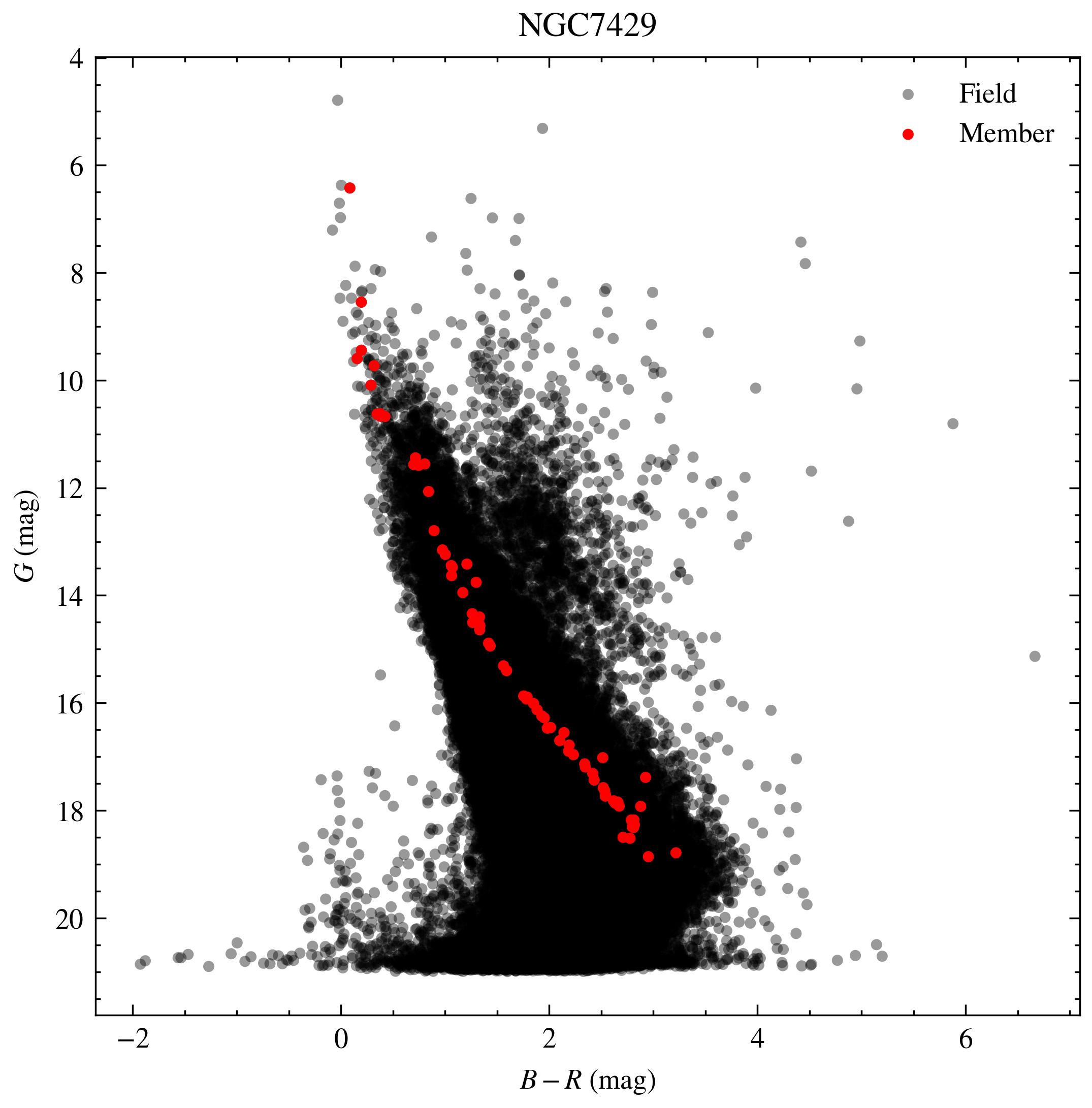} 
		
	\end{minipage}\hfill
	\begin{minipage}{0.3\textwidth}
		\includegraphics[width=\linewidth]{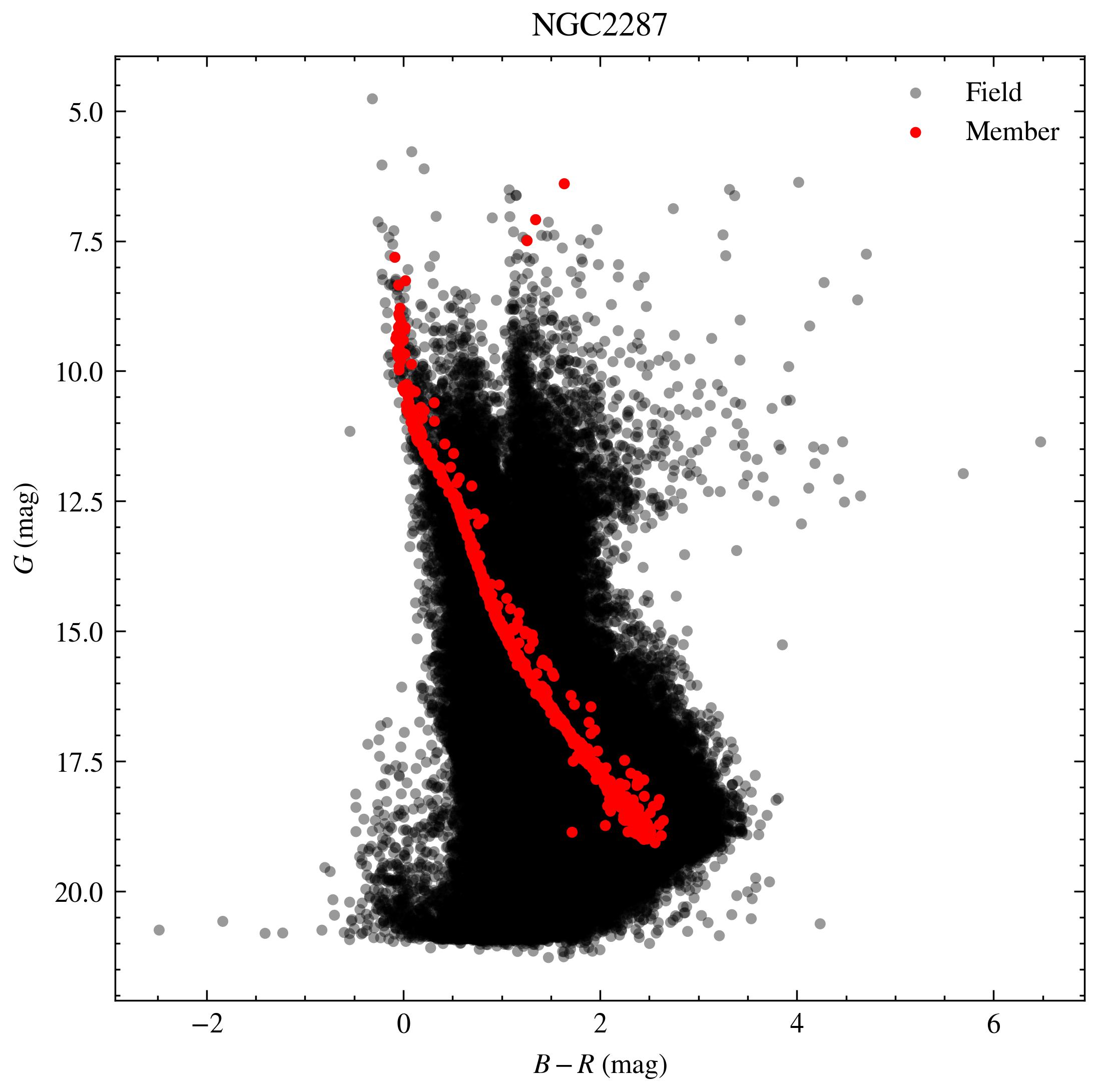} 

	\end{minipage}
	
	\caption{Color-magnitude diagrams illustrating the distribution of cluster members (red dots) and field stars (black dots) within the field of view of six open clusters: NGC 2243, IC 4756, Melotte 20, NGC 2477, NGC 7429, and NGC 2287.}
	\label{fig:cmdinfield2}
	
\end{figure}

\begin{figure}[H] 
	\centering
	\begin{minipage}{0.3\textwidth}
		\includegraphics[width=\linewidth]{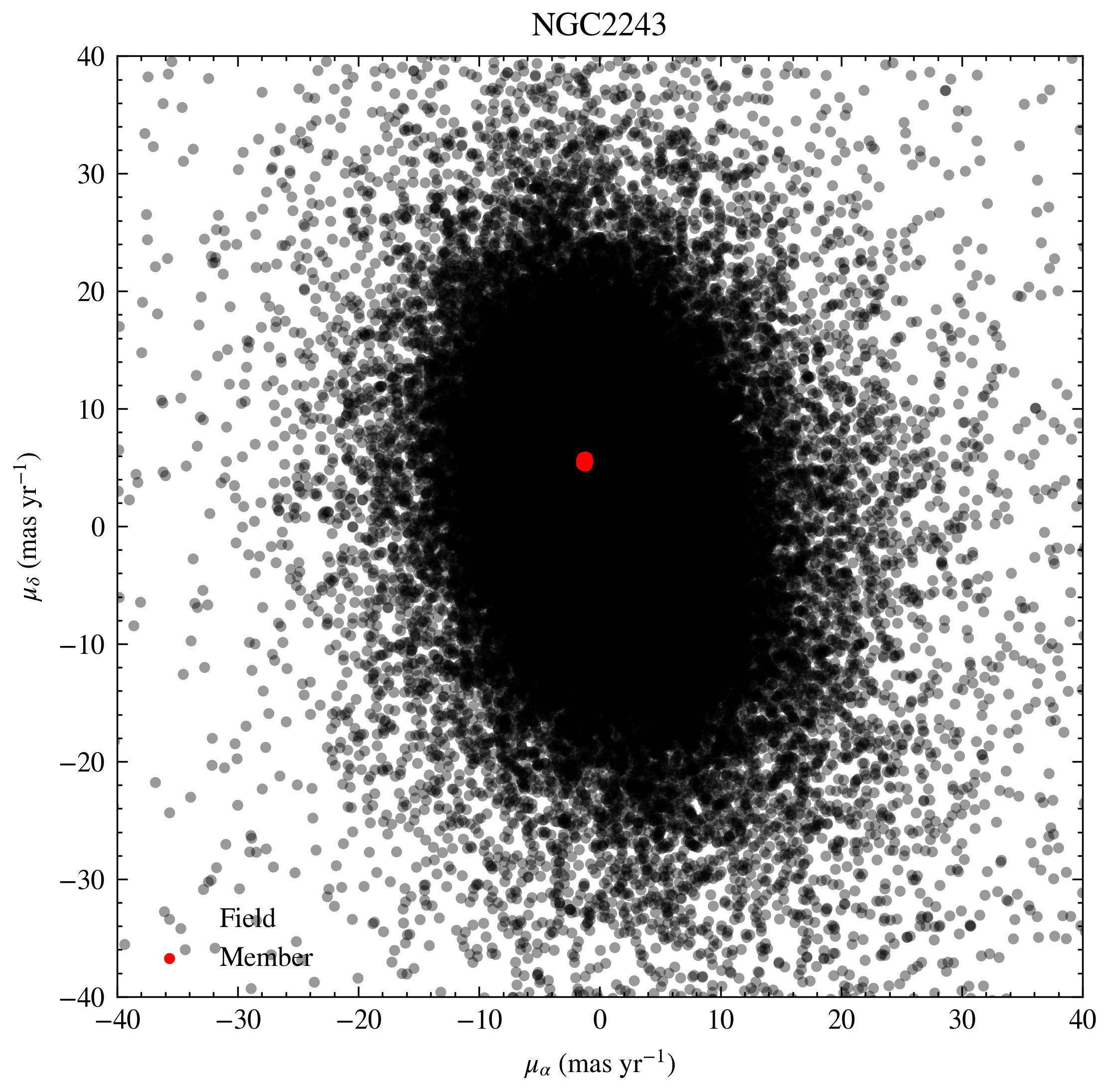} 
		
	\end{minipage}\hfill
	\begin{minipage}{0.3\textwidth}
		\includegraphics[width=\linewidth]{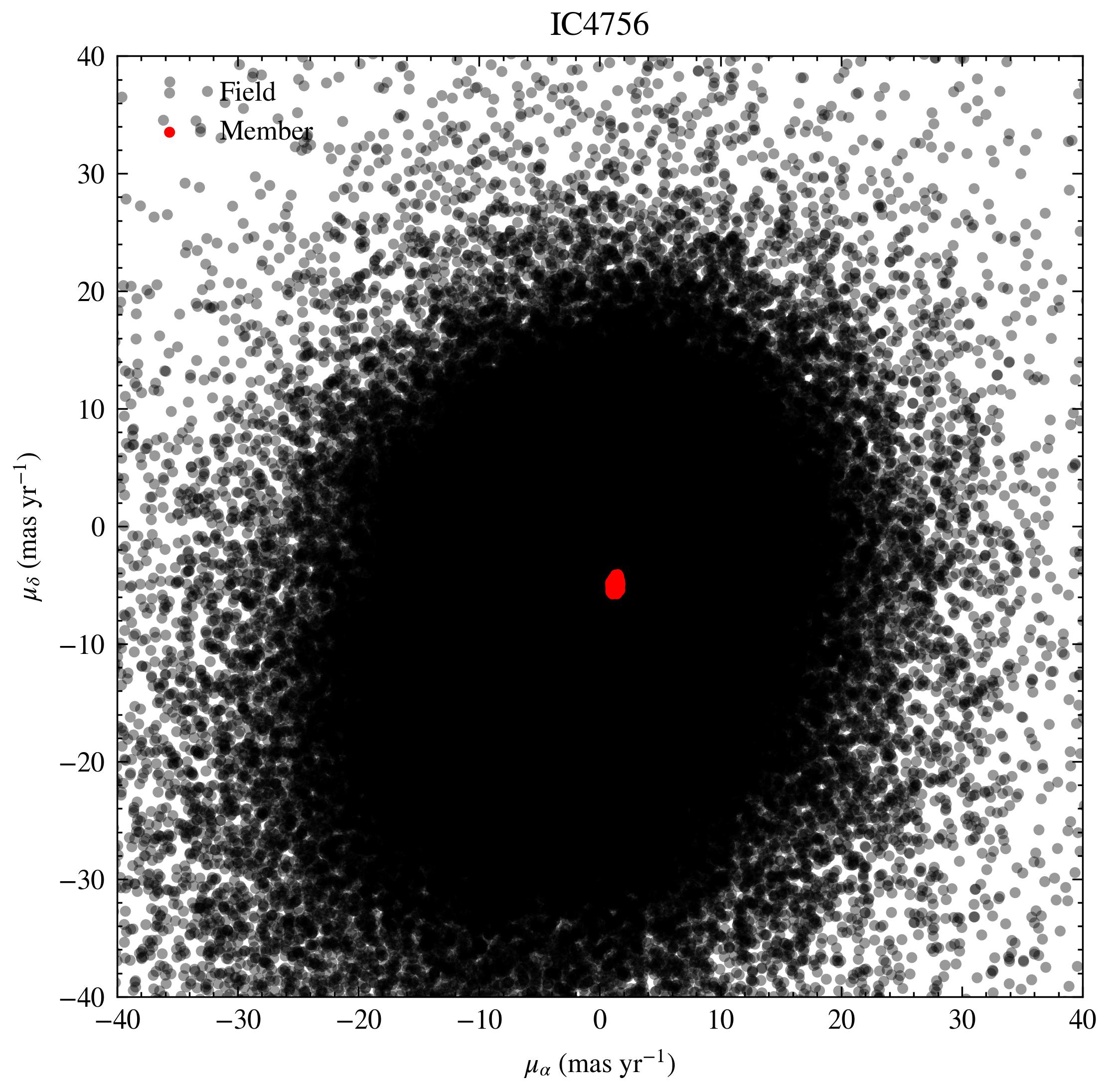} 
		
	\end{minipage}\hfill
	\begin{minipage}{0.3\textwidth}
		\includegraphics[width=\linewidth]{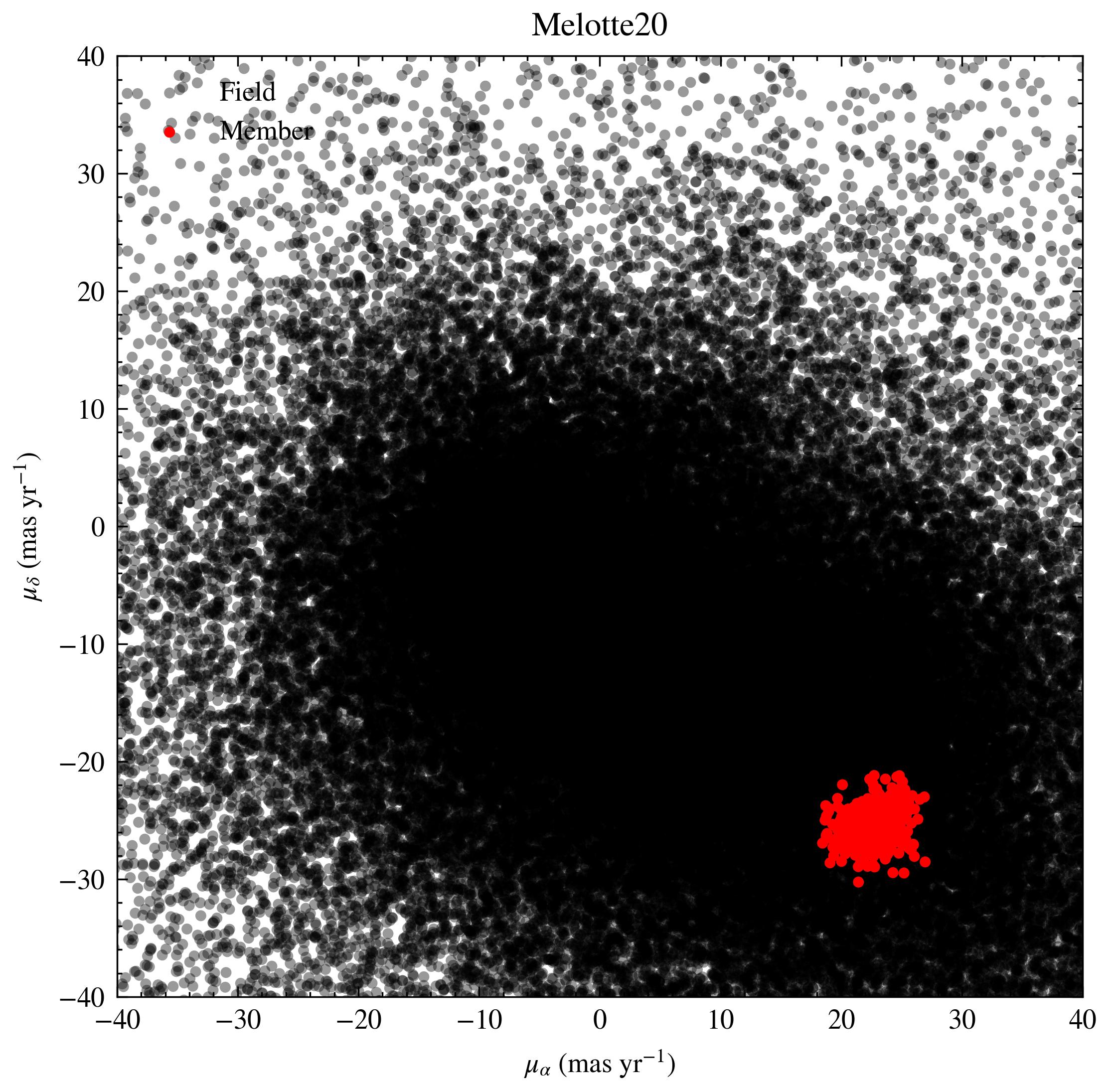} 
		
	\end{minipage}
	
	\vspace{0.5cm} 
	\begin{minipage}{0.3\textwidth}
		\includegraphics[width=\linewidth]{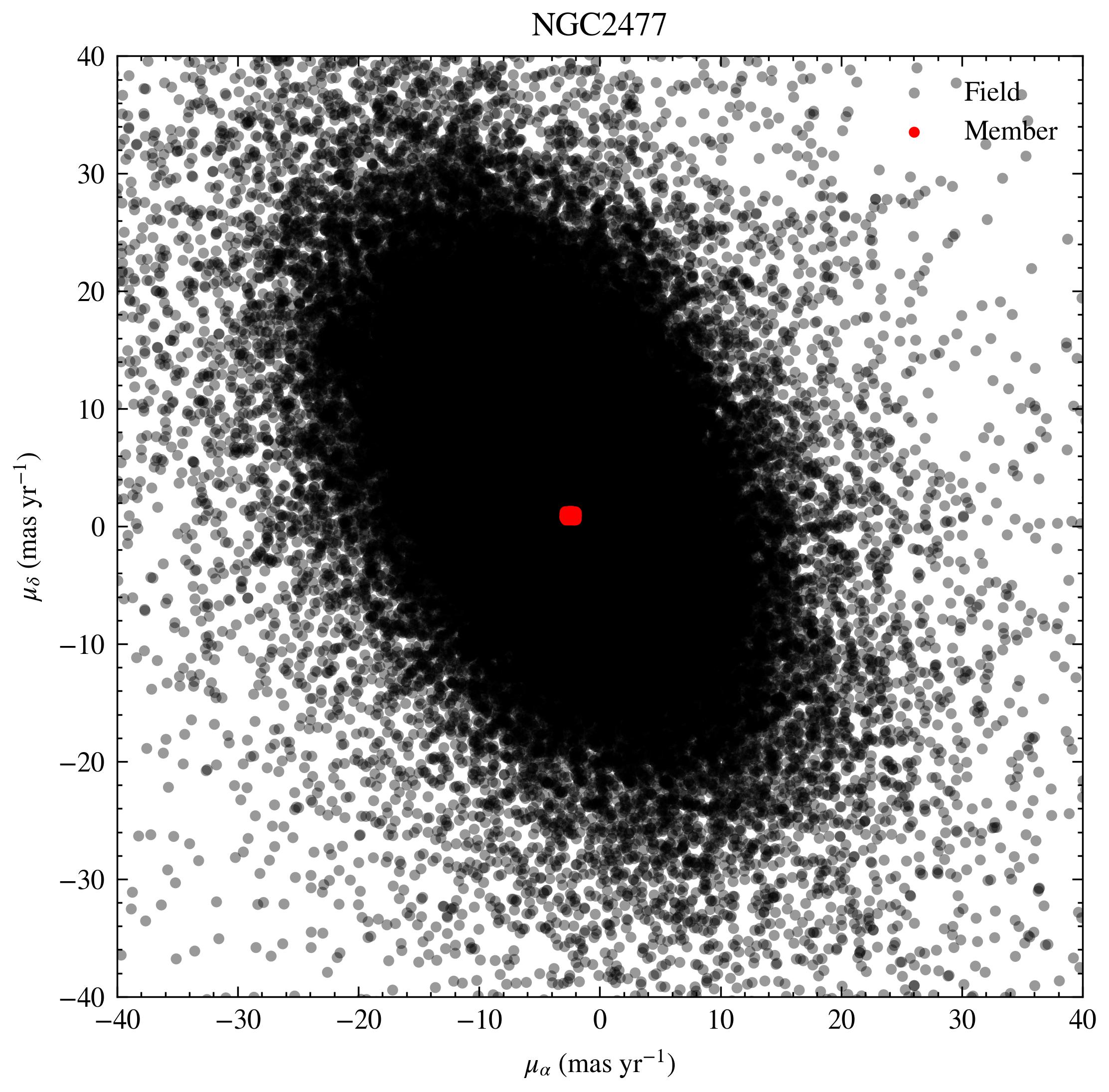} 
		
	\end{minipage}\hfill
	\begin{minipage}{0.3\textwidth}
		\includegraphics[width=\linewidth]{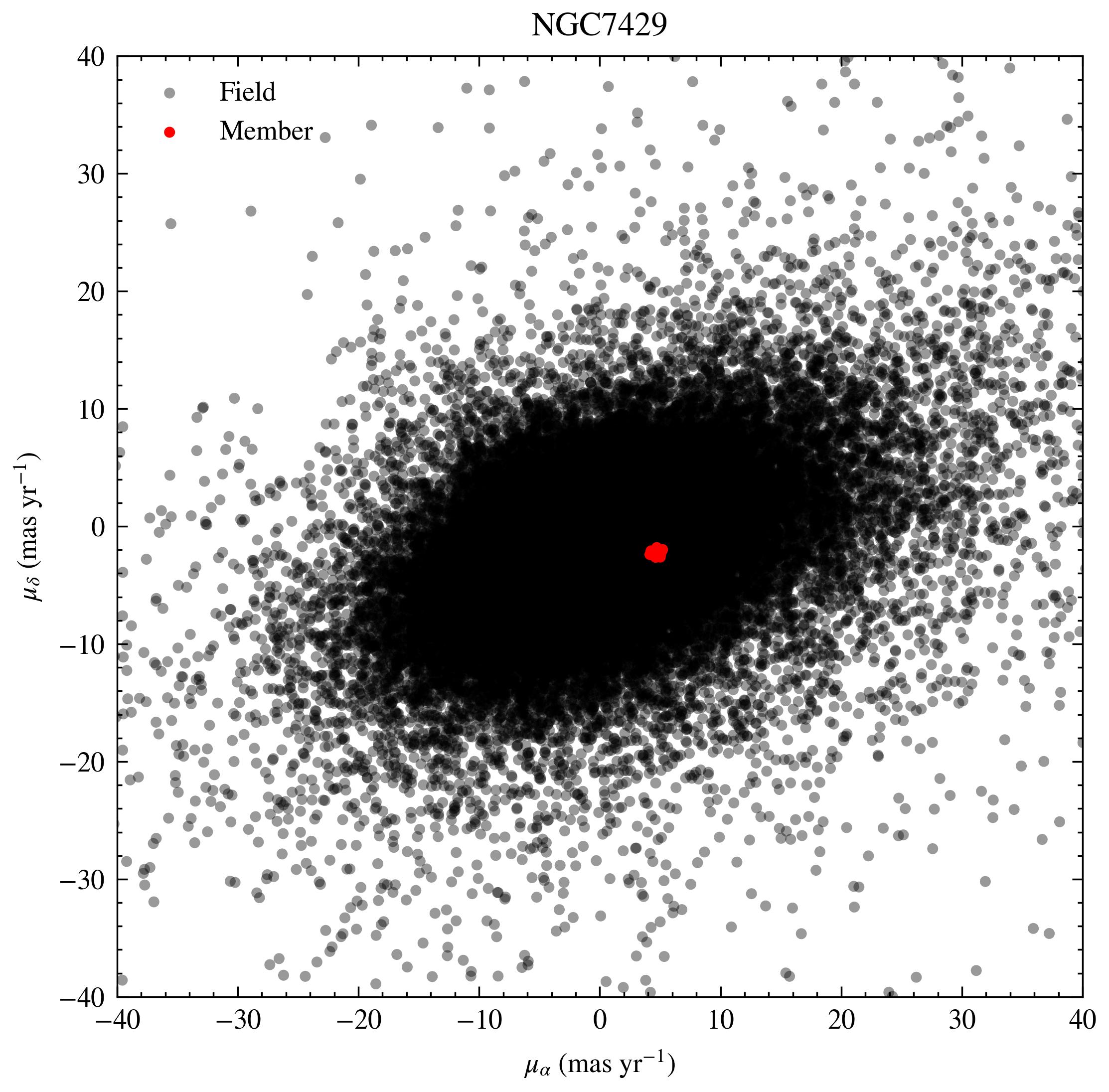} 
		
	\end{minipage}\hfill
	\begin{minipage}{0.3\textwidth}
		\includegraphics[width=\linewidth]{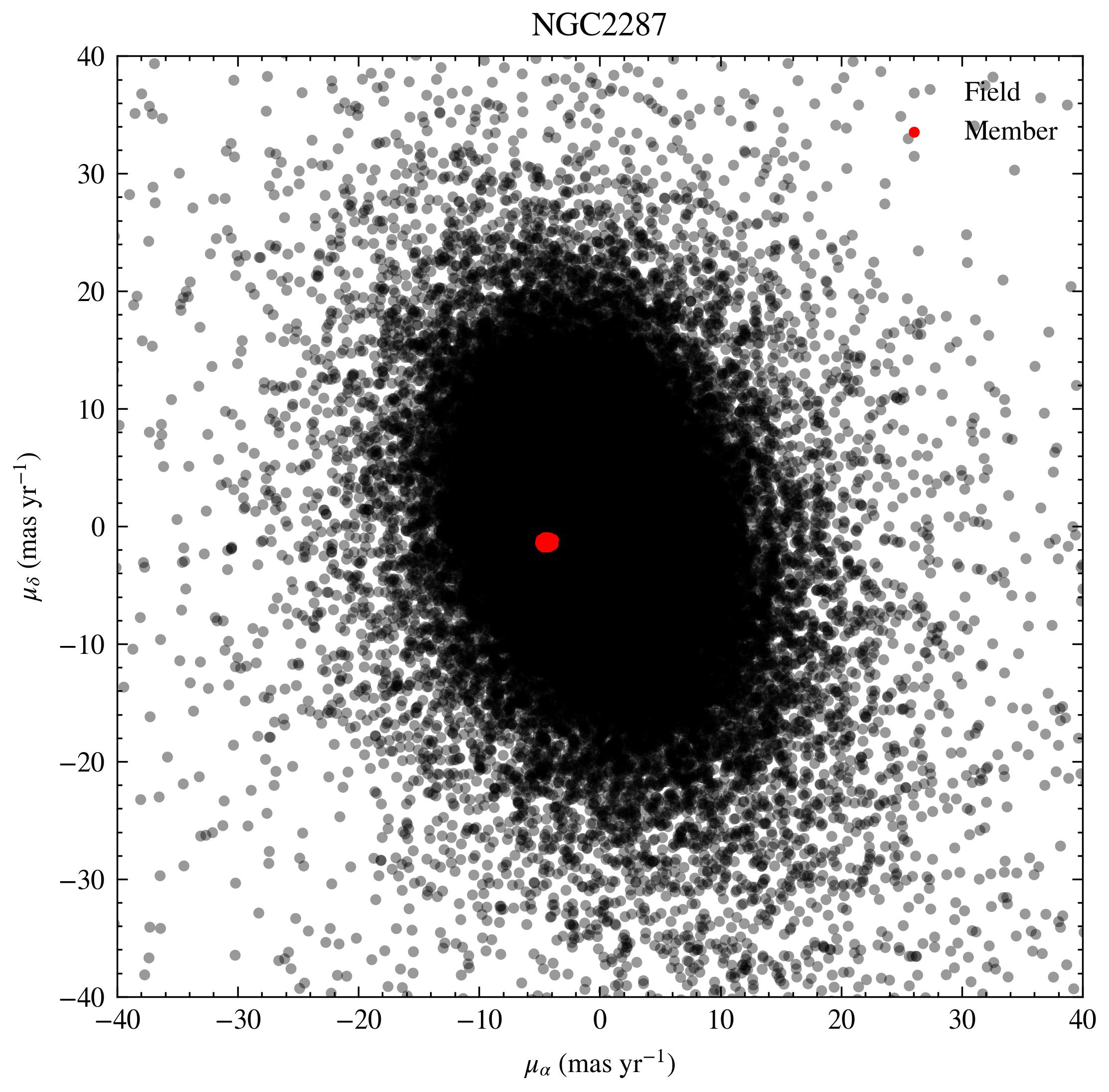} 

	\end{minipage}
	
	\caption{Proper motion diagrams illustrating the spatial distribution of cluster members (red dots) and field stars (black dots) within the field of view of six open clusters: NGC 2243, IC 4756, Melotte 20, NGC 2477, NGC 7429, and NGC 2287.}
	\label{fig:motioninfield2}
\end{figure}

\begin{figure}[H] 
	\centering
	\begin{minipage}{0.3\textwidth}
		\includegraphics[width=\linewidth]{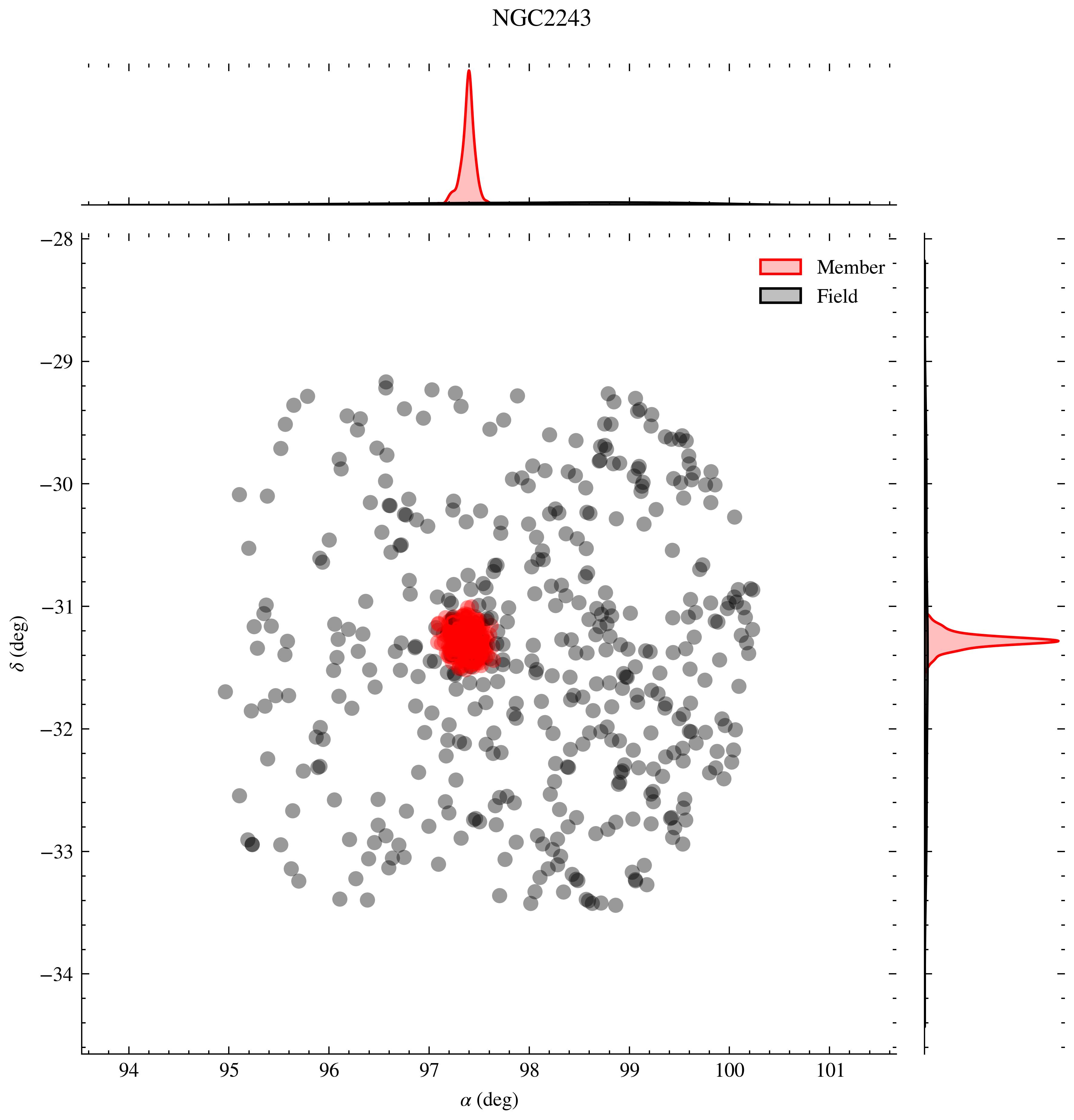} 
		
	\end{minipage}\hfill
	\begin{minipage}{0.3\textwidth}
		\includegraphics[width=\linewidth]{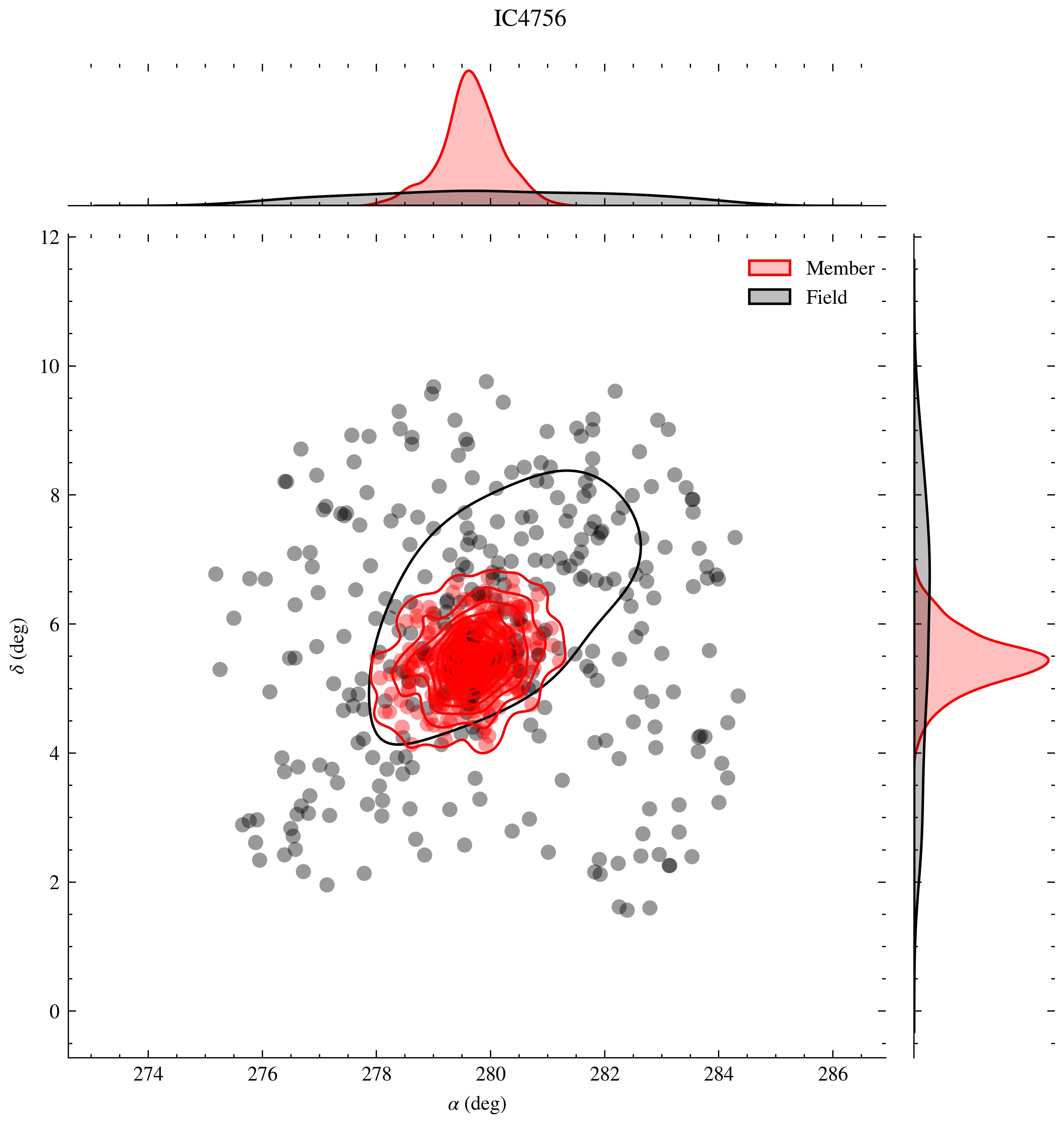} 
		
	\end{minipage}\hfill
	\begin{minipage}{0.3\textwidth}
		\includegraphics[width=\linewidth]{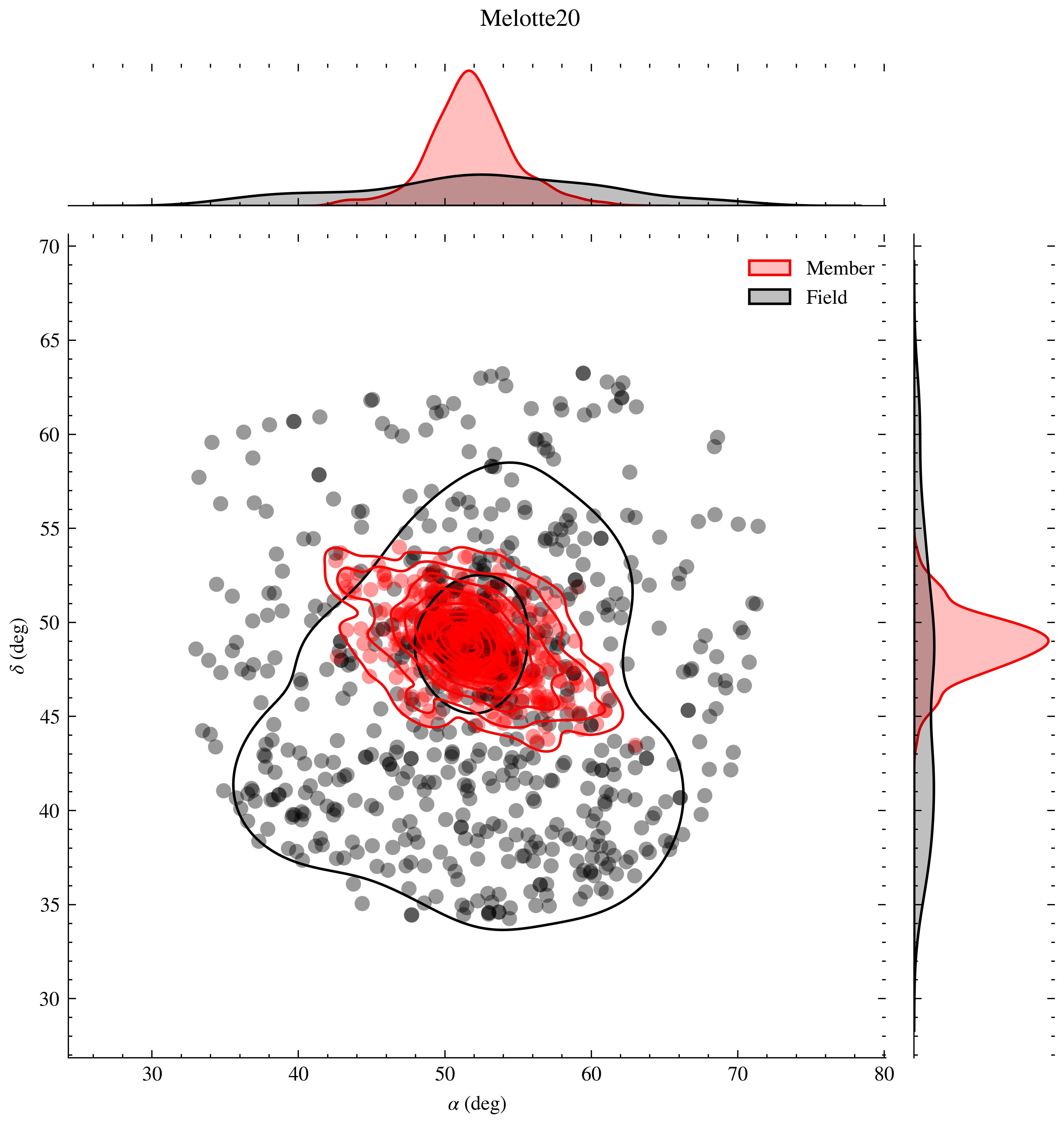} 
		
	\end{minipage}
	
	\vspace{0.5cm} 
	\begin{minipage}{0.3\textwidth}
		\includegraphics[width=\linewidth]{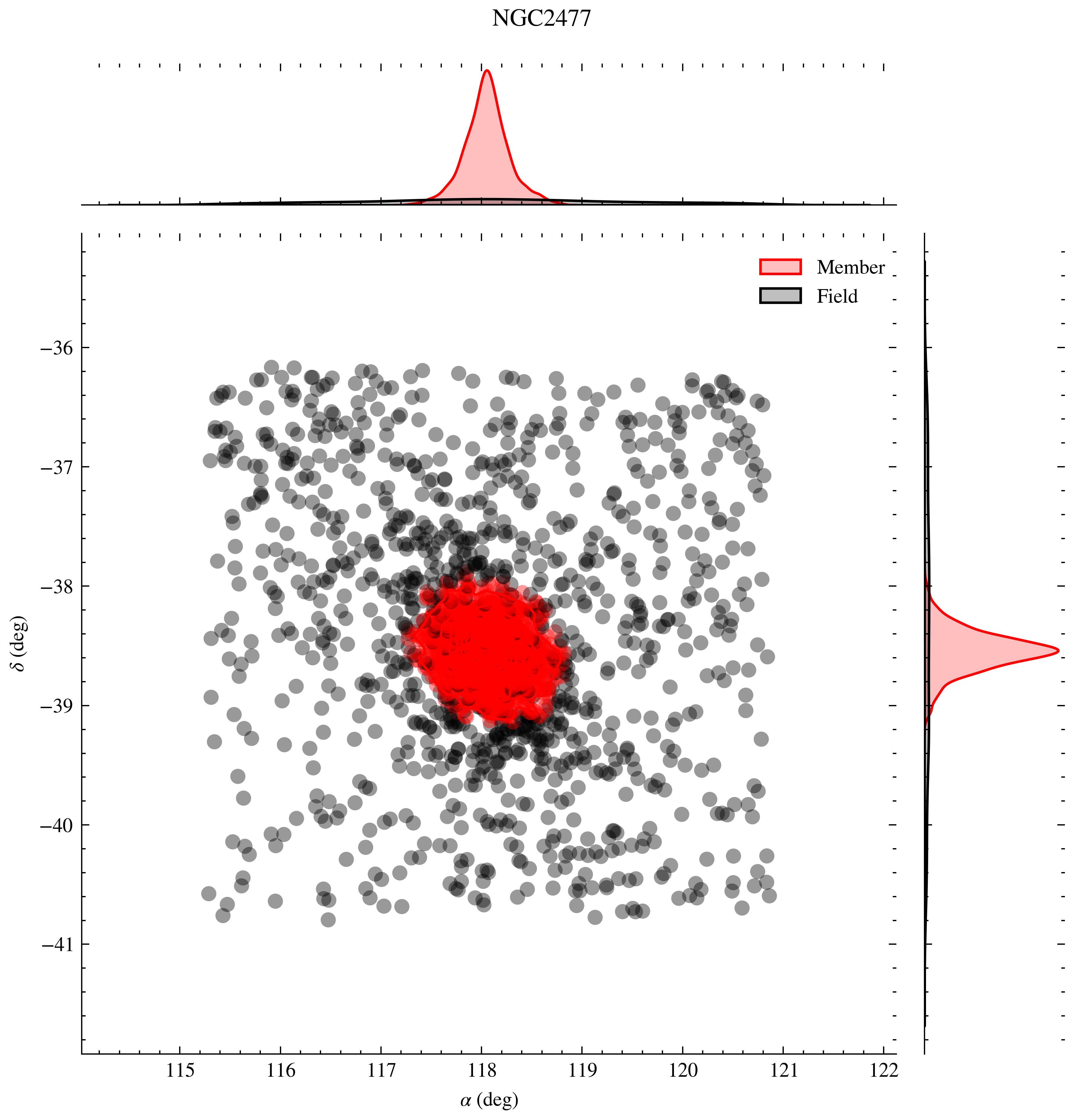} 
		
	\end{minipage}\hfill
	\begin{minipage}{0.3\textwidth}
		\includegraphics[width=\linewidth]{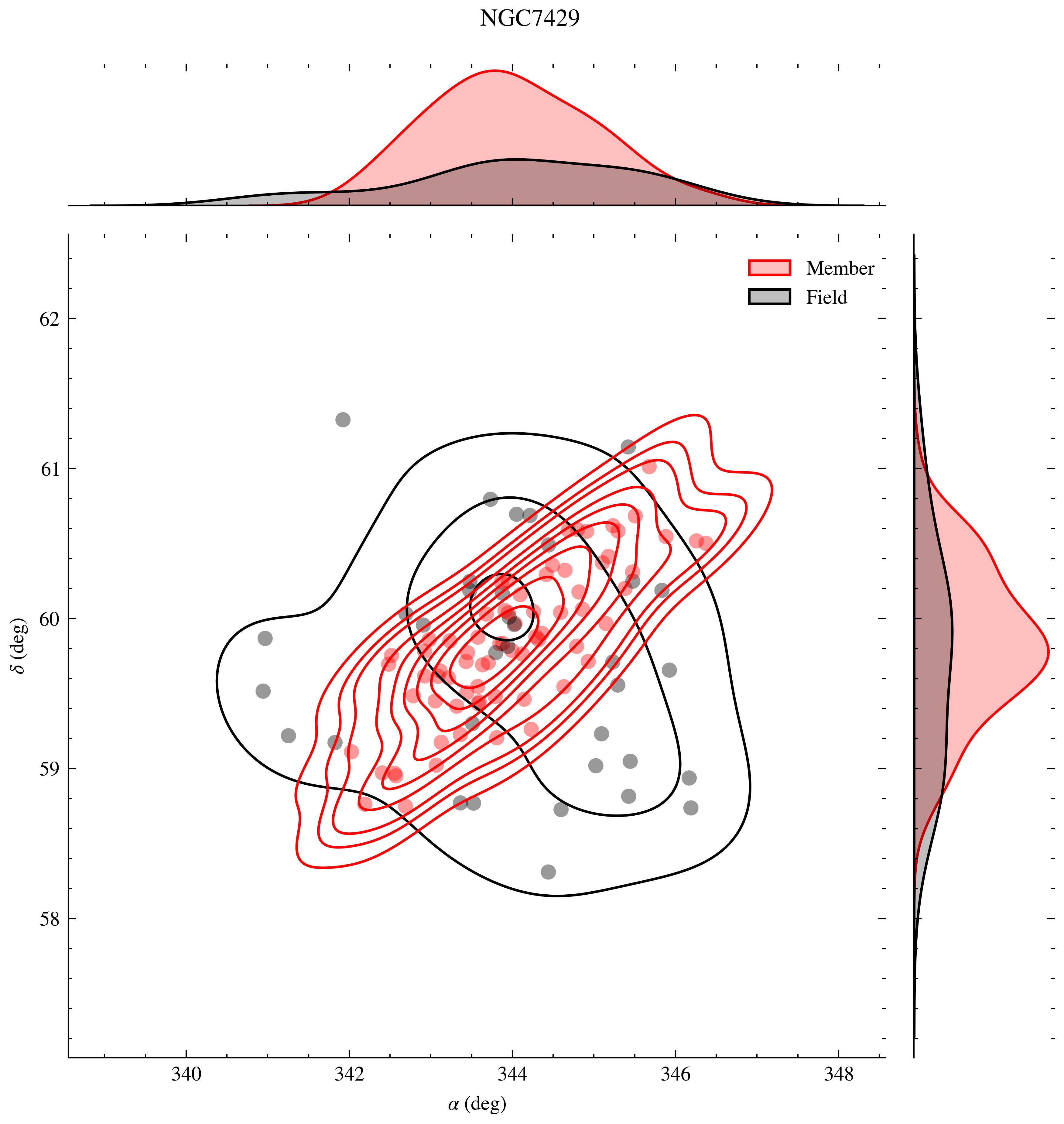} 
		
	\end{minipage}\hfill
	\begin{minipage}{0.3\textwidth}
		\includegraphics[width=\linewidth]{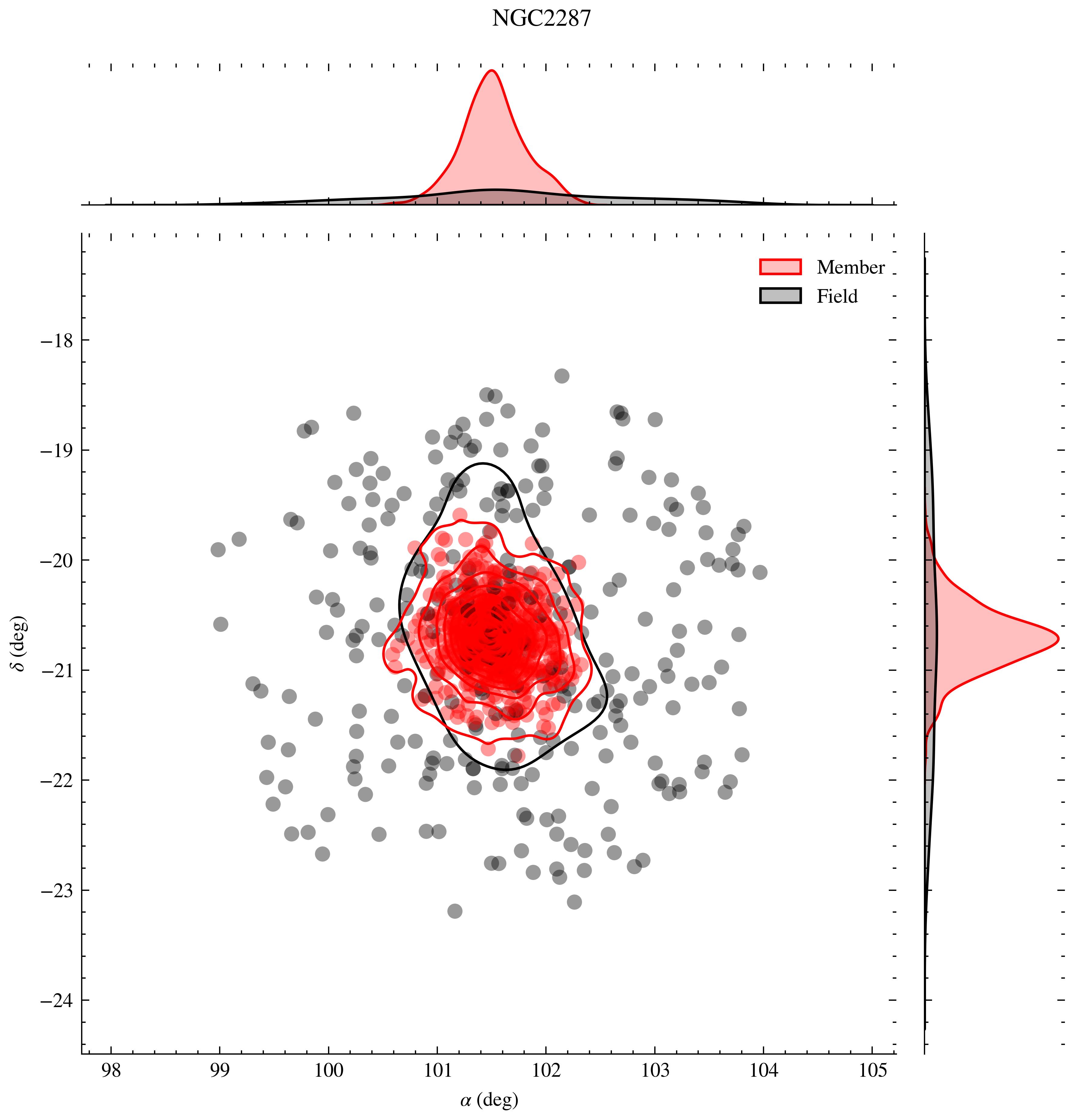} 

	\end{minipage}
	
	\caption{Kernel Density Estimator (KDE) plots illustrating the density distribution of cluster members (red dots) and field stars (black dots) within the field of view of six open clusters: NGC 2243, IC 4756, Melotte 20, NGC 2477, NGC 7429, and NGC 2287.}
	\label{fig:kde2}
	
\end{figure}

\begin{figure}[H] 
	\centering
	\begin{minipage}{0.3\textwidth}
		\includegraphics[width=\linewidth]{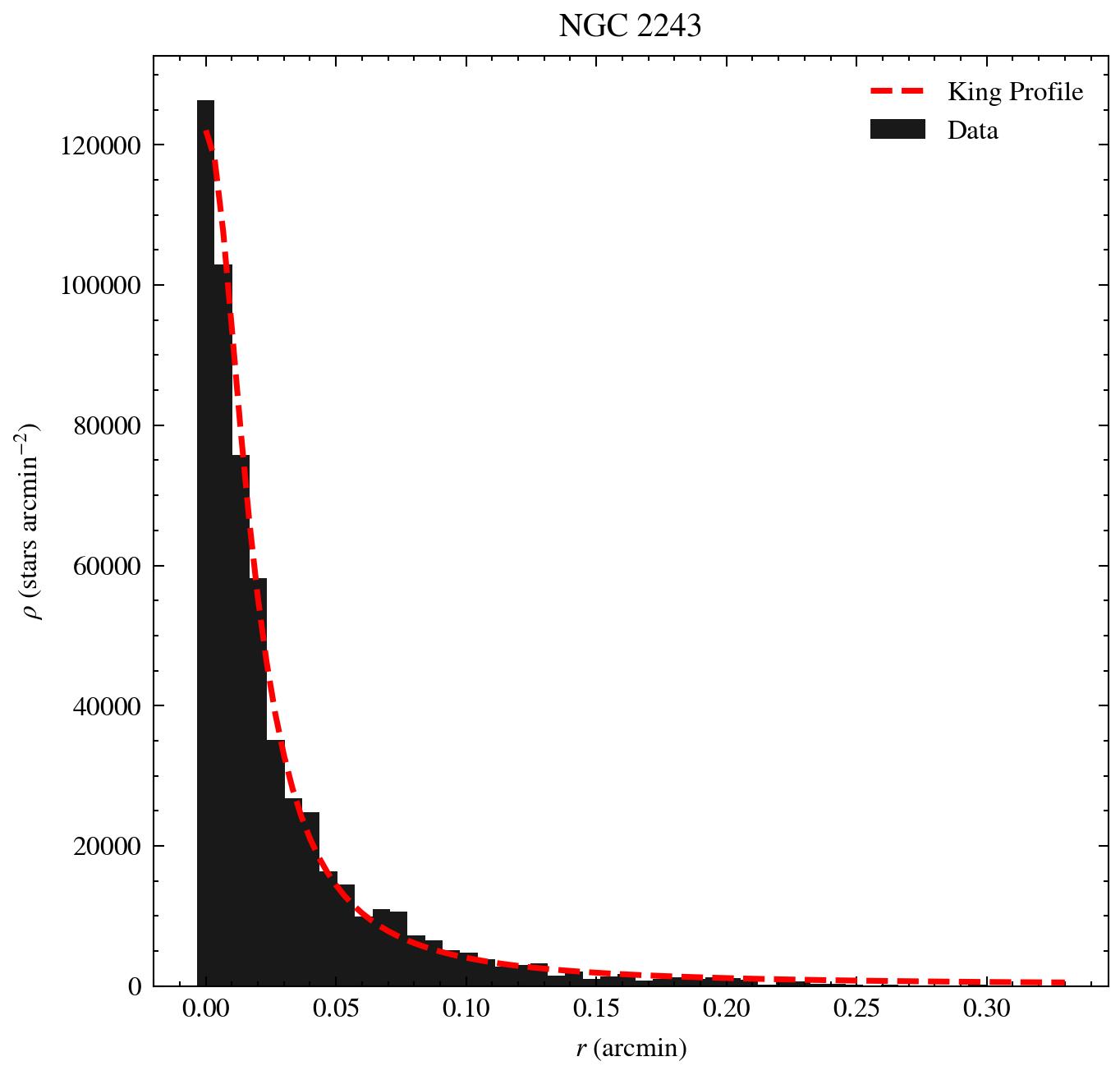} 
		
	\end{minipage}\hfill
	\begin{minipage}{0.3\textwidth}
		\includegraphics[width=\linewidth]{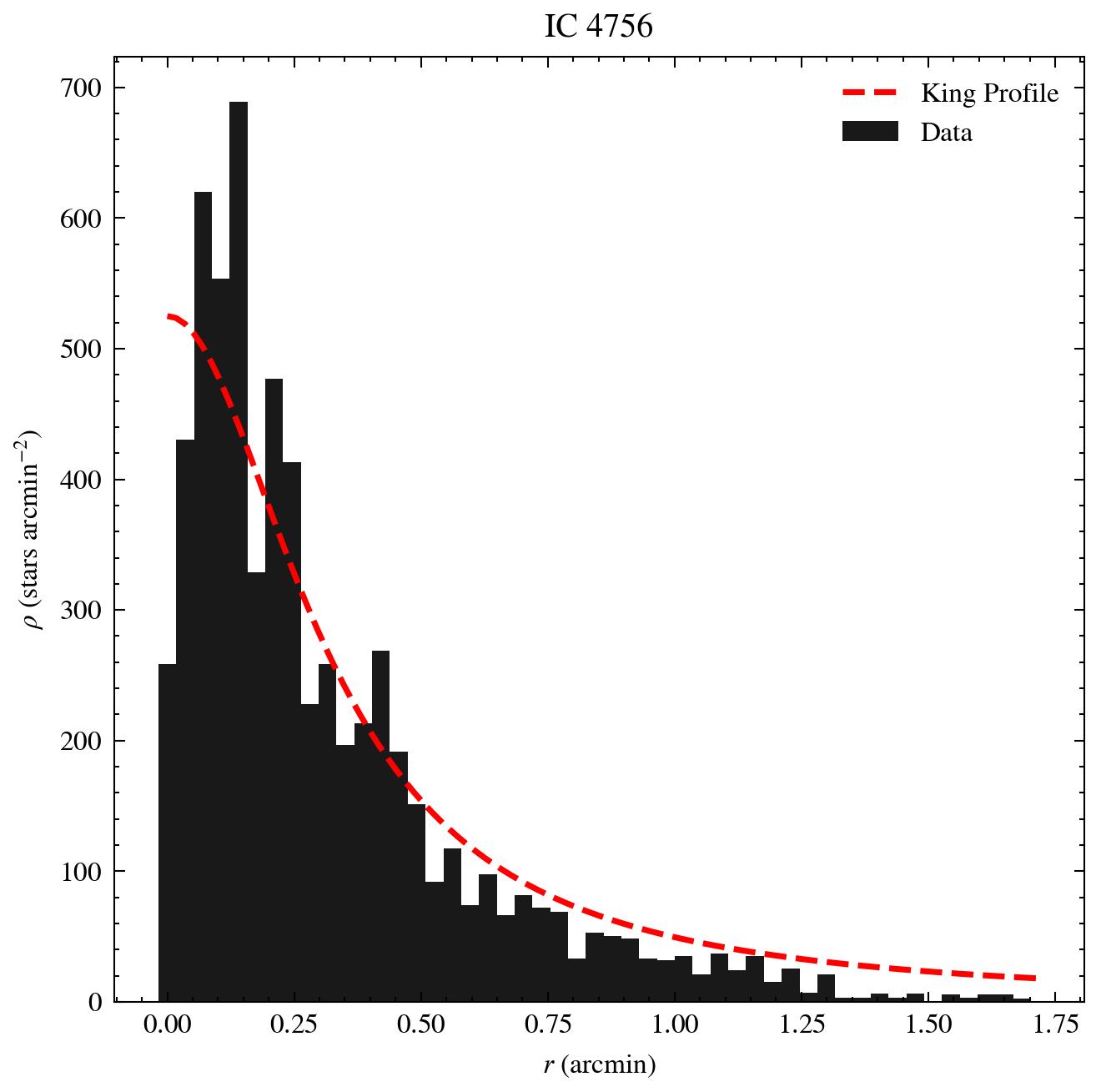} 
		
	\end{minipage}\hfill
	\begin{minipage}{0.3\textwidth}
		\includegraphics[width=\linewidth]{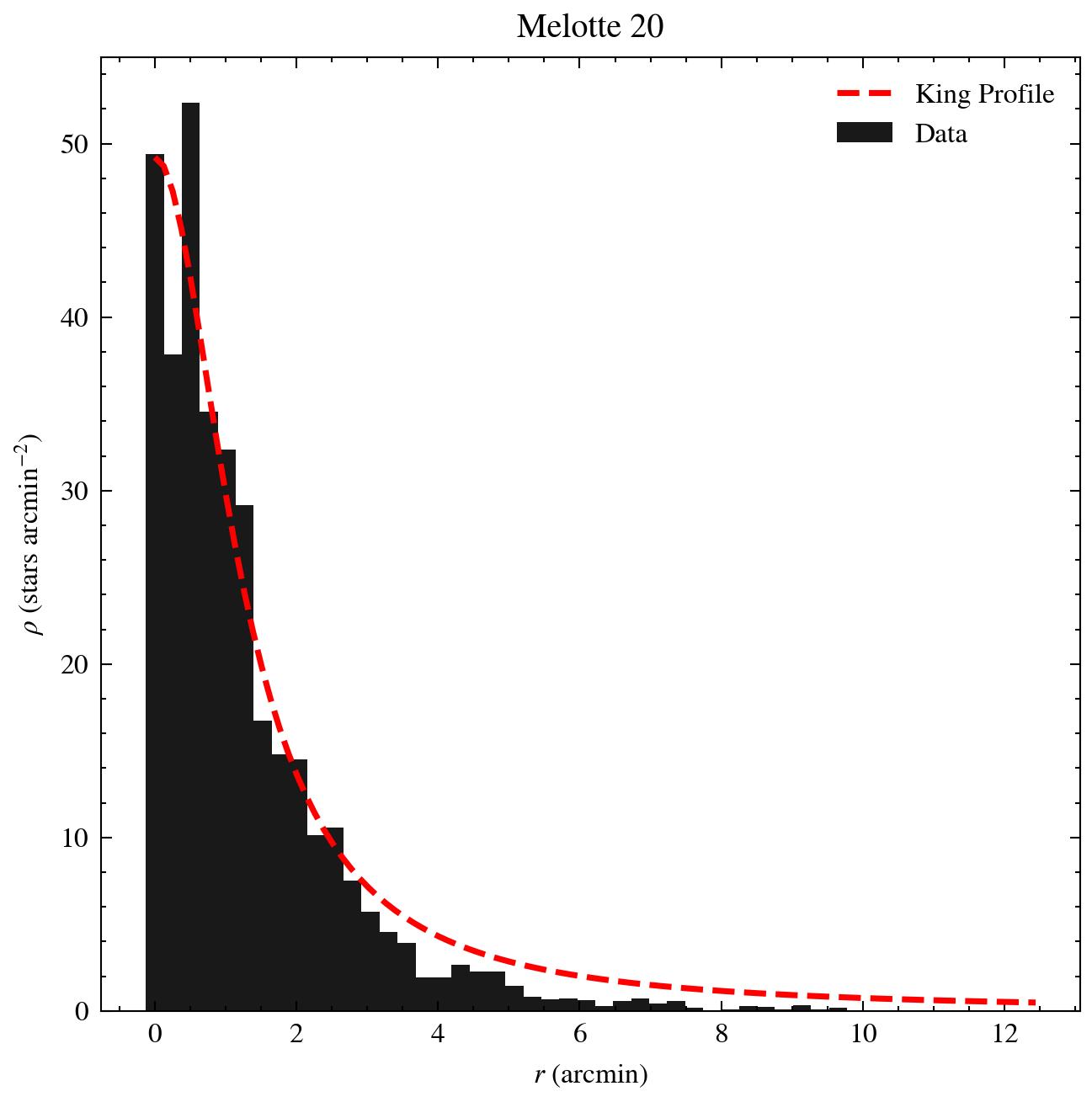} 
		
	\end{minipage}
	
	\vspace{0.5cm} 
	\begin{minipage}{0.3\textwidth}
		\includegraphics[width=\linewidth]{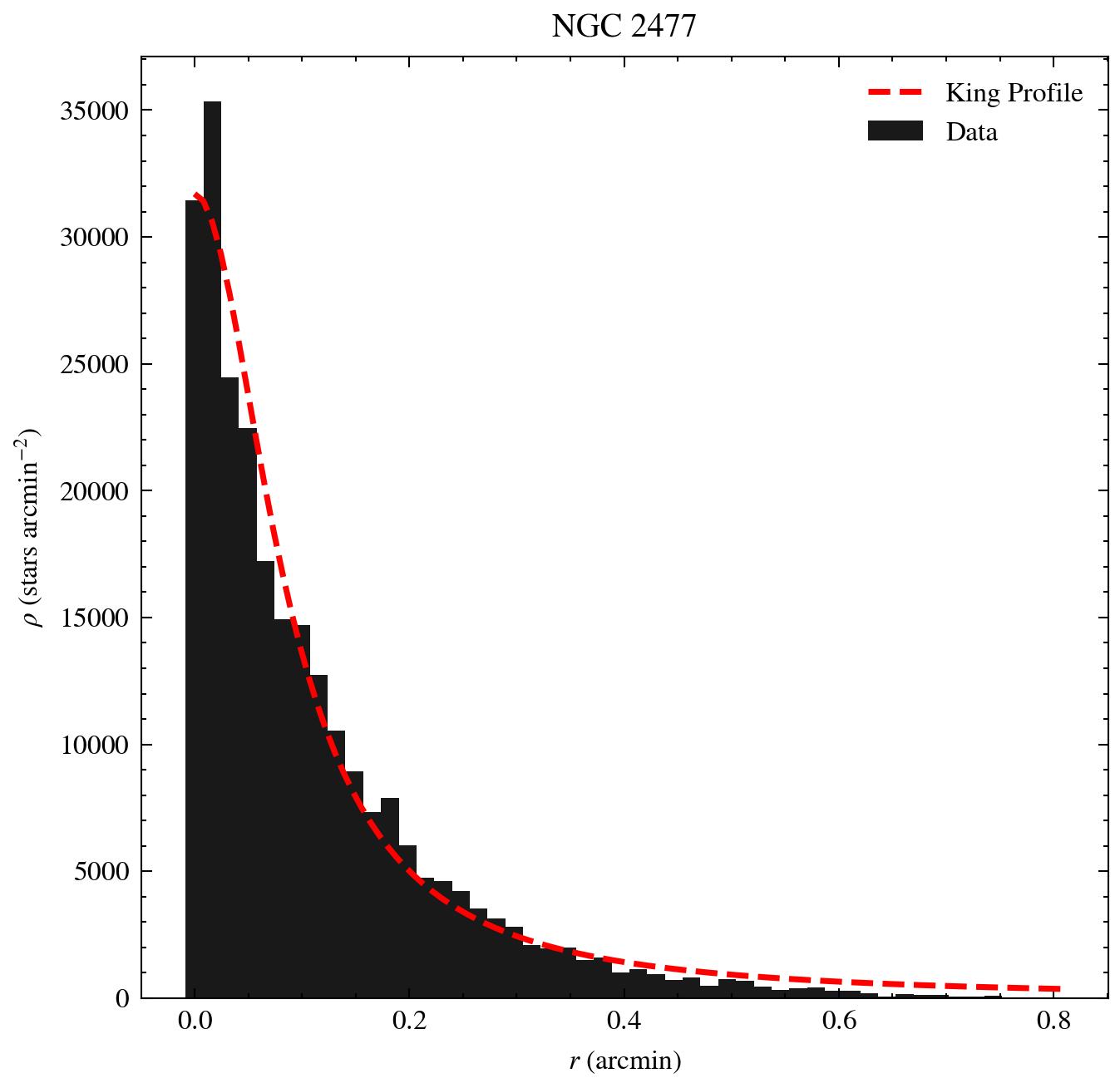} 
		
	\end{minipage}\hfill
	\begin{minipage}{0.3\textwidth}
		\includegraphics[width=\linewidth]{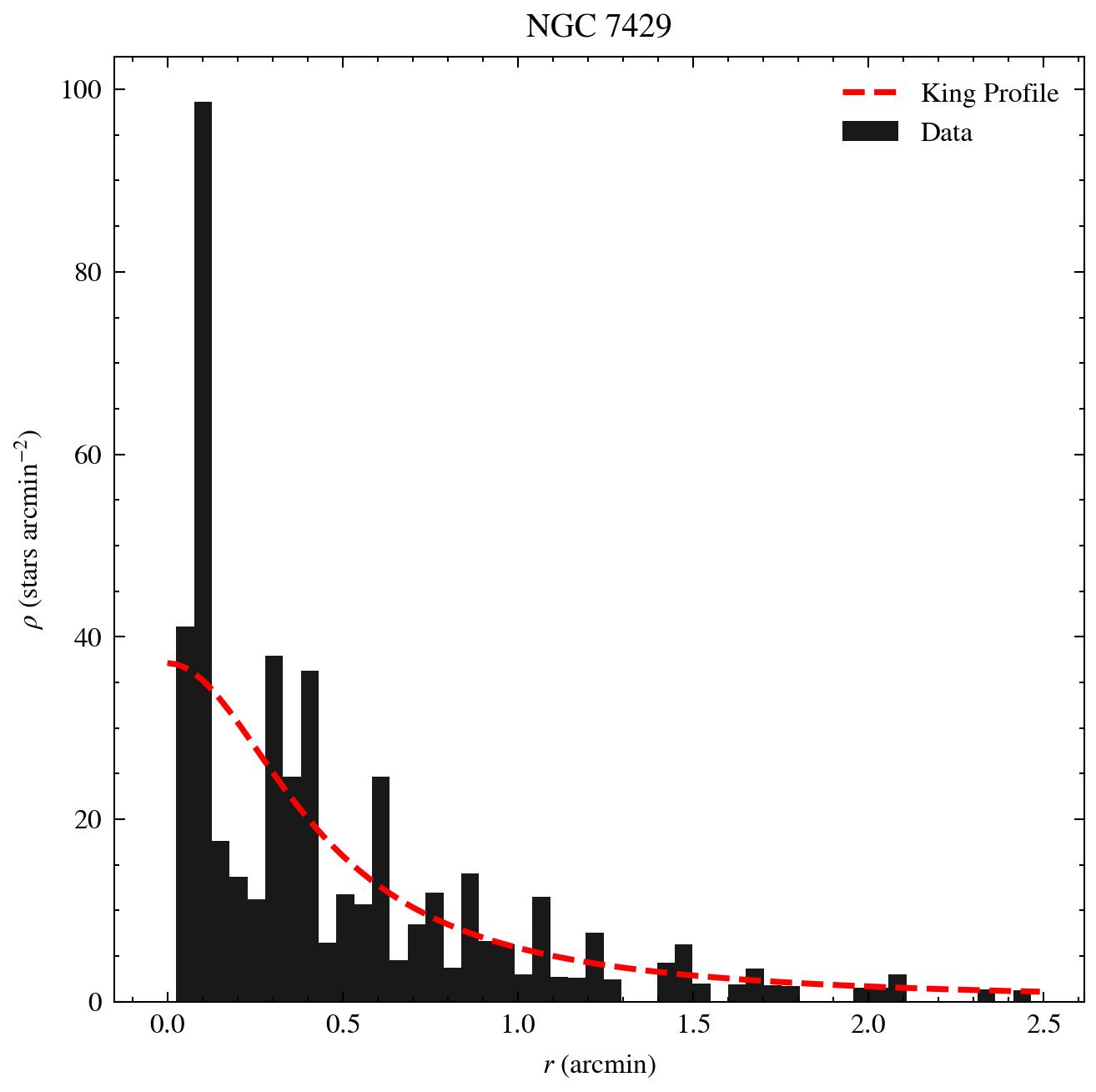} 
		
	\end{minipage}\hfill
	\begin{minipage}{0.3\textwidth}
		
		\includegraphics[width=\linewidth]{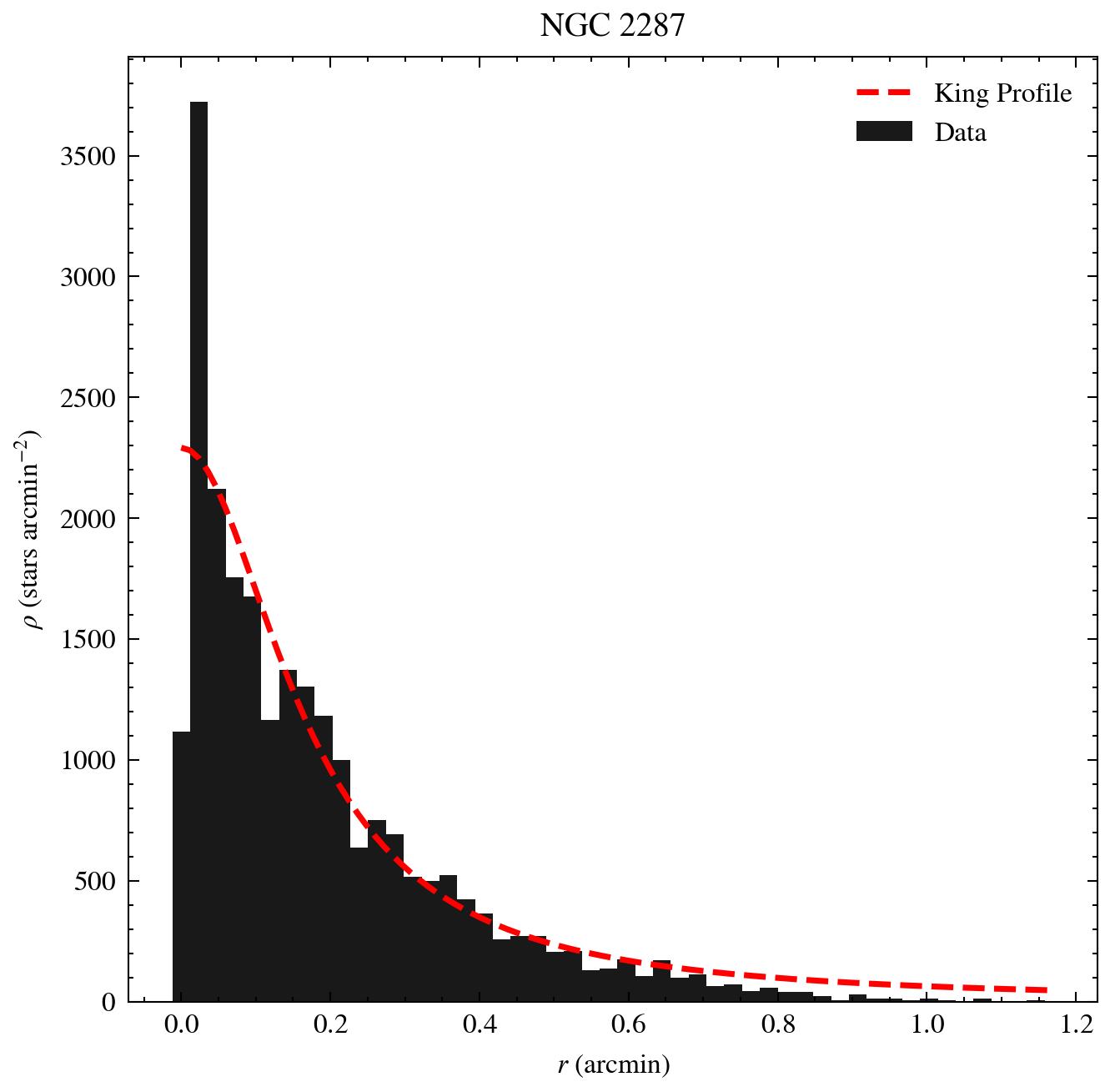} 

	\end{minipage}
	
	\caption{King profiles illustrating the radial density distribution of cluster members for six open clusters: NGC 2243, IC 4756, Melotte 20, NGC 2477, NGC 7429, and NGC 2287.}
	\label{fig:kingprofile2}
	
\end{figure}

\begin{figure}[H]
	\centering
	\begin{minipage}{0.9\textwidth}
		\includegraphics[width=\linewidth]{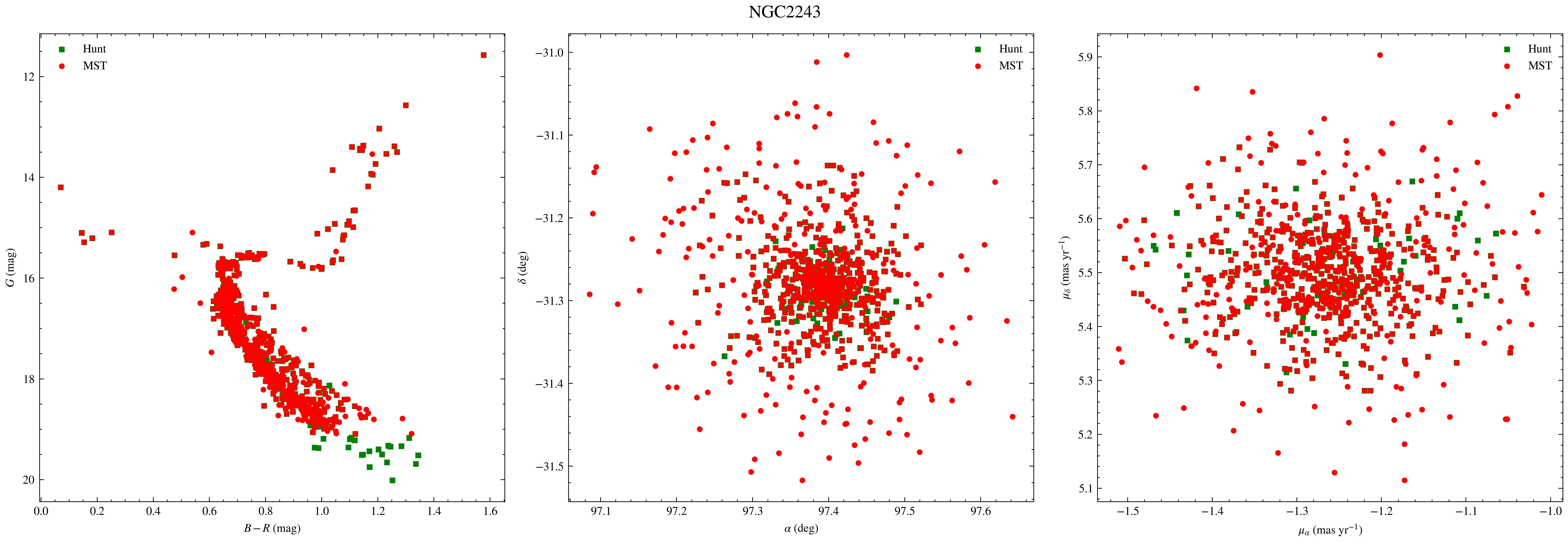} 
		
	\end{minipage}
	
	\vspace{0.5cm} 
	
	\begin{minipage}{0.9\textwidth}
		\includegraphics[width=\linewidth]{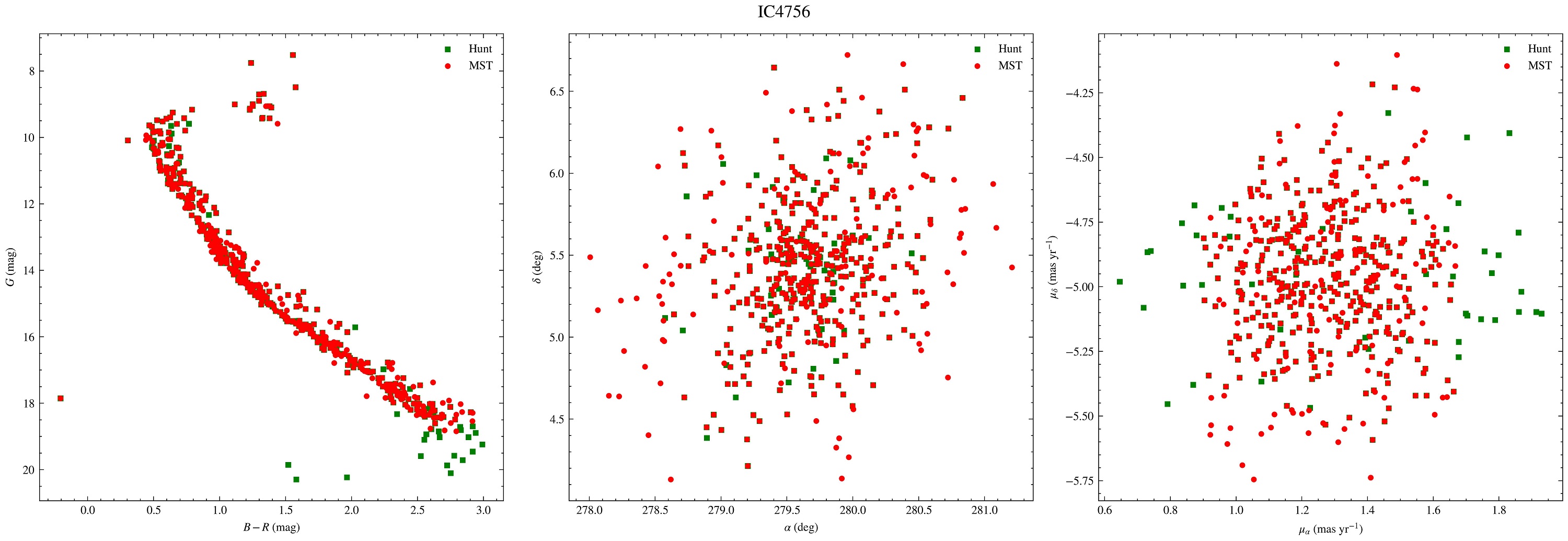} 
		
	\end{minipage}
	
	\vspace{0.5cm} 
	
	\begin{minipage}{0.9\textwidth}
		\includegraphics[width=\linewidth]{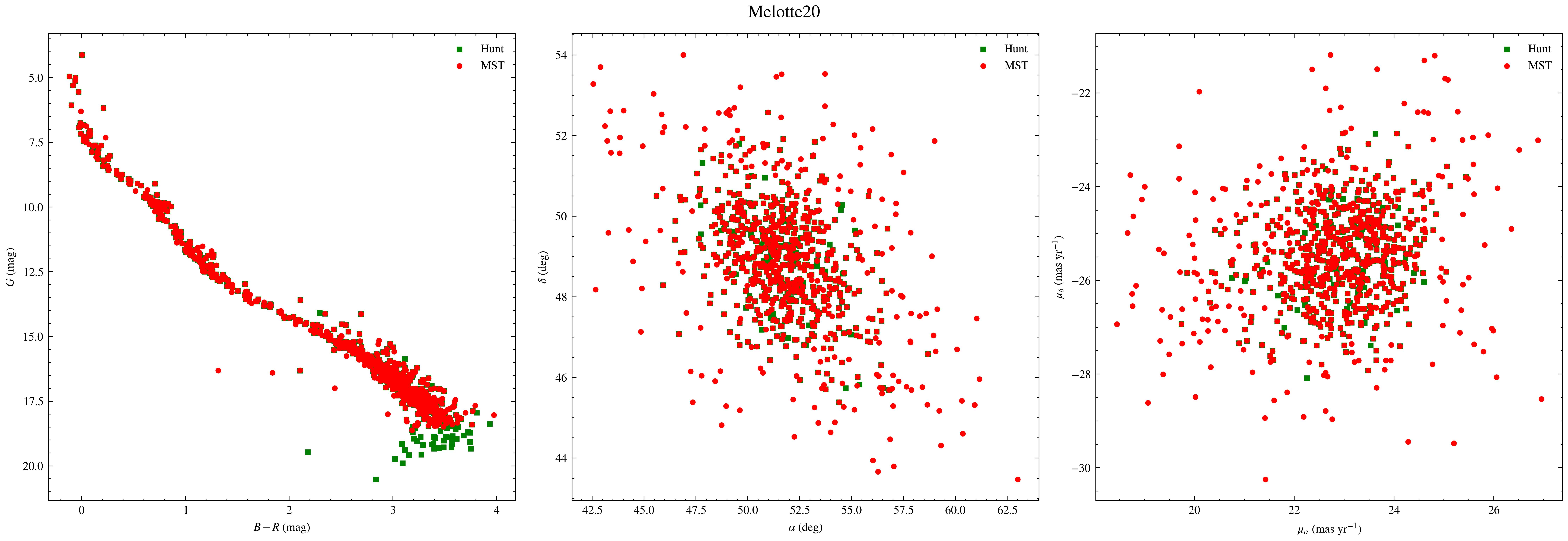} 
		
	\end{minipage}
	
	\caption{Comparison of cluster membership identification using the proposed Minimum Spanning Tree-Gaussian Mixture Model (MST-GMM) algorithm and the method of Hunt et al. (2024) for six open clusters: NGC 2243, IC 4756, Melotte 20. (Part 1)}
	\label{fig:comparison2_part1}
\end{figure}

\newpage

\begin{figure}[H]
	\centering
	
	\begin{minipage}{0.9\textwidth}
		\includegraphics[width=\linewidth]{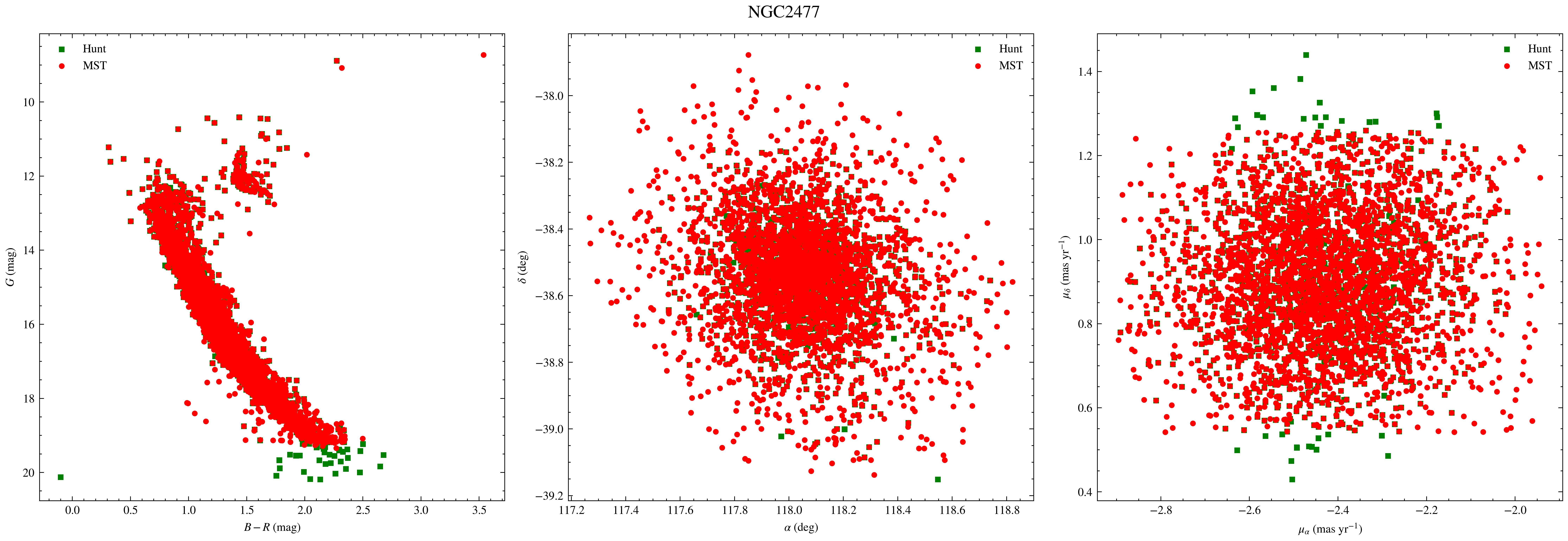} 
		
	\end{minipage}
	
	\vspace{0.5cm} 
	
	\begin{minipage}{0.9\textwidth}
		\includegraphics[width=\linewidth]{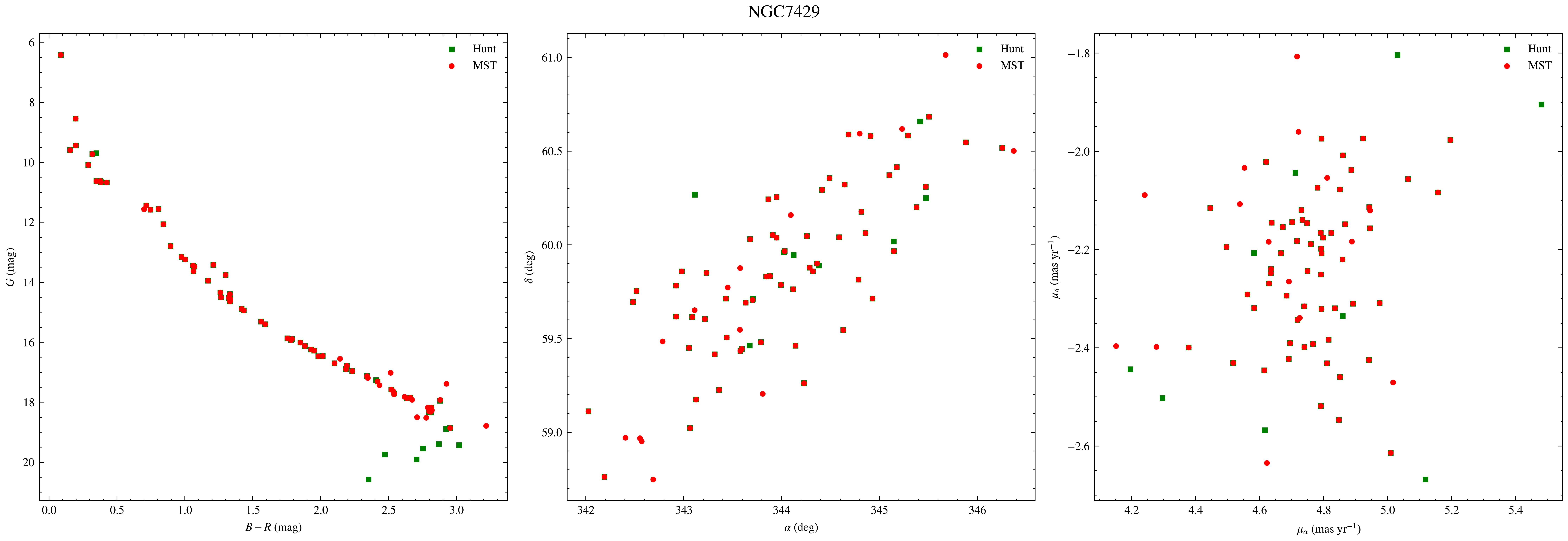} 
		
	\end{minipage}
	
	\vspace{0.5cm} 
	
	\begin{minipage}{0.9\textwidth}
		\includegraphics[width=\linewidth]{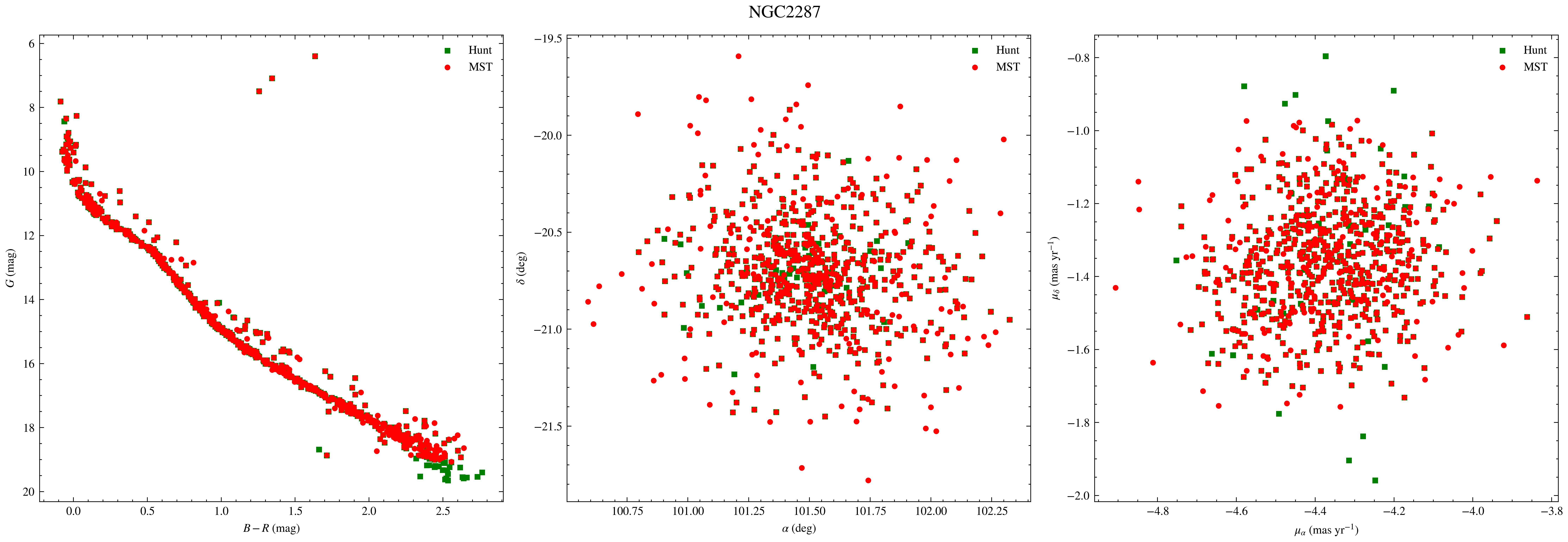} 
		
	\end{minipage}
	
	\caption{Comparison of cluster membership identification using the proposed Minimum Spanning Tree-Gaussian Mixture Model (MST-GMM) algorithm and the method of Hunt et al. (2024) for six open clusters: NGC 2477, NGC 7429, and NGC 2287. (Part 2)}
	\ContinuedFloat
	\caption*{(Continued from Figure \ref{fig:comparison2_part1})}
	
	\label{fig:comparison2_part2}
\end{figure}

\bibliographystyle{unsrtnat}
\bibliography{references}

\end{document}